%% file: beyondUniversality.tex
\documentclass[a4paper,11pt]{article}
\usepackage[T1]{fontenc}
\usepackage[utf8]{inputenc}
\usepackage[dvipsnames]{xcolor}
\usepackage{mathtools}
\usepackage{float}

\usepackage{amsmath}
\usepackage{jheppub}
\usepackage{amsthm}
\usepackage{amsmath}
\usepackage{amssymb}
\usepackage{amstext}
\usepackage{braket}
\usepackage{graphicx}
\usepackage{booktabs}
\usepackage{subfigure}
\usepackage{siunitx}
\usepackage{etoolbox}
\usepackage{array}
\usepackage{xspace}

\usepackage{tikz-feynman}
\usepackage{tikz}
\usetikzlibrary{arrows}
\usepackage[compat=1.1.0]{tikz-feynhand}

\newcolumntype{C}[1]{>{\centering\arraybackslash$}p{#1}<{$}}
\usepackage{slashed}
\usepackage[section]{placeins}
\usepackage{tabularx}
\PassOptionsToPackage{hyphens}{url}
\usepackage{hyperref}
\usepackage{makecell}	
\allowdisplaybreaks
\usepackage[normalem]{ulem}
\AtBeginEnvironment{array}{\def\arraystretch{1.3} \setlength{\arraycolsep}{8pt}}

\usepackage{etoolbox}
\makeatletter
\patchcmd{\tableofcontents}
  {%
    \@starttoc{toc}%
  }
  {%
    {\setlength{\parskip}{0pt}%
     \setlength{\itemsep}{0pt}%
     \@starttoc{toc}}%
  }
  {}{}
\makeatother

\xdefinecolor{darkred}{RGB}{153, 0, 0}
\xdefinecolor{goodgreen}{RGB}{0, 100, 0}
\xdefinecolor{tugreen}{RGB}{132, 184, 25}
\xdefinecolor{dOrange}{RGB}{251, 133, 0}
\xdefinecolor{plt_blue}{RGB}{0, 141, 169}
\xdefinecolor{plt_dark}{RGB}{0, 10, 78}
\xdefinecolor{plt_grey}{RGB}{111, 111, 111}

\usepackage{pifont}

\definecolor{laracolor}{RGB}{255, 73, 158}  
\definecolor{comcolor}{RGB}{77, 63, 186}

\newcommand{\sW}{\sin \theta_W}
\newcommand{\cW}{\cos \theta_W}
\newcommand{\szW}{\sin^2 \theta_W}

\newcommand{\aem}{\alpha_{\text{em}}}

\definecolor{lightgray}{gray}{0.91}

\preprint{CERN-TH-2025-203}

\title{Beyond Universality: Probing Lepton Flavor in the SMEFT}

\author[a]{Cornelius~Grunwald}
\author[b,a]{Gudrun~Hiller}
\author[a]{Kevin~Kr{\"o}ninger}
\author[a]{Lara~Nollen}

\affiliation[a]{TU Dortmund University, Department of Physics, Otto-Hahn-Str.4, D-44221 Dortmund, Germany}
\affiliation[b]{Theoretical Physics Department, CERN, 1211 Geneva 23, Switzerland}

\emailAdd{cornelius.grunwald@tu-dortmund.de}
\emailAdd{gudrun.hiller@cern.ch}
\emailAdd{kevin.kroeninger@tu-dortmund.de}
\emailAdd{lara.nollen@tu-dortmund.de}

  \abstract{
    We present a global analysis of lepton-flavor-specific operators in the Standard Model Effective Field Theory (SMEFT), combining data from collider and flavor physics experiments. We systematically explore various lepton-flavor scenarios, including flavor-specific, universal, and democratic patterns, while employing a minimal flavor violation (MFV) ansatz in the quark sector. We constrain a set of 17 dimension-six SMEFT operators using a Bayesian fitting approach. Our analysis yields stringent bounds on the Wilson coefficients, probing new physics scales up to $\mathcal{O}(1000)$\,TeV. The strongest constraints are obtained in lepton-flavor universal and lepton-flavor democratic scenarios, in particular for the electroweak dipole operators. We further provide posterior-based predictions for 
    $B \to K^{(*)} \nu \bar \nu$ and $B \to (\pi, \rho) \nu \bar \nu$ decays,
 highlighting their role as probes of the MFV hypothesis in upcoming Belle II measurements. Our results demonstrate the complementarity of collider and flavor observables, the impact of flavor assumptions on global SMEFT fits, and establish robust limits on  extensions of the Standard Model in the lepton sector.
}

\begin{document}
\renewcommand{\beforetochook}{\enlargethispage{2\baselineskip}}
\maketitle



\section{Introduction}
\label{sec:introduction}

The Standard Model Effective Field Theory (SMEFT) \cite{Buchmuller:1985jz,Grzadkowski:2010es} is a means to explore the physics beyond the electroweak scale in a 
systematic way.
Combining experimental information from low and high energies, such as rare  hadron decays and collider observables,
this multi-observable approach  fully exploits large data sets with complementary sensitivities to physics beyond the standard model (BSM);
it can  resolve  otherwise flat  directions in the parameter space  of new physics (NP) couplings,
eliminate  large cancellations between them, and thus  leads  in general to significantly improved results  relative  to single sector fits.
 This advantage stems not just from  more data hence better statistics,  but also from the fact that  different sectors are subject to  different theory and experimental systematics, leading to more control  in the global analysis. The increase in diagnostic power allows also to perform flavorful fits in which the fermions' generational structure is taken into account, or even probed, e.g., recently
\cite{Aoude:2020dwv,Bruggisser:2021duo,Greljo:2022jac,Grunwald:2023nli,Allwicher:2022gkm,Hiller:2025hpf}.

In this work we test the standard model (SM) in processes involving charged leptons and neutrinos. We employ the SMEFT and  work beyond lepton flavor universality (LU), performing fits in a lepton-flavor specific manner. 
We go beyond existing works by considering  ${\mathcal{O}}(170)$ decays of mesons, leptons, electroweak bosons, as well as Drell-Yan (DY) and electroweak 
observables.
In addition, we take into account the recent  contraints from the muons anomalous magnetic moment \cite{Aliberti:2025beg,Muong-2:2025xyk}, and  data from ATLAS on $tt \ell\ell$, with results available
for electrons and muons separately~\cite{ATLAS:2025yww}.
As processes which are used to extract the Cabibbo-Kobayashi-Maskawa (CKM)-matrix elements can be subjected to BSM physics we perform a joint fit to the CKM elements and SMEFT Wilson coefficients. 
To benefit from correlations between different quark generations we assume  minimal flavor violation (MFV)  in the quark sector, whose expansion we test  simultaneously in the global fit.

The paper is organized as follows.
The effective theory set up in SMEFT and the low energy effective theory (LEFT) is given in Sec.~\ref{sec:eft_setup}, together with the MFV ansatz and lepton flavor patterns considerend:
flavor-specific,  LU, and lepton flavor democracy (LD).
We present collider and flavor observables in Sec.~\ref{sec:collider_observables} and Sec.~\ref{sec:flavor_observables} respectively. 
The fit methodoloy and   a comparative discussion of SMEFT constraints  for different  operator types and
lepton flavor patterns  are given in Sec.~\ref{sec:global_analysis}.
The posterior distributions allow further to predict branching ratios of rare $B$-decays into dineutrinos. Results are presented in Sec.~\ref{sec:BKnunu}.
We conclude in Sec.~\ref{sec:conclusions}.
Numerical input is provided in App.~\ref{app:input_parameters}.
In App.~\ref{app:CKM} we give details on the CKM prefit to obtain priors for the full, global analysis.
Matching conditions from SMEFT onto LEFT are given in App.~\ref{app:matching_conditions}.
In App.~\ref{app:credible_intervals} we give the 90 \% credible intervals of the global fit.
Further predictions for $B \to (K,K^*, \pi, \rho) \nu \bar \nu$ decays are provided in App.~\ref{app:predictions_dineutrino}.

\section{Effective Field Theory Model}
\label{sec:eft_setup}

Effective field theories are a powerful tool to connect observables from various energy scales  in a joint, largely model-independent framework. We employ the 
SMEFT~\cite{Buchmuller:1985jz} to probe NP in lepton-flavor specific fits. The set up used in this work is given in Sec.~\ref{sec:smeft}.
We discuss flavor patterns in Sec.~\ref{sec:flavor} and the matching onto the LEFT in Sec.~\ref{sec:left}.

\subsection{SMEFT Setup \label{sec:smeft}}
The SMEFT Lagrangian can be written as
\begin{equation}
  \mathcal{L}_{\text{SMEFT}}\,=\, \mathcal{L}_{\text{SM}}\,+ \,\sum_{d=5}^{\infty}\,\sum_i \: \frac{{C_i^{(d)}}}{\Lambda^{d-4}}{{O}_i^{(d)}}  \,,
  \label{eqn:SMEFT_Lagrangian}
\end{equation}
where $\mathcal{L}_{\text{SM}}$  denotes the Lagrangian of the SM, $O_i^{(d)}$ are dimension-$d$ operators, and $C_i^{(d)}$  the corresponding Wilson coefficients. 
The scale $\Lambda$ denotes the scale of NP, assumed to be sufficiently above the electroweak scale. The operators are composed out of SM degrees of freedom, and respect
SM gauge and Poincare symmetries.
To cover leading effects we restrict our analysis to $d=6$ operators for which we employ the Warsaw basis~\cite{Grzadkowski:2010es}. 
As we  focus on leptonic NP, we exclude purely bosonic operators as well as operators involving only quark fields, such as quark-current and four-quark operators.
 In total, we consider 17 operators which are given in Tab.~\ref{tab:operators}.

\begin{table}[h]
  \renewcommand{\arraystretch}{1.5}
  \centering
  \begin{tabular}{|l l | l l | l l|}
    \hline
    \multicolumn{6}{|c|}{{Dipole}} \\
    \hline
    ${O}_{\underset{ij}{eB}}$ & $\bigl(\bar L_i \sigma^{\mu\nu} E_j \bigr) \varphi B_{\mu\nu}$ & 
    ${O}_{\underset{ij}{eW}}$ & $\bigl(\bar L_i \sigma^{\mu\nu} E_j \bigr) \tau^I \varphi W_{\mu\nu}^I$ & & \\
    \hline
    \multicolumn{6}{|c|}{{Higgs-Current}} \\
    \hline
    ${O}_{\underset{ij}{\varphi l}}^{(1)}$ & $\bigl(\varphi^{\dagger} i\overleftrightarrow D_{\mu} \varphi\bigr)\bigl(\bar L_i \gamma^{\mu} L_j\bigr)$ & 
    ${O}_{\underset{ij}{\varphi l}}^{(3)}$ & $\bigl(\varphi^{\dagger} i\overleftrightarrow D_{\mu}^I \varphi\bigr)\bigl(\bar L_i \tau^I \gamma^{\mu} L_j\bigr)$ &
    ${O}_{\underset{ij}{\varphi e}}$ & $\bigl(\varphi^{\dagger} i\overleftrightarrow D_{\mu} \varphi\bigr)\bigl(\bar E_i \gamma^{\mu} E_j\bigr)$ \\
    \hline
    \multicolumn{6}{|c|}{{Four-Lepton}} \\
    \hline
    ${O}_{\underset{ijkm}{ll}}$ & $\bigl(\bar L_{i} \gamma_{\mu} L_{j}\bigr)\bigl(\bar L_{k} \gamma^{\mu} L_{m}\bigr)$ &
    ${O}_{\underset{ijkm}{ee}}$ & $\bigl(\bar E_{i} \gamma_{\mu} E_{j}\bigr)\bigl(\bar E_{k} \gamma^{\mu} E_{m}\bigr)$ &
    ${O}_{\underset{ijkm}{le}}$ & $\bigl(\bar L_{i} \gamma_{\mu} L_{j}\bigr)\bigl(\bar E_{k} \gamma^{\mu} E_{m}\bigr)$ \\
    \hline
    \multicolumn{6}{|c|}{{Semileptonic Four-Fermion}} \\
    \hline
    ${O}_{\underset{ijkm}{lq}}^{(1)}$ & $\bigl(\bar L_i \gamma_{\mu} L_j \bigr)\bigl(\bar Q_k \gamma^{\mu} Q_m \bigr)$ &
    ${O}_{\underset{ijkm}{lq}}^{(3)}$ & $\bigl(\bar l_i \gamma_{\mu} \tau^I l_j \bigr)\bigl(\bar Q_k \gamma^{\mu} \tau^I Q_m \bigr)$ & ${O}_{\underset{ijkm}{qe}}$ & $\bigl(\bar Q_i \gamma_{\mu} Q_j \bigr)\bigl(\bar E_k \gamma^{\mu} E_m \bigr)$ \\
    ${O}_{\underset{ijkm}{eu}}$ & $\bigl(\bar E_i \gamma_{\mu} E_j \bigr)\bigl(\bar U_k \gamma^{\mu} U_m \bigr)$ &
    ${O}_{\underset{ijkm}{ed}}$ & $\bigl(\bar E_i \gamma_{\mu} E_j \bigr)\bigl(\bar D_k \gamma^{\mu} D_m \bigr)$ & ${O}_{\underset{ijkm}{lequ}}^{(1)}$ & $\bigl(\bar l^p_i E_j \bigr) \epsilon_{pr} \bigl(\bar q^r_k U_m \bigr)$ \\
    ${O}_{\underset{ijkm}{lu}}$ & $\bigl(\bar L_i \gamma_{\mu} L_j \bigr)\bigl(\bar U_k \gamma^{\mu} U_m \bigr)$ &
    ${O}_{\underset{ijkm}{ld}}$ & $\bigl(\bar L_i \gamma_{\mu} L_j \bigr)\bigl(\bar D_k \gamma^{\mu} D_m \bigr)$ & ${O}_{\underset{ijkm}{lequ}}^{(3)}$ & $\bigl(\bar l^p_i \sigma_{\mu\nu} E_j \bigr) \epsilon_{pr} \bigl(\bar q^r_k \sigma^{\mu\nu} U_m \bigr)$ \\
    \hline
  \end{tabular}
  \caption{Dimension-six SMEFT Wilson coefficients considered in this work.}
  \label{tab:operators}
\end{table}

We denote the left-handed lepton doublets by $L$ and the right-handed charged lepton singlets by $E$. The Higgs doublet is written as $\varphi$. Quark doublets are denoted by $Q$, while the right-handed up-type and down-type quarks are represented by $U$ and $D$, respectively. Generation indices are labeled by $i, j, k, m = 1, 2, 3$, and isospin indices by $p, r = 1, 2$. The Pauli matrices are denoted by $\tau^I$, and the antisymmetric tensor by $\epsilon_{pr}$, with $\epsilon_{12} = +1$. The gauge field strength tensors of the $U(1)$ and $SU(2)$ gauge groups are denoted by $B_{\mu\nu}$ and $W_{\mu\nu}^I$, respectively. The covariant derivative is written as $D_\mu$.

Given that our analysis involves collider processes at high energies, we set $\Lambda = 10\,\text{TeV}$ to ensure the validity of the EFT expansion. We further assume all Wilson coefficients to be real, thereby excluding additional sources of CP violation within the SMEFT framework.
To evolve the Wilson coefficients from the scale $\Lambda = 10,\text{TeV}$ to the energy scale relevant for our observables, we use the one-loop SMEFT renormalization group equations (RGEs)~\cite{Jenkins:2013zja, Alonso:2013hga, Jenkins:2013wua}, which we solve numerically with the Python package \texttt{Wilson}~\cite{Aebischer:2018bkb}.
In the following we use  rescaled  Wilson coefficients 
\begin{equation} \label{eq:scaling}
\tilde C =  \frac{v^2}{\Lambda^2}  C \, , 
\end{equation}
where $v \simeq 246 \, \text{GeV}$ denotes the Higgs vacuum expectation value. This absorbs the explicit $\Lambda$-dependence 
(\ref{eqn:SMEFT_Lagrangian})  into the definition of the Wilson coefficients $\tilde C$, making them comparable across different choices of $\Lambda$, up to  residual 
effects from   renormalization group running. 
We expand observables  in inverse powers of the NP scale and keep terms including those of the order $\Lambda^{-4}$.

For the electroweak input parameters, we adopt the ${\alpha_{\text{em}}, m_W, m_Z}$ input scheme. With this choice and given the set of operators listed in Tab.\ref{tab:operators}, the input parameters remain unaffected by SMEFT-induced shifts. For  details  see Appendix~\ref{app:input_parameters}. Since the extraction of CKM matrix elements can be influenced by  BSM physics, we treat the CKM parameters as nuisance parameters in the fit following Refs.~\cite{Descotes-Genon:2018foz}. Further details are  provided in Appendix~\ref{app:CKM}.

\subsection{Flavor Patterns  \label{sec:flavor}}

In its most general form, the SMEFT contains 2499 free parameters at dimension-six after accounting for all generation indices~\cite{Grzadkowski:2010es}. This number is far too large to be fully constrained by current experimental data, making additional assumptions about the flavor structure necessary.
In this analysis, we adopt the MFV framework in the quark-flavor sector to reduce the number of free parameters. In MFV, the Yukawa couplings $Y_{u,d}$ of the up-type and down-type quarks are the sole sources of flavor violation, treated as spurions that break the $\text{U}(3)^3$ quark flavor symmetry of the SM gauge sector  \cite{DAmbrosio:2002vsn}.
We include terms up to $\mathcal{O}(Y_i^2)$ in the MFV expansion and approximate  $Y_d \sim 0$ and $Y_u \sim \text{diag}(0, 0, y_t)$, that is, from all the quark Yukawas we only keep the one of the 
top.~\footnote{BSM effects can break the relation between fermion masses and their Yukawa couplings. While only upper limits  exist on the latter 
for the first generation and  the strange quark at the level of $\mathcal{O}(10^2)$ above their SM values, the others are more or less well constrained, and model-independently
$y_{f\neq t } \ll 1$, except for the top, for which holds $y_t \simeq 1$.}
\newpage
The quark bilinears can then be written as
\begin{equation}
  \begin{alignedat}{2}
    \bar{Q} Q  &: \quad C_{ij} = a_1 \delta_{ij} + a_2\, y_t^2\, \delta_{i3} \delta_{j3}\,, \qquad &
    \bar{U} U  &: \quad C_{ij} = b_1 \delta_{ij} + b_2\, y_t^2\, \delta_{i3} \delta_{j3}\,, \\
    \bar{Q} U  &: \quad C_{ij} = \left( c_1 + c_2\, y_t^2 \right) y_t\, \delta_{i3} \delta_{j3}\,, \qquad &
    \bar{D} D &: \quad C_{ij} = e_1 \delta_{ij}\,, \\
    \bar{Q} D &: \quad C_{ij} = 0\,, \qquad &
    \bar{U}  D  &: \quad C_{ij} = 0\,.
  \end{alignedat}
  \label{eqn:MFV_expansion}
  \end{equation}
Since we neglect the small down-type Yukawas chirality-flipping operators with a right-handed down-type quark i.e. $\bar{Q} D$ and right-handed charged currents $\bar{U} D$ vanish.
The up-type scalar and tensor operators ${O}_{{lequ}}^{(1)}$ and ${O}_{{lequ}}^{(3)}$ involving $\bar{Q} U$, on the other hand, couple only to third-generation quarks in this framework, leading to the absence of chirality-flipping interactions involving first- or second-generation quarks.

To confront the SMEFT with experimental data, the operators defined in the gauge basis must be rotated to the mass basis. In the MFV framework, this rotation yields identical results regardless of whether it is performed in the up- or down-type quark sector~\cite{Grunwald:2023nli}. In the mass basis, the parameterization of the quark bilinears reads
\begin{equation}
    \begin{alignedat}{2}
        &\bar q_L q_L :\quad &C^{\prime, \bar u u }_{ij} &= a_1 \delta_{ij} + a_2\, y_t^2\, \delta_{i3} \delta_{j3} \,,\quad C^{\prime, \bar d d}_{ij} = a_1 \delta_{ij} + a_2\, y_t^2\, V_{ti}^* V_{tj}  \,, \\
        & &C^{\prime, \bar u d }_{ij} &= a_1 V_{ij} + a_2\, y_t^2\, \delta_{i3} V_{tj} \,,\quad C^{\prime, \bar d u}_{ij} = a_1 V_{ji}^* + a_2\, y_t^2\, V_{ti}^* \delta_{j3}  \,, \\
        &\bar q_L u_R :\quad &C^{\prime, \bar u u}_{ij} &= (c_1 + c_2\, y_t^2)\, y_t\, \delta_{i3} \delta_{j3} \,,\quad C^{\prime, \bar d u}_{ij} = (c_1 + c_2\, y_t^2)\, y_t\, V_{ti}^* \delta_{j3} \,, \\
        &\bar u_R u_R : \quad &C^{\prime, \bar u u}_{ij} &= b_1 \delta_{ij} + b_2\, y_t^2\, \delta_{i3} \delta_{j3} \,, \\
        &\bar d_R d_R :\quad &C^{\prime, \bar d d}_{ij} &= b_1 \delta_{ij} \,, \\
    \end{alignedat}
    \label{eqn:MFV_mass_basis}
\end{equation}
where $V_{ij}$ are the CKM matrix elements. In this approach, down-type flavor-changing neutral currents (FCNCs) arise at second order in the MFV expansion for operators involving two left-handed quark doublets, and are proportional to the CKM factor $V_{ti}^* V_{tj}$. In contrast, up-type FCNCs are absent due to  $Y_d \sim 0$.
For further details on the MFV expansion in the SMEFT context, see e.g. Refs.~\cite{Greljo:2022cah, Grunwald:2023nli}.

To test MFV in the fit we follow Ref.~\cite{Grunwald:2023nli} and  define the following degrees of freedom 
\begin{equation}
  \begin{array}{l l l}
    \tilde C_{\bar q q} = \frac{v^2}{\Lambda^2} a_1 \,, \quad & \tilde C_{\bar u u} = \frac{v^2}{\Lambda^2} b_1 \,, \quad & 
    \tilde C_{\bar q u} = \frac{v^2}{\Lambda^2} (c_1 + y_t^2 c_2) \,, \\
    \tilde C_{\bar d d} = \frac{v^2}{\Lambda^2} d_1 \,, \quad & 
    r_{L} = y_t^2 \frac{a_2}{a_1} \,, \quad & r_R = y_t^2 \frac{b_2}{b_1} \,.
  \end{array}
  \label{eqn:MFV_degrees_of_freedom}
  \vspace{0.3cm}
\end{equation}
The ratios $r_L$ and $r_R$\footnote{These ratios are denoted by $\gamma_a$ and $\gamma_b$ in Ref.~\cite{Grunwald:2023nli}.} quantify the relative size of higher-order terms in the MFV expansion compared to the leading-order contributions. 
In our fit, each operator with a $\bar{Q} Q$ or $\bar{U} U$ structure is assigned its own MFV parameter $r^i_{L/R}$, allowing for differences in flavor structure that may arise in more intricate UV completions.

We perform lepton-flavor-specific and lepton-flavor-universal analyses to explore the impact of lepton flavor on the  bounds. In the lepton-flavor-specific fits, we assume that NP couples exclusively to a single lepton flavor. 
We further consider a lepton-flavor-universal (LU) scenario and a lepton-flavor-democratic (LD) scenario. In the LU case, we assume that NP couples equally to each lepton flavor in a flavor-diagonal manner, such that ${\tilde{C}_{ij} = \tilde{C}_{\text{LU}}\, \delta_{ij}}$ for all lepton bilinears. In contrast, the LD scenario assumes that NP couples equally to all lepton flavors, including off-diagonal terms. This implies ${\tilde{C}_{ij} = \tilde{C}_{\text{LD}}}$ for all $i, j$, thereby allowing lepton-flavor-conserving (LFC) and lepton-flavor-violating (LFV) interactions.

\subsection{Matching onto the LEFT  \label{sec:left}}

While collider observables are naturally described within the SMEFT, flavor observables are typically evaluated at lower energy scales. To account for this, we compute these observables in the LEFT, which is valid at scales below the electroweak one. The LEFT is obtained by integrating out the top quark, the $W$ and $Z$ bosons, and the Higgs boson. As a result, the gauge symmetry is reduced to $SU(3)_c \times U(1)_{\text{em}}$, in contrast to the full SM symmetry $SU(3)_c \times SU(2)_L \times U(1)_Y$ that characterizes the SMEFT.

We perform the matching of SMEFT Wilson coefficients onto the LEFT at the scale ${\mu = m_W}$ at tree level, following Ref.~\cite{Jenkins:2017jig}. Details of the matching conditions are provided in Appendix~\ref{app:matching_conditions}. The resulting LEFT Wilson coefficients are then evolved to the  low-energy scales using the one-loop RGEs of the LEFT~\cite{Jenkins:2017dyc}, implemented in the \texttt{Wilson} package~\cite{Aebischer:2018bkb}.

Tab.~\ref{tab:LEFT_operators} lists the LEFT Wilson coefficients induced by the  SMEFT operators considered in this work. Within our MFV framework, no scalar or tensor operators are generated in the LEFT, as the top quark, which is the only flavor contributing to these operators, is integrated out. To  distinguish between the two EFTs, we use $\mathcal{C}$ and $\mathcal{Q}$ to denote LEFT Wilson coefficients and operators, respectively, and $\tilde{C}$ and $O$ for their SMEFT counterparts. For processes below the electroweak scale, we denote charged leptons by $\ell$, neutrinos by $\nu$, and up- and down-type quarks by $u$ and $d$, respectively. Fermion chirality is indicated by $L$ and $R$, and the electromagnetic field strength tensor is denoted by $F_{\mu\nu}$.

\begin{table}[h]
  \renewcommand{\arraystretch}{1.5}
  \centering
  \begin{tabular}{|l l | l l | l l|}
    \hline
    \multicolumn{6}{|c|}{{Dipole}} \\
    \hline
    ${\cal{Q}}_{\underset{ij}{e\gamma}}$ & $\bigl(\bar \ell_{\underset{i}{L}} \sigma^{\mu\nu} \ell_{\underset{j}{R}} \bigr) F_{\mu\nu}$ & & & & \\
    \hline
    \multicolumn{6}{|c|}{{LL}} \\
    \hline
    ${\cal{Q}}_{\underset{ijkm}{ee}}^{V,LL}$ & $\bigl(\bar \ell_{\underset{i}{L}} \gamma_{\mu} \ell_{\underset{j}{L}}\bigr)\bigl(\bar \ell_{\underset{k}{L}} \gamma^{\mu} \ell_{\underset{l}{L}}\bigr)$ & ${\cal{Q}}_{\underset{ijkm}{\nu e}}^{V,LL}$ & $\bigl(\bar \nu_{\underset{i}{L}} \gamma_{\mu} \nu_{\underset{j}{L}}\bigr)\bigl(\bar \ell_{\underset{k}{L}} \gamma^{\mu} \ell_{\underset{l}{L}}\bigr)$ & ${\cal{Q}}_{\underset{ijkm}{\nu u}}^{V,LL}$ & $\bigl(\bar \nu_{\underset{i}{L}} \gamma_{\mu} \nu_{\underset{j}{L}}\bigr)\bigl(\bar u_{\underset{k}{L}} \gamma^{\mu} u_{\underset{l}{L}}\bigr)$ \\
    ${\cal{Q}}_{\underset{ijkm}{\nu d}}^{V,LL}$ & $\bigl(\bar \nu_{\underset{i}{L}} \gamma_{\mu} \nu_{\underset{j}{L}}\bigr)\bigl(\bar d_{\underset{k}{L}} \gamma^{\mu} d_{\underset{l}{L}}\bigr)$ & ${\cal{Q}}_{\underset{ijkm}{eu}}^{V,LL}$ & $\bigl(\bar \ell_{\underset{i}{L}} \gamma_{\mu} \ell_{\underset{j}{L}}\bigr)\bigl(\bar u_{\underset{k}{L}} \gamma^{\mu} u_{\underset{l}{L}}\bigr)$ & ${\cal{Q}}_{\underset{ijkm}{ed}}^{V,LL}$ & $\bigl(\bar \ell_{\underset{i}{L}} \gamma_{\mu} \ell_{\underset{j}{L}}\bigr)\bigl(\bar d_{\underset{k}{L}} \gamma^{\mu} d_{\underset{l}{L}}\bigr)$ \\
    ${\cal{Q}}_{\underset{ijkm}{\nu edu}}^{V,LL}$ & $\bigl(\bar \nu_{\underset{i}{L}} \gamma_{\mu} \ell_{\underset{j}{L}}\bigr)\bigl(\bar d_{\underset{k}{L}} \gamma^{\mu} u_{\underset{l}{L}}\bigr)$ +\text{h.c.} & & & & \\
    \hline
    \multicolumn{6}{|c|}{{RR}} \\
    \hline
    ${\cal{Q}}_{\underset{ijkm}{ee}}^{V,RR}$ & $\bigl(\bar \ell_{\underset{i}{R}} \gamma_{\mu} \ell_{\underset{j}{R}}\bigr)\bigl(\bar \ell_{\underset{k}{R}} \gamma^{\mu} \ell_{\underset{l}{R}}\bigr)$ & ${\cal{Q}}_{\underset{ijkm}{eu}}^{V,RR}$ & $\bigl(\bar \ell_{\underset{i}{R}} \gamma_{\mu} \ell_{\underset{j}{R}}\bigr)\bigl(\bar u_{\underset{k}{R}} \gamma^{\mu} u_{\underset{l}{R}}\bigr)$ & ${\cal{Q}}_{\underset{ijkm}{ed}}^{V,RR}$ & $\bigl(\bar \ell_{\underset{i}{R}} \gamma_{\mu} \ell_{\underset{j}{R}}\bigr)\bigl(\bar d_{\underset{k}{R}} \gamma^{\mu} d_{\underset{l}{R}}\bigr)$ \\
    \hline
    \multicolumn{6}{|c|}{{LR}} \\
    \hline
    ${\cal{Q}}_{\underset{ijkm}{ee}}^{V,LR}$ & $\bigl(\bar \ell_{\underset{i}{L}} \gamma_{\mu} \ell_{\underset{j}{L}}\bigr)\bigl(\bar \ell_{\underset{k}{R}} \gamma^{\mu} \ell_{\underset{l}{R}}\bigr)$ & ${\cal{Q}}_{\underset{ijkm}{\nu e}}^{V,LR}$ & $\bigl(\bar \nu_{\underset{i}{L}} \gamma_{\mu} \nu_{\underset{j}{L}}\bigr)\bigl(\bar \ell_{\underset{k}{R}} \gamma^{\mu} \ell_{\underset{l}{R}}\bigr)$ & ${\cal{Q}}_{\underset{ijkm}{\nu u}}^{V,LR}$ & $\bigl(\bar \nu_{\underset{i}{L}} \gamma_{\mu} \nu_{\underset{j}{L}}\bigr)\bigl(\bar u_{\underset{k}{R}} \gamma^{\mu} u_{\underset{l}{R}}\bigr)$ \\
    ${\cal{Q}}_{\underset{ijkm}{\nu d}}^{V,LR}$ & $\bigl(\bar \nu_{\underset{i}{L}} \gamma_{\mu} \nu_{\underset{j}{L}}\bigr)\bigl(\bar d_{\underset{k}{R}} \gamma^{\mu} d_{\underset{l}{R}}\bigr)$ & ${\cal{Q}}_{\underset{ijkm}{eu}}^{V,LR}$ & $\bigl(\bar \ell_{\underset{i}{L}} \gamma_{\mu} \ell_{\underset{j}{L}}\bigr)\bigl(\bar u_{\underset{k}{R}} \gamma^{\mu} u_{\underset{l}{R}}\bigr)$ & ${\cal{Q}}_{\underset{ijkm}{ed}}^{V,LR}$ & $\bigl(\bar \ell_{\underset{i}{L}} \gamma_{\mu} \ell_{\underset{j}{L}}\bigr)\bigl(\bar d_{\underset{k}{R}} \gamma^{\mu} d_{\underset{l}{R}}\bigr)$ \\
    ${\cal{Q}}_{\underset{ijkm}{ue}}^{V,LR}$ & $\bigl(\bar u_{\underset{i}{L}} \gamma_{\mu} u_{\underset{j}{L}}\bigr)\bigl(\bar \ell_{\underset{k}{R}} \gamma^{\mu} \ell_{\underset{l}{R}}\bigr)$ & ${\cal{Q}}_{\underset{ijkm}{de}}^{V,LR}$ & $\bigl(\bar d_{\underset{i}{L}} \gamma_{\mu} d_{\underset{j}{L}}\bigr)\bigl(\bar \ell_{\underset{k}{R}} \gamma^{\mu} \ell_{\underset{l}{R}}\bigr)$ & & \\
    \hline
  \end{tabular}
  \caption{LEFT operators considered in this work.}
  \label{tab:LEFT_operators}
\end{table}

While there is an intrinsic $SU(2)_L$ link between the left-handed fermions in the SMEFT, this symmetry is no longer present in the LEFT due to the breaking of the $SU(2)_L$ gauge symmetry. Following from the matching, however, the SMEFT dictates correlations among the LEFT Wilson coefficients comprising left-handed fermions, which can be tested experimentally. For instance, the operator $O_{lq}^{1}$, which couples left-handed leptons to left-handed quarks, matches onto four different LEFT operators, ${\cal{Q}}_{\nu u}^{V,LL}$, ${\cal{Q}}_{\nu d}^{V,LL}$, ${\cal{Q}}_{eu}^{V,LL}$ and ${\cal{Q}}_{ed}^{V,LL}$. Without the $SU(2)_L$ structure, these operators can be treated independently, while they are correlated within the  SMEFT. This allows to test flavor patterns and also  the SMEFT ansatz with the hypothesis of heavy NP itself.

\begin{figure}
  \centering
  \input{feynman/doubleInsertion.tex}
  \caption{Feynman diagrams illustrating double insertions of the SMEFT Higgs-current operators (left diagram)  in the matching to the LEFT four-lepton operator (right diagram).}
  \label{fig:double_insertion}
\end{figure}
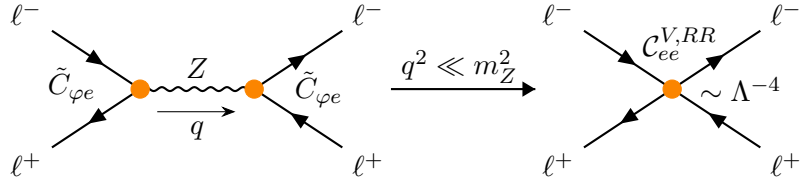

In addition to the direct contributions from SMEFT four-fermion operators, four-fermion operators in the LEFT also receive contributions from Higgs-current operators. Double insertions of these SMEFT Higgs-current operators can lead to contributions of order $\Lambda^{-4}$, as illustrated in Fig.~\ref{fig:double_insertion}. To maintain a consistent expansion in the SMEFT scale $\Lambda$, we include these double-insertion contributions only in the interference terms between the SM and the NP amplitudes. 

\section{Collider Observables}
\label{sec:collider_observables}
\enlargethispage{\baselineskip}

In this section, we discuss collider observables within the SMEFT,  $Z$-decays (Sec.~\ref{sec:Zboson}) and $W$-decays (Sec.~\ref{sec:Wboson}), dilepton production at lepton colliders
(Sec.~\ref{sec:dilepton_production}), neutral- and charged-current Drell–Yan processes (Sec.~\ref{sec:Drell_Yan}), and top-quark pair production in association with leptons (Sec.~\ref{sec:ttll}).

\subsection{$Z$-Boson Observables}
\label{sec:Zboson}

The tree-level partial width for the decay $Z \to l_i  l_j$ can be expressed as \cite{Brivio:2017vri} 
\begin{equation}
  \Gamma(Z\to l_i l_j) = \frac{G_F m_Z^3}{6\pi \sqrt{2}} \left( \lvert g^{l}_{\underset{ij}{L}} \rvert^2 + \lvert g^{l}_{\underset{ij}{R}}\rvert^2 \right) \, ,
  \label{eqn:Z_decay_tree}
\end{equation}
where $g^{l}_{\underset{ij}{L}}$ and $g^{l}_{\underset{ij}{R}}$ denote the left- and right-handed couplings of the charged leptons and neutrinos to the $Z$ boson, respectively.
In the SMEFT, the $Z$ boson couplings to fermions are modified by the Higgs-current operators, which can also induce flavor-violating decays. For the  Wilson coefficients considered in this work, the  couplings read 
\begin{alignat}{2}
  g_{\underset{ij}{L}}^{\nu, } &= \frac{1}{2} \delta_{ij} - \frac{1}{2} \tilde C_{\underset{ij}{\varphi l}}^{-}  \,, \qquad &g_{\underset{ij}{R}}^{\nu } &= 0 \,, \\
  g_{\underset{ij}{L}}^{\ell} &= \frac{1}{2}  \delta_{ij} (2\szW -1) - \frac{1}{2} \tilde C_{\underset{ij}{\varphi l}}^{+} \,, \qquad &g_{\underset{ij}{R}}^{\ell} &= \delta_{ij} \szW - \frac{1}{2} \tilde C_{\underset{ij}{\varphi e}} \,,
\label{eqn:Z_couplings_SMEFT}
\end{alignat}
with 
\begin{equation}
  \tilde C_{\varphi l}^{\pm} = \tilde C_{\varphi l}^{(1)} \pm \tilde C_{\varphi l}^{(3)} \,,
  \label{eqn:Chl_linear_combinations}
\end{equation}
where the Wilson coefficients are evaluated at the scale of the $Z$-mass,  ${\mu \sim m_Z}$.
$\theta_W$ and $G_F$ denote the weak mixing angle and  Fermi's constant, respectively.

In addition, dipole operators contribute to the $Z$ boson decay width into charged leptons. Their contribution to the partial width is given by 
\begin{equation}
  \Gamma_{\text{Dipole}} (Z\to \ell_i \ell_j) = \frac{G_F m_Z^3}{24 \pi \sqrt{2}} \left(\tilde C_{\underset{ij}{eZ}}^2 + \tilde C_{\underset{ji}{eZ}}^2 \right) \,,
  \label{eqn:Z_decay_dipole}
\end{equation}
where 
\begin{equation}
  \tilde C_{\underset{ij}{eZ}} = \sW \tilde C_{\underset{ij}{eB}} + \cW \tilde C_{\underset{ij}{eW}} \,.
  \label{eqn:Z_dipole_coefficient}
\end{equation}

The measurements of the $Z$ boson decay widths and the limits on the LFV branching fractions are listed in Tab.~\ref{tab:Z_measurements} alongside the corresponding SM predictions. For the partial widths to charged leptons, we include the correlation of the data provided by LEP~\cite{ALEPH:2005ab}. 

\begin{table}[h]
  \setlength{\tabcolsep}{15pt}
  \renewcommand{\arraystretch}{1.2}
  \centering
  \begin{tabular}{l l l}
      \toprule
      Observable & Measurement & SM prediction \\
      \midrule
      $\Gamma (Z\to e^+ e^-)$ & $83.92(12)$\,MeV \cite{ALEPH:2005ab} & $83.987(68)$\,MeV \cite{ALEPH:2005ab} \\
      $\Gamma (Z\to \mu^+ \mu^-)$ & $83.99(18)$\,MeV \cite{ALEPH:2005ab} & $83.986(68)$\,MeV \cite{ALEPH:2005ab} \\
      $\Gamma (Z\to \tau^+ \tau^-)$ & $84.08(22)$\,MeV \cite{ALEPH:2005ab} & $83.796(67)$\,MeV \cite{ALEPH:2005ab} \\ 
      $\Gamma (Z\to \nu \bar \nu)$ & $497.4(25)$\,MeV \cite{ALEPH:2005ab} & $501.435(45)$\,MeV \cite{ParticleDataGroup:2024cfk} \\
      \midrule
      Observable & 95\% CL limit & SM prediction \\
      \midrule
      ${\cal B} (Z\to e\mu)$ & $2.62\cdot 10^{-7}$~\cite{ATLAS:2022uhq} & 0 \\
      ${\cal B} (Z\to e\tau)$ & $5.04\cdot 10^{-6}$~\cite{ATLAS:2021bdj} & 0 \\
      ${\cal B} (Z\to \mu\tau)$ & $6.5 \cdot 10^{-6}$~\cite{ATLAS:2021bdj} & 0 \\
      \bottomrule
  \end{tabular}
  \caption{Measurements of the $Z$ boson decay widths and limits on the LFV branching fractions as well as the corresponding SM predictions used in this analysis.}
  \label{tab:Z_measurements}
\end{table}

\begin{figure}[h]
    \centering
    \includegraphics[width=0.6\textwidth]{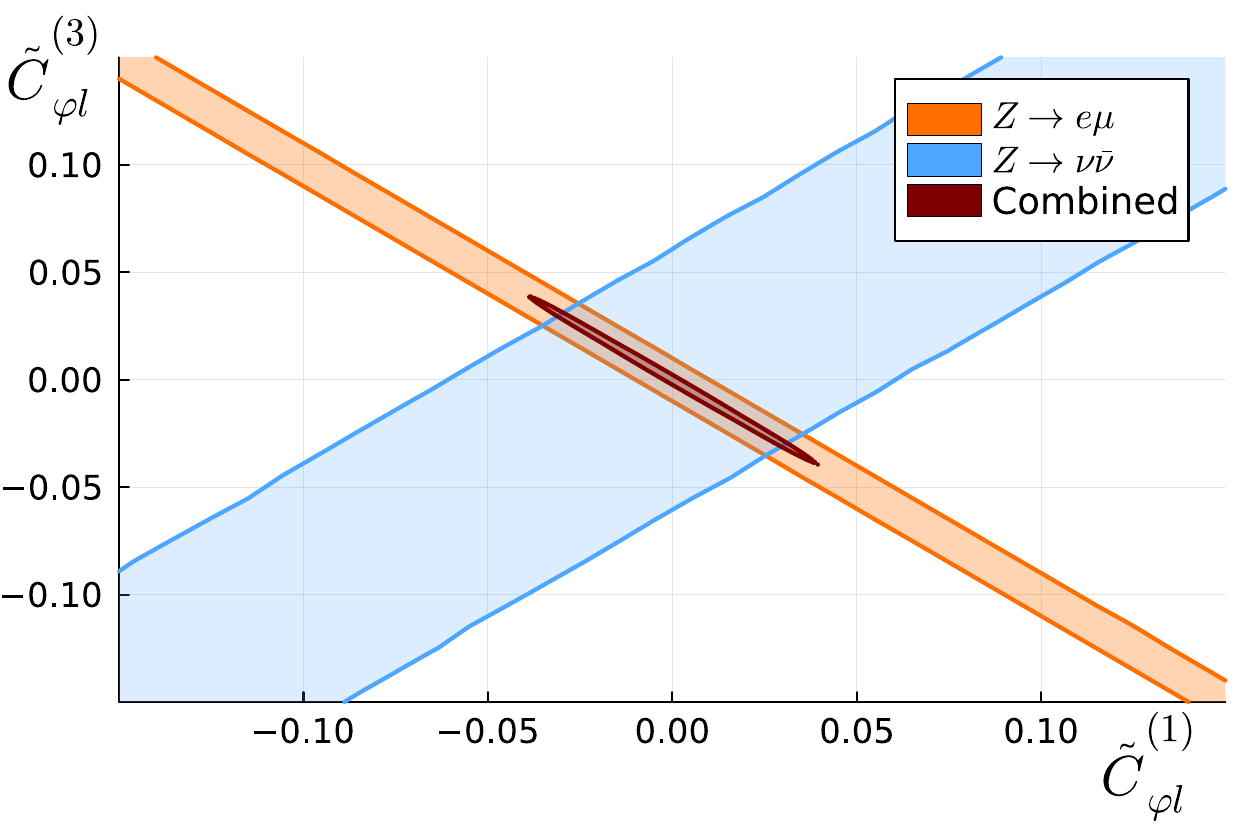}
    \caption{90\% credible regions for the two-dimensional fit in the $\tilde C^{(1)}_{\varphi l}$--$\tilde C^{(3)}_{\varphi l}$ plane. The blue (orange) region is obtained from  $Z$ decays  to neutrinos
     (charged leptons $Z \to e \mu$). The red region presents a combined fit of both decays.}
    \label{Z_synergies}
\end{figure}

While the decay widths to charged leptons constrain $\tilde C^{+}_{\varphi l}$, the decay width to neutrinos is sensitive to  $\tilde C^{-}_{\varphi l}$. A combined analysis of both decay modes therefore enables to disentangle  $\tilde C^{(1)}_{\varphi l}$ and $\tilde C^{(3)}_{\varphi l}$, which would otherwise remain degenerate in the fit.
Fig.~\ref{Z_synergies} illustrates this synergy for the example of a lepton flavor specific pattern where only the $e\mu$ coefficient is switched on. The plot shows the 90\% credible regions from a two-dimensional fit in the $\tilde C^{(1)}_{\varphi l}$--$\tilde C^{(3)}_{\varphi l}$ plane. The region in blue is obtained from a fit of $Z \to \nu \bar \nu$ decays, while the orange region is derived from 
$Z \to e \mu$ decays. The combination of both channels  (red region) breaks the degeneracy and significantly reduces the allowed parameter space.

We further consider forward--backward asymmetries  $A_{\text{FB}}^{\ell}$  at the $Z$-pole,
 defined  as  the difference  of cross sections  in the $Z$ boson center-of-mass frame
regarding the scattering angle $\theta$  normalized to the total one
\begin{equation}
  A_{\text{FB}}^{\ell} = \frac{\sigma_{\text{F}}^{\ell} - \sigma_{\text{B}}^{\ell}}{\sigma_{\text{F}}^{\ell} + \sigma_{\text{B}}^{\ell}} \,,
  \label{eqn:Z_forward_backward}
\end{equation}
where
\begin{equation}
  \sigma_{\text{F}}^{\ell} = \int_{0}^{1} d\cos \theta \frac{d\sigma (e^+ e^- \to \ell^+ \ell^-)}{d\cos \theta} \,, \quad \sigma_{\text{B}}^{\ell} = \int_{-1}^{0} d\cos \theta \frac{d\sigma (e^+ e^- \to \ell^+ \ell^-)}{d\cos \theta} \,.
  \label{eqn:Z_forward_backward_cross_sections}
\end{equation}
Defined as ratios of cross sections, $A_{\text{FB}}^{\ell}$ benefit from partial cancellation of systematic uncertainties. Moreover, they are sensitive to the chiral structure of the $Z$-couplings and thus probe different combinations of Wilson coefficients than the decay rates.
While $A_{\text{FB}}^{\ell} $ is affected by $\gamma$--exchange and interference between the $Z$ and $\gamma$ contribution, 
LEP has defined a "pole" quantity which is corrected for these effects, denoted by $A_{\text{FB}}^{0,\ell}$.
The latter is related to the $Z$-couplings as
\begin{equation}
  {\cal A}_{\ell} = \frac{(g_{{L}}^{\ell})^2-(g_{{R}}^{\ell})^2}{(g_{{L}}^{\ell})^2+(g_{{R}}^{\ell})^2} \,,
  \label{eqn:Z_asymmetry_parameters}
\end{equation}
via
\begin{equation}
  A_{\text{FB}}^{0,\ell} = \frac{3}{4} {\cal A}_e {\cal A}_{\ell} \,.
  \label{eqn:Z_pole_AFB}
\end{equation}
Since there is no interference with the SM and in view of their strong constraints from $Z$-decays, we  neglect contributions from electroweak dipole operators  in $A_{\text{FB}}^{0,\ell}$.
 This probes and hence concerns only LFC-fits. 
To maintain consistency in the SMEFT expansion, we expand the forward--backward asymmetries in powers of the NP scale  including terms  up to order $\Lambda^{-4}$. Tab.~\ref{tab:Z_asymmetries_measurements} summarizes the measurements used in this analysis, alongside their corresponding SM predictions. 

Four--lepton operators also contribute to the asymmetry parameters by modifying the $e^+ e^- \to \ell^+ \ell^-$ cross section, 
schematically,
\begin{equation}
  A_{\text{FB}}^{0,\ell} \sim \frac{3( \tilde C_{\underset{ee\ell\ell}{ee}}^2 + \tilde C_{\underset{ee\ell\ell}{ll}}^2 - \tilde C_{\underset{ee\ell\ell}{le}}^2 - \tilde C_{\underset{\ell\ell ee}{le}}^2)}{4( \tilde C_{\underset{ee\ell\ell}{ee}}^2 + \tilde C_{\underset{ee\ell\ell}{ll}}^2 + \tilde C_{\underset{ee\ell\ell}{le}}^2 + \tilde C_{\underset{\ell\ell ee}{le}}^2)} \,.
  \label{eqn:Z_AFB_SMEFT}
\end{equation}
 In the fit we include the full formula, which we again expand up to order $\Lambda^{-4}$,  including SM contributions and terms with other SMEFT coefficients.

\begin{table}[h]
  \setlength{\tabcolsep}{15pt}
  \renewcommand{\arraystretch}{1.2}
  \centering
  \begin{tabular}{l l l}
      \toprule
      Observable & Measurement & SM prediction  \\
      \midrule
      $A_{\text{FB}}^{0,e}$ & 0.0145(25) & 0.01627(27) \\
      $A_{\text{FB}}^{0,\mu}$ & 0.0169(13) & 0.01627(27) \\
      $A_{\text{FB}}^{0,\tau}$ & 0.0188(17) & 0.01627(27) \\
      \bottomrule
  \end{tabular}
  \caption{Measurements and SM predictions of the $Z$ boson forward--backward asymmetries~\cite{ALEPH:2005ab}.}
  \label{tab:Z_asymmetries_measurements}
\end{table}

\subsection{$W$-Boson Decays}
\label{sec:Wboson}

In the SMEFT  the partial decay width $\Gamma(W \to \ell_i \nu_j)$ is given by \cite{Brivio:2017vri}  
\begin{equation}
  \Gamma (W\to \ell_i \nu_j) = \frac{G_F m_W^3}{6 \pi \sqrt{2}} \left(\left(\delta_{ij} + \tilde C_{\underset{ij}{\varphi l}}^{(3)}\right)^2 + \tilde C_{\underset{ij}{eW}}^2 \right) \,,
  \label{eqn:W_decay_width_SMEFT}
\end{equation}
where the coefficients are evaluated at $\mu \sim m_W$.
Since the neutrino flavor cannot be resolved experimentally, the observed decay width corresponds to the sum over all neutrino flavors
\begin{equation}
  \Gamma (W\to \ell_i \nu) = \sum_{j} \Gamma (W\to \ell_i \nu_j) \,,
  \label{eqn:W_decay_width_summed}
\end{equation}
where only $i=j$ receives a SM contribution. As the Wilson coefficients contribute incoherently, we can nonetheless constrain all SMEFT coefficients at $\mu = m_W$ simultaneously.

\begin{table}[h]
  \centering
  \setlength{\tabcolsep}{15pt}
\renewcommand{\arraystretch}{1.2}
  \begin{tabular}{l l l }
      \toprule
      $\Gamma(W \to e\nu)$ & $\Gamma(W \to \mu\nu)$ & $\Gamma(W \to \tau\nu)$ \\
      \midrule
      0.226(5)\,GeV & 0.228(5)\,GeV & 0.225(6)\,GeV\\
      \bottomrule
  \end{tabular}
  \caption{Experimental values of the partial widths $\Gamma (W\to \ell \nu)$. They are obtained by multiplying the branching fractions stated in Ref.~\cite{CMS:2022mhs} with the total $W$ boson decay width of $\Gamma_W = 2.085(42)$\,GeV~\cite{ParticleDataGroup:2024cfk}.}
  \label{tab:W_measurements}
\end{table}

As experimental input, we employ partial decay widths rather than branching fractions. This choice avoids the need to expand the branching ratios in powers of $\Lambda$, since both, the partial and total decay widths can receive SMEFT contributions. The partial widths are obtained by multiplying the $W$ boson branching fractions to charged leptons with the total $W$ decay width, $\Gamma_W = 2.085(42)$\,GeV~\cite{ParticleDataGroup:2024cfk}. For the branching fractions, we use the most recent measurements of the CMS collaboration~\cite{CMS:2022mhs}. Results are listed in Tab.~\ref{tab:W_measurements}.

\begin{figure}[h]
    \centering
    \includegraphics[width=0.6\textwidth]{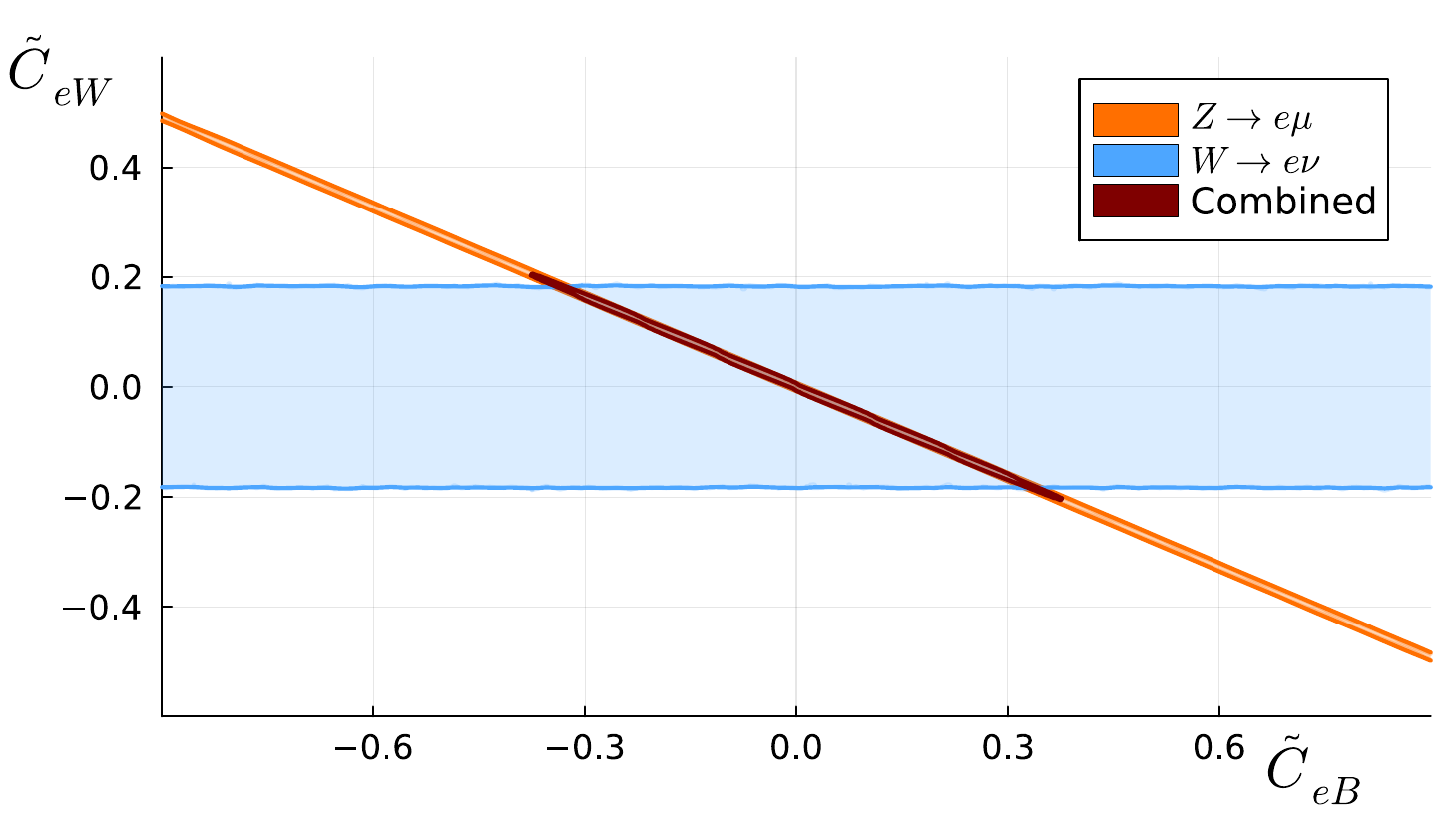}
    \caption{90\% credible regions for the two-dimensional fit in the $\tilde C_{eB}$--$\tilde C_{eW}$ plane for a lepton flavor specific analysis with only the $e\mu$ coefficient switched on. The blue region is obtained from a  fit to  $W\to e\nu$ decays, while the orange region is obtained from $Z\to e\mu$. The red region shows the combined fit. }
    \label{fig:dipole_ZW_overlay}
\end{figure}

The decays $W \to \ell \nu$ provide complementary constraints to the ones from the $Z$. Together, they resolve flat directions
 in the $\tilde C_{eB}$-$\tilde C_{eW}$ parameter space that arises in $Z$ boson observables.
This is illustrated for the example of an $e\mu$-specific fit in Fig.~\ref{fig:dipole_ZW_overlay}, where we show the 90\% credible regions resulting from a two-dimensional fit of the Wilson coefficients.
The blue (orange) region corresponds to constraints from the $W \to e \nu$ decay ($Z \to e\mu$ decay). The red region shows the result of a combined fit to both decay channels. 
The combinations  significantly reduces the allowed parameter space and enables constraints on $\tilde C_{eB}$.

\subsection{Dilepton Production at LEP}
\label{sec:dilepton_production}

The production of charged lepton pairs in electron-positron collisions allows to cleanly probe NP in the lepton sector.  In addition to the contact four-lepton interactions, the process also receives contributions from vertex corrections induced by Higgs-current operators, as well as from dipole operators.

As experimental input we employ the differential measurements of the $e^+ e^- \to \ell^+ \ell^-$ cross section for all flavors $\ell=e, \mu, \tau$  by the DELPHI collaboration at $\sqrt{s} = 207$\,GeV~\cite{DELPHI:2005wxt}. The data is binned in terms of $\cos \theta$, where $\theta$ is the angle between the outgoing lepton and the beam axis in the center-of-mass frame. 
We compute the SMEFT contributions to this process at tree-level using {\textsc{Mathematica}}~\cite{Mathematica} in combination with the \textsc{FeynCalc}~\cite{Mertig:1990an, Shtabovenko:2016sxi, Shtabovenko:2020gxv, Shtabovenko:2023idz} and {\tt SmeftFR}~v3.0~\cite{Dedes:2023zws} packages. The coefficients inducing vector couplings, such as the four-lepton coefficients and the Higgs-current operators, induce contributions proportional to $(1\pm \cos \theta)^2$, while the dipole operators induce contributions proportional to $1- \cos^2 \! \theta$. The angular distribution thus allows one to disentangle different chirality structures.

\subsection{Drell--Yan Production}
\label{sec:Drell_Yan}

The Drell--Yan (DY) process refers to lepton production  in hadron collisions, $pp \to \ell^+\ell^-$ and $pp \to \ell \nu$. In the SM, it arises at tree-level from exchange of electroweak bosons.
In contrast to the $Z$, $W$ boson decays and dilepton production at LEP, the experimental measurements of  DY  are typically reported as differential distributions of event counts rather than differential cross sections. Thus, in order to confront the SMEFT to experimental data, hadronization, parton showering and detector effects need to be taken into account. 

We compute the partonic cross section at LO using \texttt{MadGraph5\_aMC@NLO}~\cite{Alwall:2014hca}. The hadronic cross section is obtained by convoluting the partonic cross section with the parton-luminosity-functions, defined as
\begin{equation}
    \mathcal L_{q_i \bar q_j} (\tau) = \tau \int_\tau^1 \frac{\text{d}x}{x} \left[f_{q_i}(x,\mu_F) f_{\bar q_j}(\tau/x,\mu_F) + f_{\bar q_j}(x,\mu_F) f_{q_i}(\tau/x,\mu_F) \right] \,,
    \label{eq:parton_luminosity}
\end{equation}
where $f_{q_i}$ and $f_{\bar q_j}$ denote the parton-distribution-functions (PDFs) of the incoming quarks $q_i$ and $\bar q_j$, respectively, and $\tau = \frac{\hat s}{s}$, with $\hat s$ the partonic and $s$ the hadronic center-of-mass energy. 
The resulting hadronic cross section is given by
\begin{equation}
    \sigma =\sum_{ij}\int \frac{\text{d}\tau}{\tau} \: \mathcal{L}_{q_i \bar q_j}(\tau)\:  \hat \sigma (\tau s) \,.
    \label{eq:hadronic_DY_xsec}
\end{equation}

As a result, the DY process probes a wide range of initial-state quark combinations. Since the different quark flavor contribute incoherently, we can simultaneously constrain various flavor combinations of the four-fermion operators. Due to the structure of the PDFs, the process is most sensitive to the light quarks of the first generation, $u$ and $d$, which constitute the valence quarks of the proton. Second--generation quarks, $s$ and $c$, enter as sea quarks and are thus less relevant, while third-generation quarks, $b$ and $t$, are strongly suppressed by their PDFs. In particular, the top quark is too heavy to significantly affect the DY process at LHC energies, and we therefore neglect its contribution in this analysis.

We employ the leading-order Monte Carlo variant of the NNPDF4.0~PDF~sets~\cite{NNPDF:2021njg,Cruz-Martinez:2024cbz} as implemented in LHAPDF~6~\cite{Buckley:2014ana}. The hadronic cross sections are computed numerically using the \texttt{MadGraph5\_aMC@NLO}~\cite{Alwall:2014hca} framework. Parton showering and hadronization are handled by \texttt{Pythia8}~\cite{Bierlich:2022pfr}, while detector effects are simulated using the parametric detector simulation framework \texttt{Delphes}~\cite{deFavereau:2013fsa}.
The DY observables considered in this work, including the integrated luminosity, the number of bins and the corresponding experimental reference are
summarized in  Tab.~\ref{tab:DY_observables}. 

The DY process is sensitive to a variety of different SMEFT operators, including the semileptonic four--fermion operators, which particularly benefit from the energy enhancement in the tails of the kinematic distributions, scaling 
 with the partonic center-of-mass energy $\hat{s}$ as $\hat{s}/\Lambda^2$~\cite{Farina:2016rws}. Contributions from electroweak dipole operators and  Higgs-current operators scale as $\sqrt{\hat s}v/\Lambda^2$ and $v^2/\Lambda^2$, respectively~\footnote{Gluon dipole operators, which contribute to more generalized DY processes into missing energy  with full energy-enhancement  $\hat{s}/\Lambda^2$~\cite{Hiller:2025hpf} are not considered in this work, see  Tab.~\ref{tab:operators}.}.
In our MFV setup, tensor and scalar operators do not contribute to the DY process, since they couple exclusively to quark bilinears involving the top quark, which does not enter the partonic cross section at LHC energies.

\begin{table}[h]
  \centering
  \setlength{\tabcolsep}{10pt}
  \renewcommand{\arraystretch}{1.2}
  \begin{tabular}{c c c c c c}
      \toprule
      Process & Observable & Integrated Luminosity & \# Bins &  Experiment & Reference \\
        \midrule
      $pp \to e e$ & $\frac{\text{d}\sigma}{\text{d} m_{ee}}$ & 137 fb$^{-1}$ & 117 & CMS & \cite{CMS:2021ctt}\\
      $pp \to \mu\mu$ & $\frac{\text{d}\sigma}{\text{d} m_{\mu\mu}}$ & 140 fb$^{-1}$ & 48 & CMS & \cite{CMS:2021ctt} \\
      $pp \to \tau \tau$ & $\frac{\text{d}\sigma}{\text{d} m_T^{\text{tot}}}$ & 139 fb$^{-1}$ & 4 & ATLAS & \cite{ATLAS:2020tre} \\
      \midrule 
      $pp \to e \mu$ & $\frac{\text{d}\sigma}{\text{d} m_{e\mu}}$ & 139 fb$^{-1}$ & 4 & ATLAS & \cite{ATLAS:2020tre} \\
      $pp \to e \tau$ & $\frac{\text{d}\sigma}{\text{d} m_{e\tau}}$ & 139 fb$^{-1}$ & 4 & ATLAS & \cite{ATLAS:2020tre} \\
      $pp \to \mu \tau$ & $\frac{\text{d}\sigma}{\text{d} m_{\mu\tau}}$ & 139 fb$^{-1}$ & 4 & ATLAS & \cite{ATLAS:2020tre} \\
      \midrule
      $pp \to e \nu$ & $\frac{\text{d}\sigma}{\text{d} m_T}$ & 139 fb$^{-1}$ & 65 & ATLAS & \cite{ATLAS:2019lsy} \\
      $pp \to \mu \nu$ & $\frac{\text{d}\sigma}{\text{d} m_T}$ & 139 fb$^{-1}$ & 54 & ATLAS & \cite{ATLAS:2019lsy} \\
      $pp \to \tau \nu$ & $\frac{\text{d}\sigma}{\text{d} m_T}$ & 138 fb$^{-1}$ & 18 & CMS & \cite{CMS:2022ncp} \\
      \bottomrule
  \end{tabular}
  \caption{Drell--Yan observables considered in this work.}
  \label{tab:DY_observables}
\end{table}

\subsection{Top-Quark Pair-Production in Association with Leptons}
\label{sec:ttll}

We incorporate constraints from top--quark pair--production in association with a charged lepton pair, $pp \to t\bar{t} \ell^+\ell^-$. This process provides unique sensitivity to semileptonic four--fermion operators involving right-handed couplings to the top quark. While left-handed couplings can be probed through other observables via the $SU(2)_L$ relation between the top and bottom quarks, the $t\bar{t} \ell \ell$ process constitutes the only direct collider probe of third--generation right-handed up-type quark Wilson coefficients. It therefore tests the MFV-expansion beyond leading order in the right-handed sector, parameterized by the ratio $r_R$  in Eq.~\eqref{eqn:MFV_degrees_of_freedom}.
Moreover, within the MFV framework, the $t\bar t \ell \ell$ process is the only observable that  constrains directly, at tree level, the scalar and tensor coefficients $\tilde C_{lequ}^{(1)}$ and $\tilde C_{lequ}^{(3)}$. As we neglect light fermion Yukawas, the couplings of these coefficients to the light quarks vanish. The scalar and tensor operators, however,  are indirectly  constrainted by their RG mixing onto  the dipole operators $O_{eW}$ and $O_{eB}$, e.g.~see \cite{Aebischer:2021uvt}. 

As experimental input, we use two-dimensional likelihood scans of Wilson coefficients provided by the ATLAS collaboration~\cite{ATLAS:2025yww}, based on global fits to 
$t\bar{t} ee$  and $t\bar{t} \mu \mu$ data. We interpolate these likelihood scans using radial basis function (RBF) interpolation to obtain continuous likelihood functions.

\section{Flavor Observables}
\label{sec:flavor_observables}

We discuss constraints from  the anomalous magnetic moments of leptons (Sec.~\ref{sec:anomalous_magnetic_moments}), various lepton decays
(Sec.~\ref{sec:lepton_decays}),  and leptonic and semileptonic  meson decays (Sec.~\ref{sec:meson_decays}).

\subsection{Anomalous Magnetic Moments of Leptons}
\label{sec:anomalous_magnetic_moments}

The anomalous magnetic moments (AMMs), denoting the deviation of the magnetic moment from the Dirac value $a_{\ell} = \frac{g_{\ell}-2}{2}$, are among the most precisely measured quantities in particle physics for the light leptons. The AMM of the $\tau$ lepton, on the other hand, is less precisely known due to the short lifetime of the $\tau$  and  resulting experimental challenges, so that 
at present only limits on the $\tau$ AMM exist.  Recent determinations and SM predictions of the AMMs are given in
Tab.~\ref{tab:Anomalous_magnetic_moments}.

\begin{table}[h]
  \setlength{\tabcolsep}{10pt}
  \renewcommand{\arraystretch}{1.2}
  \centering
  \begin{tabular}{c l l}
      \toprule
      Observable & Measurement & SM Prediction \\
      \midrule
      $a_{e}$ & $1159.65218062(12) \cdot 10^{-6}$ \cite{ParticleDataGroup:2024cfk} & $1159.65218161(23) \cdot 10^{-6}$ \cite{Aoyama:2019ryr}$^\dagger$  \\
      $a_{\mu}$ & $1165.92071(13) \cdot 10^{-6}$ \cite{Muong-2:2025xyk} & $1165.92033(62) \cdot 10^{-6}$ \cite{Aliberti:2025beg} \\
      $a_{\tau}$ & $-0.057 < a_{\tau} < 0.024$ $(95\% \text{CL})$ \cite{ATLAS:2022ryk} & $1177.21(5) \cdot 10^{-6}$ \cite{Eidelman:2007sb} \\
      \bottomrule
  \end{tabular}
  \caption{Measurements of the lepton AMMs and  SM predictions.
   $^\dagger$ From measurement of $\alpha_e$ in Caesium. The Rubidium-based one differs in 10th digit.}
  \label{tab:Anomalous_magnetic_moments}
\end{table}

In the LEFT, the contribution to the AMM of a charged lepton is given by
\begin{equation}
  a_{\ell}^{\text{LEFT}} = \frac{2\,m_{\ell}}{\sqrt{\pi \aem}} {\cal{C}}_{\underset{\ell \ell}{e\gamma}} \,,
  \label{eqn:amm_LEFT}
\end{equation}
where the Wilson coefficient is evaluated at the scale $\mu = 2$\,GeV. This choice serves as a lower limit for the RGE in order to avoid potential QCD-related divergences. 

The tree level matching of the LEFT dipole coefficient ${\cal{C}}_{{e\gamma}}$ onto the SMEFT  reads
\begin{equation}
  {\cal{C}}_{\underset{ij}{e\gamma}} =\frac{1}{\sqrt{2}v} \left( -\sW \tilde C_{\underset{ij}{eW}} + \cW \tilde C_{\underset{ij}{eB}} \right)  \,.
  \label{eqn:Gamma_dipole_coefficient}
\end{equation}
This implies a flat direction in the parameter space, as contributions from the dipole operators $O_{eW}$ and $O_{eB}$ can  cancel each other. This degeneracy can be lifted by combining
constraints from  additional observables, such as $Z$ and $W$ boson decays, which probe different linear combinations of Wilson coefficients.
The scalar and tensor operators also contribute to the AMMs via their RG-mixing and at 1-loop. In total, the impact of the tensor $O_{lequ}^{(3)}$   is bigger than the one of the scalar operator $O_{lequ}^{(1)}$   \cite{Aebischer:2021uvt}.

\subsection{Lepton Decays}
\label{sec:lepton_decays}

We discuss constraints from LFV radiative lepton decays, LFV decays involving four leptons, and charged-current lepton decays. 

\subsubsection{Radiative LFV Lepton Decays}
\label{sec:radiative_decays}

While the AMMs constrain the lepton-flavor-conserving (LFC) electroweak dipole coefficients, the corresponding LFV couplings are bounded by radiative lepton decays. In the LEFT framework, the branching fraction for the decay $\ell_i \to \ell_j \gamma$ is given by
\begin{equation}
\mathcal{B}(\ell_i \to \ell_j \gamma) = \frac{m_{\ell_i}^3}{4 \pi \Gamma_{\ell_i}} \left( \lvert \mathcal{C}_{\underset{ij}{e\gamma}} \rvert^2 + \lvert \mathcal{C}_{\underset{ji}{e\gamma}} \rvert^2 \right) \,,
\label{eqn:rld_SMEFT}
\end{equation}
where $\Gamma_{\ell_i}$ denotes the total decay width of the initial-state lepton. The SM contribution to this decay is negligible, as it is highly suppressed by the smallness of neutrino masses. Tab.~\ref{tab:RLD} summarizes the current experimental limits on the branching fractions.
\begin{table}[h]
  \setlength{\tabcolsep}{12pt}
  \renewcommand{\arraystretch}{1.2}
  \centering
  \begin{tabular}{c l l l}
      \toprule
      & $ {\cal B} (\mu \to e \gamma) $ & $ {\cal B} (\tau \to e \gamma) $ & $ {\cal B} (\tau \to \mu \gamma) $ \\
      \midrule
      90\% CL limit \quad & $ 4.2 \cdot 10^{-13} $~\cite{MEG:2016leq} \qquad & $ 3.3 \cdot 10^{-8} $~\cite{BaBar:2009hkt} \qquad & $ 4.2 \cdot 10^{-8} $~\cite{Belle:2021ysv} \qquad \\
      \bottomrule
  \end{tabular}
  \caption{Current limits on the branching fractions of radiative, LFV lepton decays, all at 90\% CL.  \label{tab:RLD}}
\end{table}

\subsubsection{LFV  Decays involving four charged leptons}
\label{sec:lfv_decays}

 LFV decays of the form $\ell_i \to \ell_j \ell_k \ell_l$ are  null tests of the SM. 
In the LEFT, the decay $\ell_i \to \ell_j \bar \ell_k \ell_l$ is induced by the four-lepton operators ${\cal{Q}}_{ee}^{V,LL}$, ${\cal{Q}}_{ee}^{V,RR}$, and ${\cal{Q}}_{ee}^{V,LR}$, as well as by photon exchange via the LFV components of the dipole operator ${\cal{Q}}_{e\gamma}$ for $j = k$. The four-lepton operators receive matching contributions from the SMEFT four-lepton operators ${O}_{ee}$, ${O}_{ll}$ and ${O}_{le}$ as well as from the Higgs-current operators ${O}_{\varphi l}^{(1)}$, ${O}_{\varphi l}^{(3)}$, and ${O}_{\varphi e}$ at tree--level.
We compute the LEFT contributions  following Refs.~\cite{Crivellin:2013hpa, Ali:2023kua, Plakias:2023esq}. 
Tab.~\ref{tab:4LLFV} summarizes the current experimental limits on the branching fractions of LFV lepton decays at 90\% CL.
Note, the decays $\tau^- \to e^+ \mu^- \mu^-$ and $\tau^- \to \mu^+ e^- e^- $ violate two units of lepton flavor.
\begin{table}[h]
  \centering
    \setlength{\tabcolsep}{5pt}
    \renewcommand{\arraystretch}{1.2}
  \begin{tabular}{c c c @{\,\hspace{40pt}\,} c c c}
      \toprule
      Observable & Limit & Ref. & Observable & Limit & Ref. \\
      \midrule
      $ {\cal B} (\tau^- \to e^- e^+ e^-) $ & $ 2.7 \cdot 10^{-8} $ & \cite{Hayasaka:2010np} & $ {\cal B} (\tau^- \to \mu^- \mu^+ \mu^-) $ & $ 2.1 \cdot 10^{-8} $ & \cite{Hayasaka:2010np}  \\ 
      $ {\cal B} (\tau^- \to e^- \mu^+ \mu^-) $ & $ 2.7 \cdot 10^{-8} $ & \cite{Hayasaka:2010np} & $ {\cal B} (\tau^- \to e^+ \mu^- \mu^-) $ & $ 1.7 \cdot 10^{-8} $ & \cite{Hayasaka:2010np} \\
      $ {\cal B} (\tau^- \to \mu^- e^+ e^-) $ & $ 1.8 \cdot 10^{-8} $ & \cite{Hayasaka:2010np} & $ {\cal B} (\tau^- \to \mu^+ e^- e^-) $ & $ 1.5 \cdot 10^{-8} $ & \cite{Hayasaka:2010np} \\ $ {\cal B} (\mu^- \to e^- e^+ e^-) $ & $ 1.0 \cdot 10^{-12} $ & \cite{SINDRUM:1987nra} & & & \\
      \bottomrule
  \end{tabular}
  \caption{Experimental upper  90\% CL limits on the branching fractions of LFV lepton decays. \label{tab:4LLFV}}
\end{table}

\subsubsection{Semileptonic $\tau$ Decays}
\label{sec:sl_tau_decays}

The decay rate of the $\tau$ lepton for a (charged-current) process involving a pseudoscalar meson $P$ and a neutrino is given by
\begin{equation}
  \Gamma (\tau \to \nu_i P) = \frac{m_\tau^3}{128\pi} f_P^2 \left(1 - \frac{m_P^2}{m_\tau^2} \right)^2 \lvert {\cal C}_{\underset{i3kl}{\nu edu}}^{V,LL} \rvert^2 \,,
  \label{eqn:tau_nuP_decay_width}
\end{equation}
where $f_P$ is the meson decay constant and $m_P$ the meson mass. 
In the SM, ${\cal C}_{\nu edu}^{V,LL}$ receives contributions from $W$ boson exchange.
The SM contribution can be read off from the matching conditions given in App.~\ref{app:matching_conditions}.
 In the computation of the SM predictions we  follow Refs.~\cite{Erler:2002mv, Arroyo-Urena:2021nil, HeavyFlavorAveragingGroupHFLAV:2024ctg}. 
Meson masses are taken from Ref.~\cite{ParticleDataGroup:2024cfk}, while the corresponding decay constants are listed in Appendix~\ref{app:input_parameters}, Tab.~\ref{tab:meson_decay_constants}.
The resulting SM predictions are presented in Tab.~\ref{tab:tau_SL_neutrinos}, together with the current experimental values.

\begin{table}[h]
  \setlength{\tabcolsep}{15pt}
  \renewcommand{\arraystretch}{1.2}
  \centering
  \begin{tabular}{l l l}
      \toprule
      Observable & Measurement & SM Prediction \\
      \midrule
      $ \Gamma (\tau^- \to \pi^- \bar \nu) $ & $2.453(12)\cdot 10^{-13}\,\text{GeV} $~\cite{HeavyFlavorAveragingGroupHFLAV:2024ctg} & $2.451(33)\cdot 10^{-13}\,\text{GeV} $ \\ 
      $ \Gamma (\tau^- \to K^- \bar \nu) $ & $1.578(23)\cdot 10^{-14}\,\text{GeV} $~\cite{HeavyFlavorAveragingGroupHFLAV:2024ctg} & $1.618(13)\cdot 10^{-14}\,\text{GeV} $ \\
      \hline
    \end{tabular}
    \caption{Experimental values and SM predictions of the partial widths of semileptonic $\tau$ decays.}
    \label{tab:tau_SL_neutrinos}
  \end{table}

We compute the LFV (neutral-current) semileptonic $\tau$ decay rates, $\Gamma(\tau \to \ell_i M)$, following Refs.~\cite{Angelescu:2020uug, Plakias:2023esq}. The decay  to a pseudoscalar  $P$ probes 
$\lvert \mathcal{C}_{eq}^{V,RR} - \mathcal{C}_{qe}^{V,LR} \rvert^2$ and $\lvert \mathcal{C}_{qe}^{V,LL} - \mathcal{C}_{eq}^{V,RL} \rvert^2$, while the decay rate to a vector meson $V$ probes the orthogonal combinations $\lvert \mathcal{C}_{eq}^{V,RR} + \mathcal{C}_{qe}^{V,LR} \rvert^2$ and $\lvert \mathcal{C}_{qe}^{V,LL} + \mathcal{C}_{eq}^{V,RL} \rvert^2$. 
Combining both types of final state mesons  thus allows to disentangle left- and right-handed  operator contributions.
\begin{table}[h]
  \setlength{\tabcolsep}{10pt}
  \renewcommand{\arraystretch}{1.2}
  \centering
  \begin{tabular}{c c @{\,\hspace{30pt}\,} c c @{\,\hspace{30pt}\,} c c}
        \toprule
      $\ell M$ & Limit & $\ell M$ & Limit & $\ell M$ & Limit \\
      \midrule
      $e \pi^0 $ & $ 8.0 \cdot 10^{-8} $~\cite{Belle:2007cio} & $ \mu \pi^0$ & $ 1.1 \cdot 10^{-7}$~ \cite{BaBar:2006jhm} & $ e\eta $ & $ 9.2 \cdot 10^{-8} $~\cite{Belle:2007cio} \\
      $ \mu \eta $ & $ 6.5 \cdot 10^{-8} $~\cite{Belle:2007cio} & 
      $ e \eta^{\prime} $ & $ 1.6 \cdot 10^{-7} $~\cite{Belle:2007cio} & $\mu \eta^{\prime} $ & $ 1.3 \cdot 10^{-7} $~\cite{Belle:2007cio} \\
      $  e \rho $ & $ 2.2 \cdot 10^{-8} $~\cite{Belle:2023ziz} & $ \mu \rho $ & $ 1.7 \cdot 10^{-8} $~\cite{Belle:2023ziz} & 
      $ e \omega $ & $ 2.4 \cdot 10^{-8} $~\cite{Belle:2023ziz} \\ 
      $\mu \omega $ & $ 3.9 \cdot 10^{-8} $~\cite{Belle:2023ziz} & 
      $ e K^{*} $ & $ 1.9 \cdot 10^{-8} $~\cite{Belle:2023ziz} & $\mu K^* $ & $ 2.9 \cdot 10^{-8} $~\cite{Belle:2023ziz} \\
      $ e\bar K^* $ & $ 1.7 \cdot 10^{-8} $~\cite{Belle:2023ziz} & $\mu \bar K^* $ & $ 4.3 \cdot 10^{-8} $~\cite{Belle:2023ziz} & 
      $e \phi $ & $ 2.0 \cdot 10^{-8} $~\cite{Belle:2023ziz} \\ $\mu \phi $ & $ 2.3 \cdot 10^{-8} $~\cite{Belle:2023ziz} & & & & \\
      \bottomrule
  \end{tabular}
  \caption{Experimental 90\% CL. limits on the branching fractions of the decay $\tau \to \ell M$.}
  \label{tab:tau_decay_limits}
\end{table}
Tab.~\ref{tab:tau_decay_limits} gives the  limits on the semileptonic LFV $\tau$ branching fractions at 90\% CL used in this analysis.

\subsubsection{Lepton Decays to dineutrinos}
\label{sec:cc_lepton_decays}

Charged lepton decays of the form $\ell_i \to \ell_j \nu_k \bar{\nu}_m$ are mediated by $W$-boson exchange in the SM. In the LEFT framework, the partial decay widths are given by
\begin{equation}
\Gamma (\ell_i \to \ell_j \nu_k \bar{\nu}_m) = \frac{m_{\ell_i}^5}{1536 \pi^3} \sum_{k,m}  \left(  \left| \mathcal{C}_{\underset{kmij}{\nu e}}^{V,LL} \right|^2+\left| \mathcal{C}_{\underset{kmij}{\nu e}}^{V,LR} \right|^2 \right)  \,,
\label{eqn:tau_decay_width_LEFT}
\end{equation}
where the sum is over all neutrino flavors since they  are not flavor-tagged. Their contributions add incoherently, allowing several flavor components  to be constrained simultaneously. In the SM, only the LFC terms with $i = k$ and $j = m$ contribute.
Tab.~\ref{tab:tau_decay_measurements} summarizes the current experimental values and SM predictions for the decay rates  of the charged lepton decays to dineutrinos.
\begin{table}[h]
  \setlength{\tabcolsep}{15pt}
  \renewcommand{\arraystretch}{1.2}
  \centering
  \begin{tabular}{l l l}
      \toprule
      Observable & Measurement & SM Prediction \\
      \midrule
      $ \Gamma (\mu^- \to e^- \nu \bar \nu) $ & $2.995984(3)\cdot 10^{-19}\,\text{GeV}$~\cite{HeavyFlavorAveragingGroupHFLAV:2024ctg} & $2.994(5)\cdot 10^{-19}\,\text{GeV}$ \\ 
      $ \Gamma (\tau^- \to e^- \nu \bar \nu) $ & $4.045(11)\cdot 10^{-13}\,\text{GeV} $~\cite{HeavyFlavorAveragingGroupHFLAV:2024ctg} & $4.028(7)\cdot 10^{-13}\,\text{GeV}$ \\ 
      $ \Gamma (\tau^- \to \mu^- \nu \bar \nu) $ & $3.936(11)\cdot 10^{-13}\,\text{GeV} $~\cite{HeavyFlavorAveragingGroupHFLAV:2024ctg} & $3.918(7)\cdot 10^{-13}\,\text{GeV}$ \\
      \hline
    \end{tabular}
    \caption{Experimental values and SM predictions of the  $\ell_i \to \ell_j \nu_k \bar{\nu}_m$  decay rates.}
    \label{tab:tau_decay_measurements}
  \end{table}

\subsection{Meson Decays}
\label{sec:meson_decays}

We discuss leptonic and semileptonic meson decays and their sensitivity to different  combinations of Wilson coefficients.

\subsubsection{Leptonic Meson Decays}

The purely leptonic decays of a pseudoscalar meson $P \to \ell_i \ell_j$ probe the combinations $\lvert \mathcal{C}_{eq}^{V,RR} - \mathcal{C}_{qe}^{V,LR} \rvert^2$ and $\lvert \mathcal{C}_{qe}^{V,LL} - \mathcal{C}_{eq}^{V,RL} \rvert^2$, analogous to semileptonic $\tau$ decays. In contrast, the charged-current process $P \to \ell \nu$ is sensitive to the coefficient $\mathcal{C}_{\nu edu}^{V,LL}$.
Since neutral meson decays involve FCNC transitions, they are strongly suppressed in the SM by the GIM mechanism. With the additional lepton mass suppression from the chiral flip, most leptonic decays of neutral mesons are currently constrained only by experimental upper limits, with the exception of $B_s \to \mu^+ \mu^-$, which is in good agreement with the SM prediction.
Tab.~\ref{tab:leptonic_modes} lists the leptonic decay modes of pseudoscalar mesons considered in this analysis. We do not include LFC decays of light mesons, as they are typically dominated by long distance contributions.
Recall that  in our MFV setup  no up-type FCNCs are induced so no BSM contribution to FCNC $D$-meson decays is present.

\begin{table}[h]
  \setlength{\tabcolsep}{15pt}
  \renewcommand{\arraystretch}{1.2}
  \centering
  \begin{tabular}{c c c c}
      \toprule
      Meson & final states & Meson & final states \\
        \midrule
        $\pi^0$ & $e\mu$ & $\eta$ & $e\mu$ \\
        $\eta^\prime$ & $e\mu$ & $K_L$ & $e\mu$ \\
        $B^0$ & $ee$, $\mu\mu$, $\tau \tau$, $e\mu$, $e\tau$, $\mu\tau$ & $B_s$ & $ee$, $\mu\mu$, $\tau \tau$, $e\mu$, $e\tau$, $\mu\tau$ \\
        \midrule 
        $\pi^+$ & $e\nu$, $\mu\nu$ & $K^+$ & $e\nu$, $\mu\nu$ \\
        $D^+$ & $e\nu$, $\mu\nu$, $\tau \nu$ & $D_s^+$ & $e\nu$, $\mu\nu$, $\tau \nu$ \\
        $B^+$ & $e\nu$, $\mu\nu$, $\tau \nu$ & & \\
        \bottomrule
    \end{tabular}
    \caption{Leptonic decay modes of pseudoscalar mesons considered in this analysis. The upper part lists the neutral mesons, while the lower part lists the charged mesons.}
    \label{tab:leptonic_modes}
\end{table}

Leptonic decays of quarkonium states, $V \to \ell \ell$, probe the orthogonal combinations of Wilson coefficients, $\lvert \mathcal{C}_{eq}^{V,RR} + \mathcal{C}_{qe}^{V,LR} \rvert^2$ and $\lvert \mathcal{C}_{qe}^{V,LL} + \mathcal{C}_{eq}^{V,RL} \rvert^2$, complementary to pseudoscalar meson decays. 
These decays also provide access to the quark-flavor-diagonal components of the Wilson coefficients, which dominate in the MFV setup.

Unlike pseudoscalar meson decays, vector meson decays are not affected by helicity suppression, such that the decay $V \to \nu \bar{\nu}$ is allowed in the SM. 
The partial width reads~\cite{Li:2020lba}
\begin{equation}
\Gamma (V \to \bar{\nu}_i \nu_j) = \frac{f_V^2 m_V^3}{96 \pi} \left| \mathcal{C}_{\underset{ij}{\nu q}}^{V,LL} + \mathcal{C}_{\underset{ij}{\nu q}}^{V,LR} \right|^2 \,,
\label{eqn:V_ll_LFC_decay_width}
\end{equation}
which probes the semileptonic dineutrino operators of the LEFT and thereby complements SMEFT studies into charged leptons, as the processes are related in SMEFT by  $SU(2)_L$.
SM contributions can be read off from the matching conditions given in App.~\ref{app:matching_conditions}.

The quarkonium decay modes considered in this analysis are given in Tab.~\ref{tab:quarkonia_decays}. 
While the LFC modes with charged leptons are measured with high precision, the dineutrino and LFV modes are currently constrained only by  upper limits, see Tab.~\ref{tab:quarkonia}.
\begin{table}
  \setlength{\tabcolsep}{15pt}
  \renewcommand{\arraystretch}{1.2}
  \centering
  \begin{tabular}{c c}
      \toprule
      Meson & Final States \\
      \midrule
        $\Phi$ & $e^+ e^-$, $\mu^+ \mu^-$, $e \mu$, $\nu \bar \nu$ \\ 
        $J/\Psi$ & $e^+ e^-$, $\mu^+ \mu^-$, $e \mu$, $e \tau$, $\mu \tau$, $\nu \bar \nu$ \\
        $\Upsilon(1S)$ & $e^+ e^-$, $\mu^+ \mu^-$, $\tau^+ \tau^-$, $e \mu$, $e \tau$, $\mu \tau$, $\nu \bar \nu$ \\
      \bottomrule
  \end{tabular}
  \caption{Leptonic decay modes of vector mesons considered in this analysis.}
    \label{tab:quarkonia_decays}
\end{table}
\begin{table}[h]
  \setlength{\tabcolsep}{15pt}
  \renewcommand{\arraystretch}{1.2}
  \centering
  \begin{tabular}{l l l}
      \toprule
      Observable & Measurement & SM Prediction \\
      \midrule
      $ {\mathcal{B}}(\Phi\to  \nu \bar \nu) $ & $1.7 \cdot 10^{-4}$~\cite{ParticleDataGroup:2024cfk} & $3.22 \cdot 10^{-10}$ \\ 
      $ {\mathcal{B}}(J/\Psi  \to \nu \bar \nu) $ & $7.0 \cdot 10^{-4}$~\cite{ParticleDataGroup:2024cfk}  & $2.22\cdot 10^{-8}$ \\ 
      $ {\mathcal{B}} (\Upsilon(1S) \to  \nu \bar \nu) $ & $3.0 \cdot 10^{-4}$~\cite{ParticleDataGroup:2024cfk}  & $9.28\cdot 10^{-6}$ \\
      \hline
    \end{tabular}
    \caption{90\% CL limits  and SM predictions 
     of  $(q \bar q)  \to \nu \bar{\nu}$  branching ratios.}
    \label{tab:quarkonia}
  \end{table}

\subsubsection{Semileptonic Meson Decays}

Semileptonic $P \to M \ell \nu$  and $P \to M \ell \ell$  decays are sensitive to different combinations of Wilson coefficients compared to purely leptonic decays, allowing
to disentangle chirality structures. As three-body decays, they offer also to measure angular distributions as functions of the  hadronic momentum transfer $q^2$. 

Tab.~\ref{tab:semileptonic_P_modes_cc} lists the charged-current decays $P \to M \ell \nu$ considered in this work. For these processes, we compute the SMEFT and SM predictions following Refs.~\cite{Korner:1990ri, Wu:2006rd, HeavyFlavorAveragingGroupHFLAV:2024ctg}. The form factors used in the computation are listed in Appendix~\ref{app:input_parameters}, Tab.~\ref{tab:form_factors}. 

\begin{table}[htbp]
    \setlength{\tabcolsep}{15pt}
    \renewcommand{\arraystretch}{1.2}
    \centering
    \begin{tabular}{c c c c}
        \toprule
        Meson & $e\nu$ & $\mu\nu$ & $\tau\nu$ \\
        \midrule
        $D^0$ & $K^-$, $K^{*-}$, $\pi^-$, $\rho^-$ & $K^-$, $K^{*-}$, $\pi^-$, $\rho^-$ & \\
        $D^+$ & $\bar K^0$, $\bar K^{*0}$, $\pi^0$, $\eta$, $\rho$, $\omega$ & $\bar K^0$, $\bar K^{*0}$, $\pi^0$, $\eta$, $\rho$, $\omega$ & \\
        $D_s$ & $\phi$, $K^{*0}$ & $\phi$ & \\
        $B^0$ & $D^-$, $D^{*-}$, $\pi^-$, $\rho^-$ & $D^-$, $D^{*-}$, $\pi^-$, $\rho^-$ & $D^-$, $D^{*-}$, $\pi^-$ \\
        $B^+$ & $\bar D^0$, $D^{*}$, $\pi^0$, $\rho^0$ & $\bar D^0$, $D^{*}$, $\pi^0$, $\rho^0$ & $\bar D^0$, $D^{*}$ \\
        $B_s$ & & $D_s^-$, $D_s^{*-}$, $K^-$ & \\
        \bottomrule
    \end{tabular}
    \caption{Semileptonic charged-current decays included in this analysis. The decay channels are grouped according to the lepton flavor of the final state.}
    \label{tab:semileptonic_P_modes_cc}
  \end{table}

The decay rate for $P \to P^{\prime} \ell \ell$ processes  is typically significantly larger than that of purely leptonic decays $P \to \ell \ell$, as the former is not chirally-suppressed. The resulting higher statistics allow for differential measurements, which provide enhanced sensitivity to the Wilson coefficients. Due to the increased complexity in computing LEFT contributions here, we utilize the Python package \texttt{flavio}~\cite{Straub:2018kue} to evaluate the SMEFT contributions, and we adopt the corresponding SM predictions from the literature.
The observables used in this analysis are given in Tab.~\ref{tab:semileptonic_P_modes_P}.

\begin{table}[h]
    \setlength{\tabcolsep}{15pt}
    \renewcommand{\arraystretch}{1.2}
    \centering
    \begin{tabular}{l l c c}
        \toprule
        Observables & Data & Ref & SM Ref \\
        \midrule
        $\Gamma (B^0 \to \pi^0 e^+ e^-)$ & 90\% CL limit & \cite{BaBar:2013qaj} & \cite{Soni:2020bvu}\\
        $\Gamma (B^0 \to \pi^0 \mu^+ \mu^-)$ & 90\% CL limit & \cite{BaBar:2013qaj} & \cite{Soni:2020bvu} \\
        $\Gamma (B^+ \to \pi^+ e^+ e^-)$ & 90\% CL limit & \cite{Belle:2008tjs} & \cite{Soni:2020bvu} \\
        $\bigl\langle\tfrac{d\sigma}{dq^2}\bigr\rangle(B^+ \to \pi^+ \mu^+ \mu^-)$ & 9 bins & \cite{LHCb:2015hsa} & \cite{FermilabLattice:2015cdh, Bause:2022rrs} \\ 
        $\Gamma (B^0 \to K^0 e^+ e^-)$ & full kinematic range & \cite{ParticleDataGroup:2024cfk} & \cite{Parrott:2022zte}\\
        $\bigl\langle\tfrac{d\sigma}{dq^2}\bigr\rangle(B^0 \to K^{0} \mu^+ \mu^-)$ & 6 bins & \cite{LHCb:2014cxe} & \cite{Parrott:2022zte} \\
        $\Gamma (B^+ \to K^+ e^+ e^-)$ & full kinematic range & \cite{ParticleDataGroup:2024cfk} & \cite{Parrott:2022zte} \\
        $\bigl\langle\tfrac{d\sigma}{dq^2}\bigr\rangle(B^+ \to K^+ \mu^+ \mu^-)$ & 11 bins & \cite{LHCb:2014cxe} & \cite{Gubernari:2022hxn, Parrott:2022zte}\\ 
        $\Gamma(B^+ \to K^+ \tau^+ \tau^-)$ & 90\% CL limit & \cite{BaBar:2016wgb} & \cite{Parrott:2022zte} \\
        \bottomrule 
    \end{tabular}
    \caption{Observables of $P \to P^{\prime} \ell_i^+ \ell_i^-$ decays considered in this analysis.}
    \label{tab:semileptonic_P_modes_P}
\end{table}

The semileptonic decay $P \to V \ell \ell$, involving a vector meson in the final state, features additional polarization degrees of freedom and thus offers a richer angular distribution.
The various helicity amplitudes are commonly expressed in terms of observables such as the longitudinal polarization fraction $F_L(q^2)$, the forward--backward asymmetry $A_{\text{FB}}(q^2)$, and the normalized angular coefficients $S_i(q^2)$~\cite{Altmannshofer:2008dz}. Many analyses further improve sensitivity by using optimized observables $P_i^{(\prime)}(q^2)$~\cite{Matias:2012xw, Descotes-Genon:2013vna}, which are designed to reduce both experimental and theoretical uncertainties. For the computation of the SMEFT contributions, we employ the \texttt{flavio} package~\cite{Straub:2018kue}, while SM predictions are taken from the literature.
We list the observables for the LFC semileptonic decays $P \to V \ell_i^+ \ell_i^-$ in Tab.~\ref{tab:semileptonic_P_modes_Vll}.

\begin{table}[h]
    \setlength{\tabcolsep}{15pt}
    \renewcommand{\arraystretch}{1.2}
    \centering
    \begin{tabular}{l l c c}
        \toprule
        Process & Observables & Ref & SM Ref \\
        \midrule 
        $B^0 \to K^{*0} e^+ e^-$ & $F_L$, $P_i^{(\prime)}$ (1 bin) & \cite{LHCb:2025pxz} & \cite{Alguero:2023jeh} \\
        $B^0 \to K^{*0} \mu^+ \mu^-$ &  $F_L$, $P_i^{(\prime)}$ (8 bins) & \cite{LHCb:2020lmf} & \cite{Alguero:2023jeh,Straub:2018kue} \\
        $B^0 \to K^{*0} \tau^+ \tau^-$ & 90\% CL limit on ${\cal B}$ & \cite{Belle-II:2025lwo} & \cite{Straub:2018kue} \\
        $B^+ \to K^{*+} e^+ e^-$ & $\Gamma(B^+ \to K^{*+} e^+ e^-)$ & \cite{ParticleDataGroup:2024cfk} & \cite{Straub:2018kue} \\
        $B^+ \to K^{*+} \mu^+ \mu^-$ &  $F_L$, $P_i^{(\prime)}$ (8 bins) & \cite{LHCb:2020gog} & \cite{Alguero:2023jeh,Straub:2018kue} \\
        $B_s \to \phi e^+ e^-$ & $F_L$, $S_3$ (3 bins) & \cite{LHCb:2025rfy} & \cite{Straub:2018kue} \\
        $B_s \to \phi \mu^+ \mu^-$ & $F_L$, $S_3$, $S_4$, $S_7$ (5 bins) & \cite{LHCb:2021xxq} &  \cite{Straub:2018kue} \\
        $B_s \to K^* \mu^+ \mu^-$ & $\Gamma(B_s \to K^* \mu^+ \mu^-)$ & \cite{LHCb:2018rym} & \cite{Kindra:2018ayz} \\
        \bottomrule 
    \end{tabular}
    \caption{Observables of $P \to V \ell_i^+ \ell_i^-$ decays considered in this work.}
    \label{tab:semileptonic_P_modes_Vll}
\end{table}

Searches have also been conducted for LFV decays of the form $P \to P^{\prime} \ell_i \ell_j$ and $P \to V \ell_i \ell_j$ with $i\neq j$. As there is no signal in the SM, only upper limits on the branching fractions are available. The channels included in our analysis are listed in Tab.~\ref{tab:semileptonic_P_modes_LFV}.

\begin{table}[htbp]
    \setlength{\tabcolsep}{15pt}
    \renewcommand{\arraystretch}{1.2}
    \centering
    \begin{tabular}{c c c c}
        \toprule
        Meson & $e\mu$ & $e\tau$ & $\mu\tau$ \\
        \midrule
        $B^0$ & $\pi^0$, $K^0$, $K^{*0}$ & $K^{*0}$ & $K^{*0}$ \\
        $B^+$ & $\pi^+$, $K^+$, $K^{*+}$ & $\pi^+$, $K^+$, $K^{*+}$ & $\pi^+$, $K^+$ \\
        $B_s$ & $\phi$ & & \\
        \bottomrule
    \end{tabular}
    \caption{Semileptonic LFV meson decays included in this work. }
    \label{tab:semileptonic_P_modes_LFV}
  \end{table}

We also consider semileptonic decays into dineutrinos, $P \to M \nu \bar \nu$. In our MFV setup, FCNC decays   of this type are sensitive only to the  coefficient $\mathcal{C}_{\nu q}^{V,LL}$, since no right-handed FCNCs are induced. We compute the SM and SMEFT predictions following Ref.~\cite{Felkl:2021uxi}.

The FCNC dineutrino decays are currently constrained only by upper limits, with the exceptions of $K^+ \to \pi^+ \nu \bar{\nu}$ and $B^+ \to K^+ \nu \bar{\nu}$. The latter has recently been measured by the Belle II collaboration~\cite{Belle-II:2023esi}, yielding ${\cal{B}}(B^+ \to K^+ \nu \bar \nu) = 2.3(0.7) \cdot 10^{-5}$ which deviates from the SM prediction of 
$5.58(0.37) \cdot 10^{-6}$~\cite{Parrott:2022zte} at the $2.7\sigma$ level. The relevant observables are listed in Tab.\ref{tab:semileptonic_P_modes_dineutrinos}.\footnote{The BESIII collaboration has set an upper limit of ${\mathcal{B}(D^0 \to \pi^0 \nu \bar{\nu}) < 2.1 \cdot 10^{-4}}$~\cite{BESIII:2021slf}. As our MFV setup does not induce $c \to u$ FCNCs, this bound gives no constraint in our analysis.}

\begin{table}[htbp]
    \setlength{\tabcolsep}{15pt}
    \renewcommand{\arraystretch}{1.2}
    \centering
    \begin{tabular}{l l l}
        \toprule
        Observable & Measurement [GeV] & SM Prediction [GeV] \\
        \midrule
        $\Gamma (K^+ \to \pi^+ \nu \bar \nu)$ & $6.1(2.1) \cdot 10^{-27}$~\cite{ParticleDataGroup:2024cfk} & $4.18(0.32) \cdot 10^{-27}$~\cite{DAmbrosio:2022kvb} \\
        $\Gamma (B^0 \to \pi^0 \nu \bar \nu)$ & $< 3.9 \cdot 10^{-18}$~\cite{Belle:2017oht} & $2.56(0.55) \cdot 10^{-20}$ \\
        $\Gamma (B^0 \to K^0 \nu \bar \nu)$ & $< 1.1 \cdot 10^{-17}$~\cite{Belle:2017oht} & $1.78(0.10) \cdot 10^{-18}$ \\
        $\Gamma (B^+ \to \pi^+ \nu \bar \nu)$ & $< 5.6 \cdot 10^{-18}$~\cite{Belle:2017oht} & $5.1(1.1) \cdot 10^{-20}$ \\
        $\Gamma (B^+ \to K^+ \nu \bar \nu)$ & $9.2(2.8) \cdot 10^{-18}$~\cite{Belle-II:2023esi} & $2.08(0.10) \cdot 10^{-18}$ \\
        \midrule
        $\Gamma (B^0 \to \rho^0 \nu \bar \nu)$ & $< 1.7 \cdot 10^{-17}$~\cite{Belle:2017oht} & $7.4(1.1) \cdot 10^{-20}$ \\
        $\Gamma (B^0 \to K^{*0} \nu \bar \nu)$ & $< 7.8 \cdot 10^{-18}$~\cite{Belle:2017oht} & $3.95(0.36) \cdot 10^{-18}$ \\
        $\Gamma (B^+ \to \rho^+ \nu \bar \nu)$ & $< 1.2 \cdot 10^{-17}$~\cite{Belle:2017oht} & $1.48(0.21) \cdot 10^{-19}$ \\
        $\Gamma (B^+ \to K^{*+} \nu \bar \nu)$ & $< 1.6 \cdot 10^{-17}$~\cite{Belle:2017oht} & $3.95(0.36) \cdot 10^{-18}$ \\
        $\Gamma (B_s \to \phi \nu \bar \nu)$ & $< 2.3 \cdot 10^{-15}$~\cite{DELPHI:1996ohp} & $4.2(0.3) \cdot 10^{-18}$ \\
        \bottomrule
    \end{tabular}
    \caption{Measurements and  90\% CL limits on the decay rates of semileptonic meson decays $P \to M \nu \bar \nu$, alongside the corresponding SM predictions. 
     The upper (lower) part presents measurements into  pseudoscalar (vector) mesons. 
    Resonance contributions from $B^+ \to M^+ (\to \tau \nu) \nu$ affecting  decays of charged $B$-mesons, see Sec.~\ref{sec:BKnunu}, are not included  in the SM predictions. }
     \label{tab:semileptonic_P_modes_dineutrinos}
\end{table}

\section{Global Analysis}
\label{sec:global_analysis}

In this section, we present the results of the  global analysis. We outline the fit methodology and the underlying statistical framework in Sec.~\ref{sec:fitM}. We discuss the outcome of the fit, comparing results for individual lepton flavors with those from lepton-flavor-universal and lepton-flavor-democratic scenarios for different types of operators in Secs.~\ref{sec:HCO}-\ref{sec:ST}. We examine the posterior distributions of the MFV ratios  $r_L$ and $r_R$ (Sec.~\ref{sec:MFV}) and the fit results of the CKM matrix (Sec.~\ref{sec:CKM}). In Sec.~\ref{sec:BKnunu} we use the posteriors from the global analysis  to make predictions for the branching fractions of $B \to P \nu \bar \nu$ and $B \to V \nu \bar\nu$ decays.

\subsection{Fit Methodology \label{sec:fitM}}

We perform global fits to the Wilson coefficients of the SMEFT operators listed in Tab.~\ref{tab:operators}, using rescaled Wilson coefficients $\tilde C$, see (\ref{eq:scaling}).
For LFV interactions, the dipole, scalar and tensor coefficients $\tilde C_{ij}$ and $\tilde C_{ji}$ are treated as two independent degrees of freedom, whereas for vector operators, hermiticity imposes the relation ${\tilde C_{ij} = \tilde C_{ji}^*}$.
We further include the four Wolfenstein parameters of the CKM matrix as nuisance parameters, along with the  five MFV ratios  defined in Eq.~\eqref{eqn:MFV_degrees_of_freedom}, for  $O_{lq}^{(1/3)}$, $O_{qe}$, $O_{lu}$, and $O_{eu}$. In total, this results in 26 degrees of freedom for the LFC or universal fits, and 30  for LFV scenarios. 

To constrain the high-dimensional parameter space, we adopt a Bayesian framework using the \textit{EFTfitter} package~\cite{Castro:2016jjv}, based on the Bayesian Analysis Toolkit (BAT.jl)~\cite{Caldwell:2008fw, Schulz:2020ebm}. As a preparatory step, we perform prefits using Reactive Nested Sampling, as implemented in the \texttt{UltraNest} package~\cite{Buchner:2021cql}, to determine suitable prior ranges for the Wilson coefficients.
For the final fit, we employ uniform priors within the optimized ranges and use the Robust Adaptive Metropolis--Hastings (RAM) algorithm~\cite{Vihola:2012} to sample the posterior distribution of the Wilson coefficients. We run $N_{\text{chains}} = 10$ independent chains, each with $2\cdot 10^6$ iterations, and discard the first 30\% of samples as burn-in.

To present  the  outcome of the global fit to the  full set of  the 17 Wilson coefficients we discuss the constraints separately for the five, phenomenologically distinct classes:
   Higgs--current operators (Sec.~\ref{sec:HCO}),  semileptonic four--fermion operators with vector couplings (Sec.~\ref{sec:SL4F}), four--lepton operators (Sec.~\ref{sec:4L}),  leptonic dipole operators (Sec.~\ref{sec:dipole}),  and  scalar and tensor operators (Sec.~\ref{sec:ST}).

\subsection{Higgs-Current Operators  \label{sec:HCO}}

\begin{figure}[h]
    \centering
    \includegraphics[width=\textwidth]{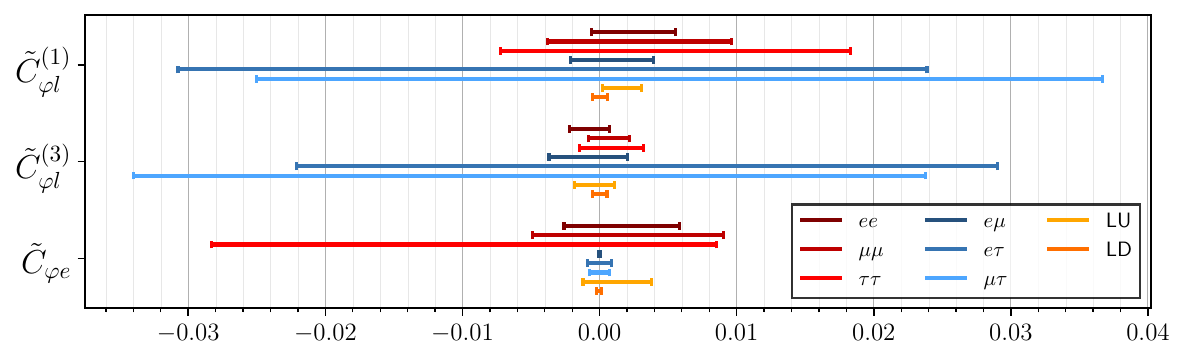}
    \caption{90\% credible intervals of the Wilson coefficients $\tilde C_{\varphi l}^{(1)}$, $\tilde C_{\varphi l}^{(3)}$ and $\tilde C_{\varphi e}$ from lepton-flavor specific global SMEFT fits.}
    \label{fig:credible_higgs_current}
\end{figure}

\begin{figure}[h]
    \centering
    \includegraphics[width=\textwidth]{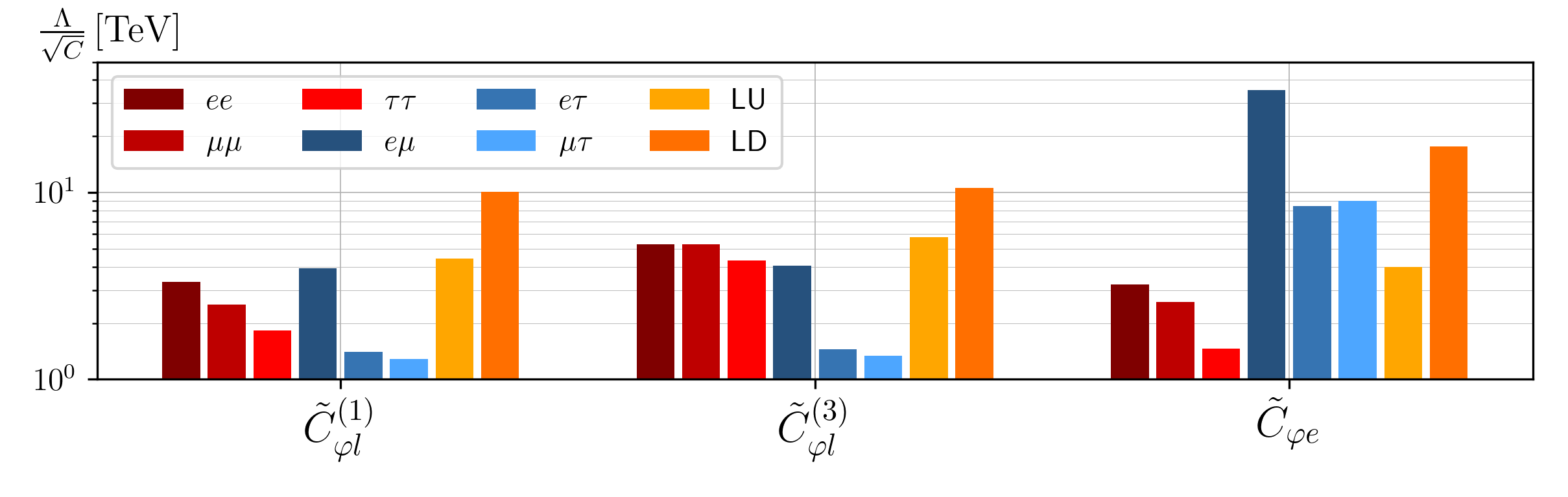}
    \caption{Lower bounds on $\Lambda/\sqrt{C}$ at the 90\% credible level for the Wilson coefficients $\tilde C_{\varphi l}^{(1)}$, $\tilde C_{\varphi l}^{(3)}$ and $\tilde C_{\varphi e}$ from lepton-flavor specific global SMEFT fits.}
    \label{fig:scale_higgs_current}
\end{figure}

We consider the  leptonic Higgs-current coefficients  $\tilde C_{\varphi l}^{(1)}$, $\tilde C_{\varphi l}^{(3)}$ and $\tilde C_{\varphi e}$. Fig.~\ref{fig:credible_higgs_current} shows the 90\% credible intervals for these Wilson coefficients for the different lepton-flavor scenarios.
LFC coefficients are shown in red, LFV coefficients in blue, and lepton-flavor-universal scenarios in orange. All Wilson coefficients are consistent with zero, corresponding to the SM expectation. In addition, Fig.~\ref{fig:scale_higgs_current} shows the resulting lower bounds on the new-physics scale $\Lambda/\sqrt{C}$. Numerical values of the 90\% credible intervals for the Higgs-current Wilson coefficients are given in Tab.~\ref{tab:credible_intervals_hl}.

The most stringent bounds are obtained in the $e\mu$ scenario, with a 90\% credible interval of $[-4.8,4.3]\cdot 10^{-5}$ for $\tilde C_{\varphi e}$, driven by the tight experimental limits on LFV decays. This corresponds to a reach of $\Lambda/\sqrt{C} \gtrsim 30\,\text{TeV}$. In contrast, the weakest bounds are found for the $e\tau$ and $\mu\tau$ couplings, where the upper bounds on $\tilde C_{\varphi l}^{(1)} $ and $\tilde C_{\varphi l}^{(3)}$ are $ \lvert \tilde C \rvert \lesssim 0.3$, testing scales of order $1\,\text{TeV}$.
In general, coefficients involving $\tau$ leptons are less constrained due to the experimental challenges associated with reconstructing $\tau$ decay products.

\begin{figure}[h]
    \centering
    \includegraphics[width=0.8\textwidth]{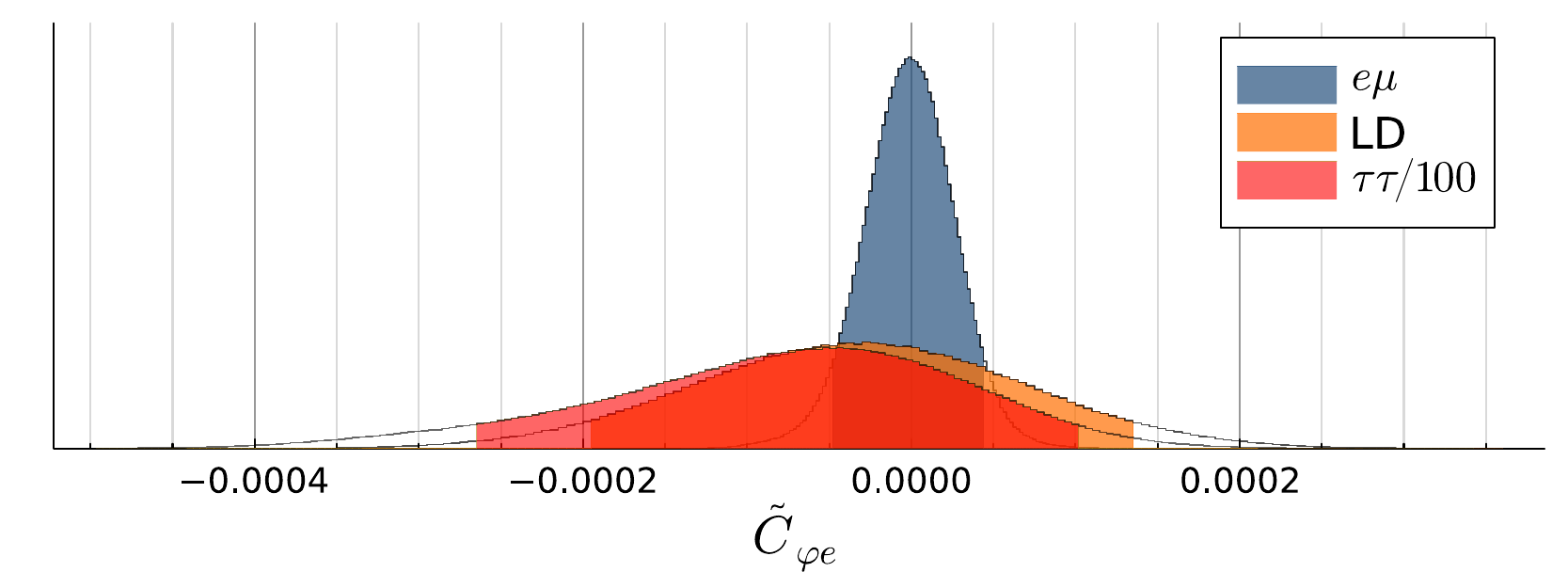}
    \caption{Posterior distributions in the $e\mu$ fit (blue), the combined LD scenario (orange), and the $\tau\tau$ channel (red,  rescaled by a factor of 100 for visibility). While the $e\mu$ fit yields the strongest constraint, the additional channels in the LD combination broadens the interval due to small, non-significant deviations, in particular from $\tau\tau$ resulting in a weaker overall bound.}
    \label{fig:che_posterior}
\end{figure}

In the LD scenario, the resulting bound on $\tilde C_{\varphi e}$ is weaker than in the $e\mu$ fit alone. This effect arises because several channels included in the combination, such as $Z\to \tau\tau$, display small but statistically non-significant deviations from the SM expectation. These fluctuations broaden the overall posterior distribution and thereby relax the constraints compared to the more precise $Z\to e\mu$ channel. This behavior is illustrated in Fig.~\ref{fig:che_posterior}, where we compare the posteriors of the $e\mu$ and LD fits alongside the $\tau\tau$ contribution (rescaled by a factor of 100 for visibility).

For the LFC couplings $ee$, $\mu\mu$, and $\tau\tau$, we find that $\tilde C_{\varphi l}^{(3)}$ is better constrained than $\tilde C_{\varphi l}^{(1)}$. The reason is that the triplet operator $O_{lq}^{(3)}$ also contributes to charged-current processes, whereas the singlet operator only contributes to neutral-currents. The coefficient $\tilde C_{\varphi e}$ is constrained to a similar extend as $\tilde C_{\varphi l}^{(1)}$, since both affect the observable to a similar magnitude. The differences mainly arise from the interference term between the SM and SMEFT amplitudes, since the $Z$ boson does not couple equally to left- and right-handed leptons. 

In case of the LFV couplings $e\mu$, $e\tau$, and $\mu\tau$, the bounds on $\tilde C_{\varphi l}^{(1)}$ and $\tilde C_{\varphi l}^{(3)}$ are of similar magnitude, whereas the constraints on $\tilde C_{\varphi e}$ are significantly stronger. This distinction arises because the  combination $\tilde C_{\varphi l}^{-}$, governing the $Z\to \nu \bar \nu$ vertex, is only weakly constrained compared to the couplings of charged leptons proportional to $\tilde C_{\varphi l}^{+}$. Moreover, the LFV components of $\tilde C_{\varphi l}^{(3)}$ do not interfere with the SM amplitude, rendering them less constrained. Consequently, the weak bounds on $\tilde C_{\varphi l}^{-}$ remain unbroken by additional limits on $\tilde C_{\varphi l}^{(3)}$, allowing for a larger viable parameter space for both, $\tilde C_{\varphi l}^{(1)}$ and $\tilde C_{\varphi l}^{(3)}$.

In the LU and LD scenarios, the bounds on $\tilde C_{\varphi l}^{(1)}$ and $\tilde C_{\varphi l}^{(3)}$ are significantly tighter than in the flavor-specific fits. This improvement results from the strong constraints on $\tilde C_{\varphi l}^{(3)}$, particularly from the $\mu\mu$ coupling, which help to lift the degeneracy associated with $\tilde C_{\varphi l}^{-}$ and thus enable the stringent limits on $\tilde C_{\varphi l}^{+}$, e.g. from $Z \to e\mu$, to be fully exploited. 

\begin{figure}[h]
    \centering
    \includegraphics[width=0.7\textwidth]{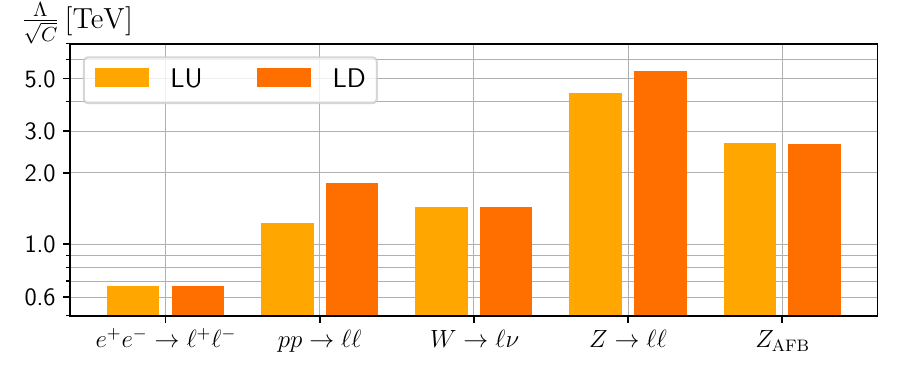}
    \caption{Lower bounds on $\Lambda/\sqrt{C}$ at the 90\% credible level for the Wilson coefficient $\tilde C_{\varphi l}^{(3)}$ resulting from single-coefficient fits to individual sectors of collider observables. We compare fits in the LU scenario (yellow) to the LD scenario (orange).}
    \label{fig:single_chl3_collider}
\end{figure}

\begin{figure}[h]
    \centering
    \includegraphics[width=0.9\textwidth]{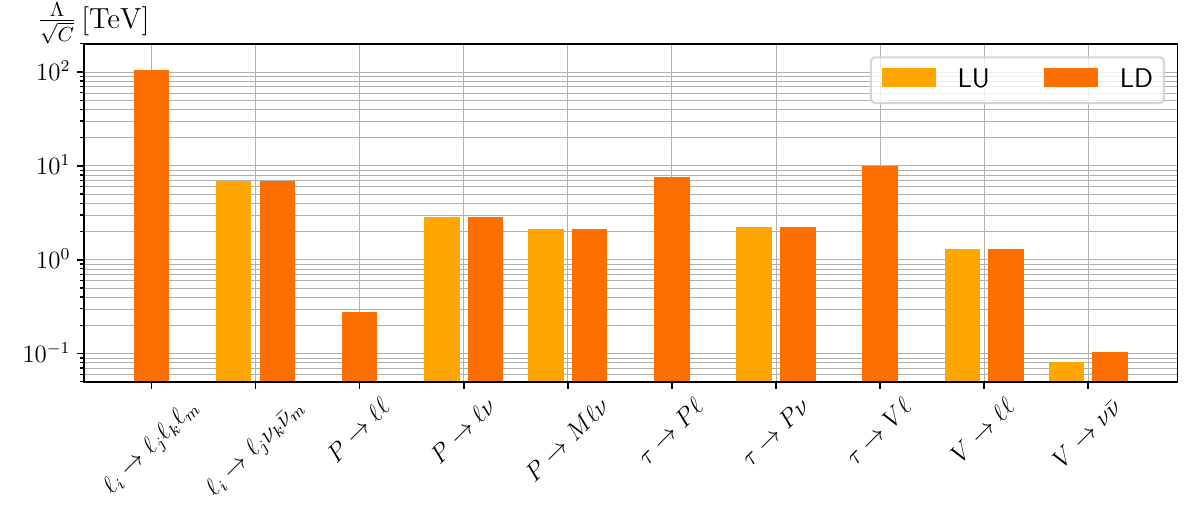}
    \caption{Lower bounds on $\Lambda/\sqrt{C}$ at the 90\% credible level for the Wilson coefficient $\tilde C_{\varphi l}^{(3)}$ resulting from single-coefficient fits to individual sectors of flavor observables. We compare fits in the LU scenario (yellow) to the LD scenario (orange).}
    \label{fig:single_chl3_flavor}
\end{figure}

To assess the impact of the different sectors, we perform single-coefficient fits in which all Wilson coefficients except for $\tilde C_{\varphi l}^{(3)}$ are set to zero, and we consider one sector at a time. The resulting sensitivities are displayed in Fig.~\ref{fig:single_chl3_collider} for collider observables and in Fig.~\ref{fig:single_chl3_flavor} for flavor data. We compare the lepton-flavor universal and the lepton-flavor democratic scenarios, with the latter including LFV observables.
Overall, the strongest constraints arise from LFV lepton decays, which are probed with very  high sensitivity. Since these processes are essentially null tests, the corresponding bounds are generally much tighter than those from lepton-flavor conserving observables. Among the latter, leptonic $\ell_i \to \ell_j \nu \bar \nu$ decays and $Z$-pole measurements provide the most stringent constraints.

\subsection{Semileptonic Four-Fermion Operators \label{sec:SL4F}}

\begin{figure}[h]
    \centering
    \includegraphics[width=\textwidth]{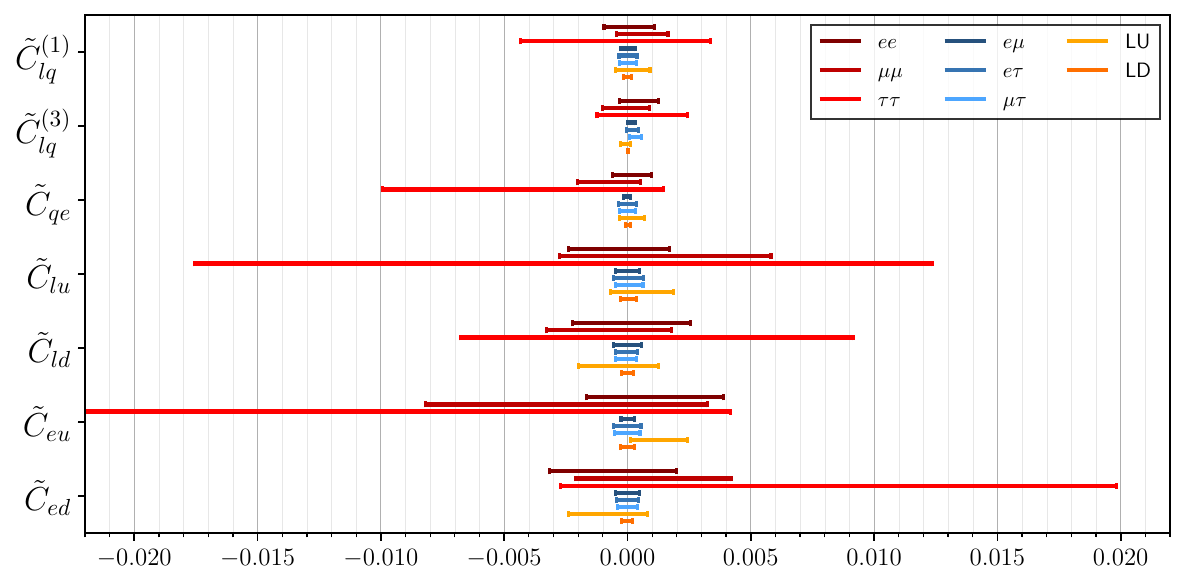}
    \caption{90\% credible intervals for the semileptonic four-fermion coefficients with  vector structure. }
      \label{fig:credible_lq}
\end{figure}

\begin{figure}[h]
    \centering
    \includegraphics[width=\textwidth]{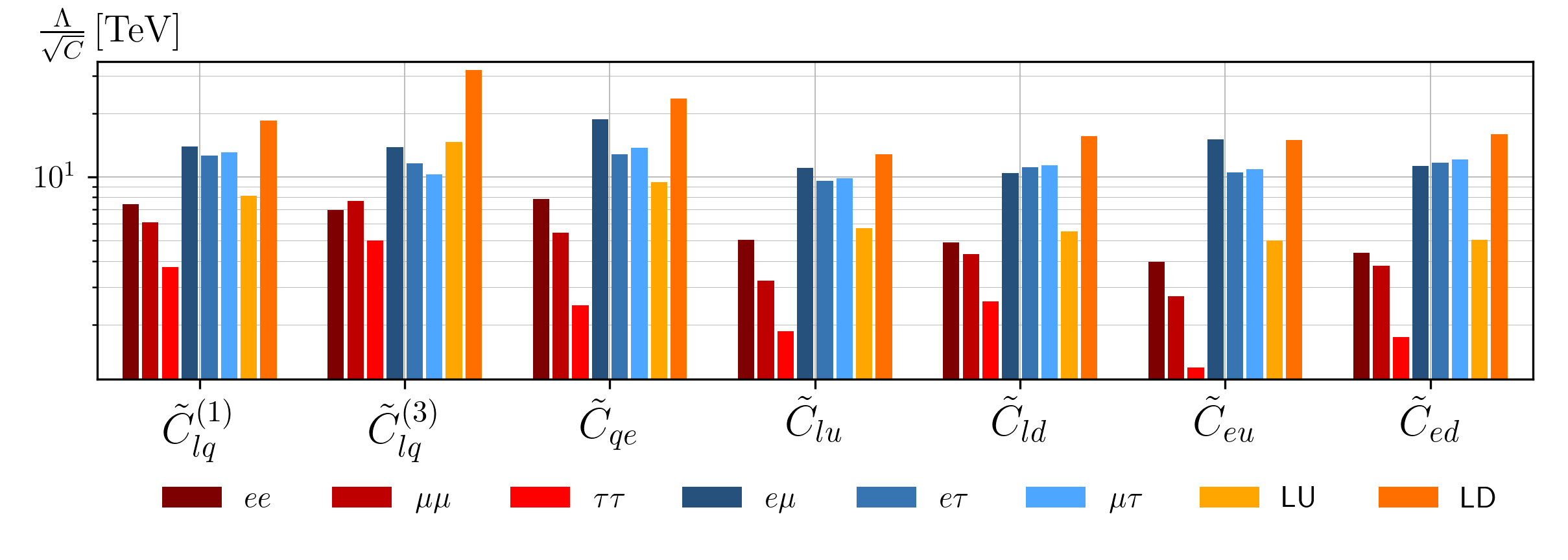}
    \caption{Lower bounds on $\Lambda/\sqrt{C}$ at 90\% obtained from the 90\% credible intervals for the semileptonic four-fermion coefficients with vector structure.}
    \label{fig:scale_lq}
\end{figure}

We consider semileptonic four-fermion coefficients with vector couplings $\tilde C_{lq}^{(1)}$, $\tilde C_{lq}^{(3)}$, $\tilde C_{qe}$, $\tilde C_{lu}$, $\tilde C_{eu}$, $\tilde C_{ld}$ and $\tilde C_{ed}$. 
Fig.~\ref{fig:credible_lq} shows the 90\% credible intervals for these Wilson coefficients obtained from our global SMEFT fit. The numerical values are provided in Appendix~\ref{app:credible_intervals}, Tabs.~\ref{tab:credible_intervals_lq_l} and~\ref{tab:credible_intervals_lq_r}. The corresponding lower bounds on the NP scale $\Lambda$ are shown in Fig.~\ref{fig:scale_lq}.

We find that the bounds on LFV coefficients are particularly stringent, probing NP scales of order $\Lambda / \sqrt{C} \sim 10$\,TeV. These limits arise primarily from flavor observables such as LFV $\tau$ or meson decays, as well as from the analyses of LFV DY processes $p p \to \ell_i \ell_j$. The bounds on LFV coefficients are of similar strength for the different operators and we see only minor variations across the LFV lepton flavor scenarios. 

In contrast, the bounds on LFC coefficients are generally weaker due to larger SM backgrounds in the corresponding observables. In particular the $\tau\tau$ channel has the  weakest constraints.
 For the LFC coefficients, we observe that operators involving left-handed quark fields are more strongly constrained than their right-handed counterparts, as only the left-handed operators induce FCNCs in our MFV setup. Moreover, the operators with left-handed quarks and left-handed leptons, $O_{lq}^{(1)}$ and $O_{lq}^{(3)}$, are particularly well constrained as they also contribute to dineutrino decays. The operator $O_{qe}$ with right-handed lepton singlets, on the other hand, contributes only to processes involving charged leptons.

The limits on the Wilson coefficients in the LU and LD scenarios are slightly tighter than those in the lepton-flavor-specific fits, demonstrating the synergy across different flavor channels. This is particularly pronounced for the coefficients $\tilde C_{lq}^{(1)}$ and $\tilde C_{lq}^{(3)}$, which are most affected by potential flat directions since they contribute via the linear combinations $\tilde C_{lq}^{\pm}$ to many observables. 
The strongest bound in this class of coefficients is obtained for $\tilde C_{lq}^{(3)}$ in the LD scenario with a 90\% credible interval of $[-0.86,5.91]\cdot 10^{-5}$, corresponding to a reach of $30\,\text{TeV}$ for $\Lambda / \sqrt{C}$, far exceeding the direct limits from the LHC.

While most coefficients are consistent with zero within the 90\% credible intervals, we find slight preferences for nonzero values in the fit results for $\tilde C_{lq}^{(3)}$ in the $\mu\tau$ scenario and for $\tilde C_{eu}$ in the LU scenario. The corresponding 90\% credible intervals are $[0.78,5.74]\cdot 10^{-4}$ and $[0.12, 2.44]\cdot 10^{-3}$, respectively.
The shift in the LFV coefficient can plausibly be traced to mild enhancements in the highest $p_T$ bin of $pp \to \mu\tau$ production measured by ATLAS~\cite{ATLAS:2020tre}, together with upward fluctuations in the high-$p_T$ bins of $pp \to \mu \nu$~\cite{ATLAS:2019lsy}, non-significant excesses in $P \to \tau \nu$ and $\tau \to K \nu$ decays, and the $2.7\sigma$ deviation observed in the $B^+ \to K^+ \nu \bar \nu$ measurement by Belle II~\cite{Belle-II:2023esi}. In contrast, for $\tilde C_{eu}$ no single observable shows a comparable deviation, suggesting that the effect originates from the interplay of multiple observables and statistical fluctuations. This underlines the additional sensitivity that can be achieved through combined fits.
Nevertheless, all coefficients remain compatible with zero at the $3\sigma$ level.

\begin{figure}[h]
    \centering
    \includegraphics[width=\textwidth]{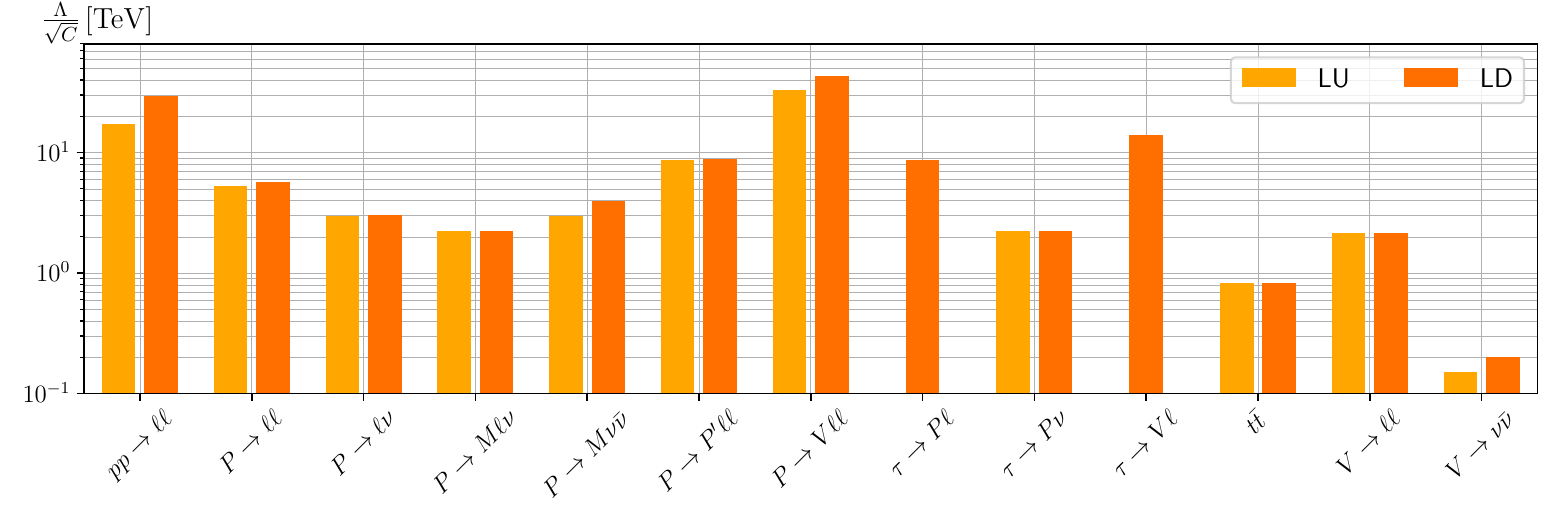}
    \caption{Lower bounds on $\Lambda/\sqrt{C}$ at the 90\% credible level for the Wilson coefficient $\tilde C_{lq}^{(3)}$ resulting from single-coefficient fits to individual sectors. We compare fits in the LU scenario (yellow) to the LD scenario (orange).}
    \label{fig:single_clq3}
\end{figure}

In addition, we perform single-coefficient fits for $\tilde C_{lq}^{(3)}$ within individual sectors to assess the relative impact of different data sets. The resulting limits on $\Lambda / \sqrt{C}$ are shown in Fig.~\ref{fig:single_clq3}. The most stringent bounds arise from decays of the type $P \to V \ell\ell$, such as $B\to K^*\mu\mu$, 
reaching scales above $40\,\text{TeV}$ in the LD scenario. The DY process $pp \to \ell\ell$ also provides strong sensitivity, with limits above $10\,\text{TeV}$.
   In contrast, the weakest bounds are obtained from quarkonia decays to dineutrinos, as expected since only upper limits on their branching ratios are currently available.

\subsection{Four-Lepton Operators \label{sec:4L}}

We briefly  discuss  flavor possibilities  of  four-lepton coefficients $\tilde C_{ll}$, $\tilde C_{le}$, and $\tilde C_{ee}$, each with four  flavor indices $ijkl$, and how they enter the global fits.

In general, all  flavors can be equal, $i=j=k=l$, which is LFC, and contributes to the $ii$-fit, the LU-fit, and the LD-fit.
If two different  flavors appear twice, abbreviated as $2+2$, and $i=j \neq k=l$
or
$i=l \neq j=k$,~\footnote{For $\mathcal{O}_{ee}$ the coefficients with $ijkl$ are identical to the ones with $ilkj$ due to Fierzing rules.} these are LFC, and contribute to the LU-fit and the LD-fit.
Examples for  such $iikk$- or  $ijji$-type induced processes are $e^+ e^- \to \mu^+ \mu^-$ or $ \mu^- \to e^- \bar \nu_e \nu_\mu$.
Another 2+2-possibility are decays such as  $ \mu^- \to e^- \bar \nu_\mu  \nu_e$, which violate
lepton flavor by two units ($i=k\neq j=l$), that is,  with $ijij$ indices.  It contributes to the $ij$-fit, and the combined, LD-fit.
The  $3+1$ case, that is $i \neq j=k=l$, or  $2+1+1$, corresponding to $i \neq j \neq k=l$,   these are  LFV, and contribute only to the LD-fit.
This classification is a choice, as we like to stick to our definition that in the $ij$-fit only a single combination, $ij$, is switched on by NP.
For bilinears this is of course unambiguous. 
As with two  leptons,   the concept  of democracy with four leptons
refers to  all lepton flavor combinations contributing with a single coefficient.
Since all operator flavors contribute to the LD-fit, we expect here the  strongest bounds, followed by those  from  the LU-fit.
 
The four-lepton coefficients 
are directly constrained by  LFC dilepton production at LEP and $\ell \to \ell^\prime ll$ decays into charged leptons
$l=\ell^{\pm}$ ($\tilde C_{ll}$, $\tilde C_{le}$,$\tilde C_{ee}$) and with neutrinos  $l= \nu$ ($\tilde C_{ll}$, $\tilde C_{le}$).
Hence, in the LFV $ijij$-case, $O_{ee}$ is  unconstrained by the observables considered in this analysis.

\begin{figure}[h]
    \centering
    \includegraphics[width=\textwidth]{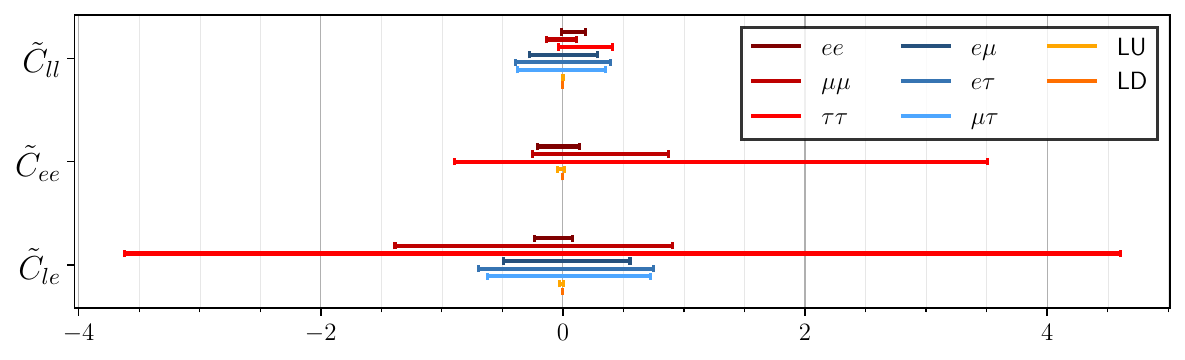}
    \caption{90\% credible intervals for the four-lepton coefficients. 
     $\tilde C_{ee}$ is not constrained in the LFV $ij$-case, see text, and no bounds are shown here.}
    \label{fig:credible_ll}
\end{figure}

\begin{figure}[h]
    \centering
    \includegraphics[width=0.95\textwidth]{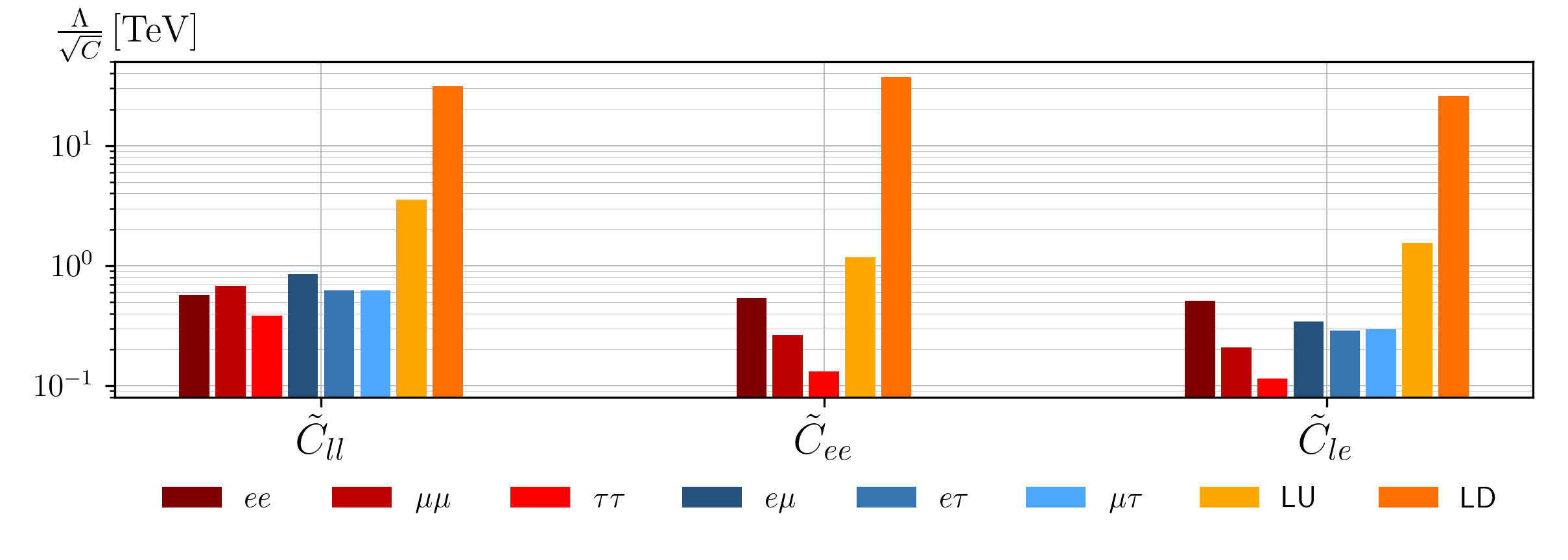}
    \caption{Lower bounds on $\Lambda/\sqrt{C}$ for the four-lepton operators,
    see Fig.~\ref{fig:credible_ll}.}
     \label{fig:scale_ll}
\end{figure}

Fig.~\ref{fig:credible_ll} shows the 90\% credible intervals for the Wilson coefficients across the different lepton-flavor scenarios;  the corresponding lower bounds on  $\Lambda / \sqrt{C}$ are shown in Fig.~\ref{fig:scale_ll}. 
Numerical values of the credible intervals are provided in Tab.~\ref{tab:credible_intervals_ll}.
Constraints on operators with four muons, or four taus arise predominantly  from their RG-mixing onto  leptonic penguin and semileptonic four-fermion operators.
$\mathcal{O}_{le}$ also enters the AMMs \cite{Aebischer:2021uvt}.

We find, as anticipated,  that the bounds in the LD and LU scenarios are significantly stronger than in  the others. In particular,
  the LD-fit   reaches  scales of $\Lambda / \sqrt{C} \sim 30\,\text{TeV}$, whereas in the lepton-flavor-individual fits they are  below $1\,\text{TeV}$.

\subsection{Dipole Operators \label{sec:dipole}}

The dipole coefficients $\tilde C_{eB}$ and $\tilde C_{eW}$ contribute to a variety of observables, including the AMMs of charged leptons, $Z$ and $W$ boson decays, and radiative lepton decays. Since the dipole operators are not Hermitian, the flavor components $ij$ and $ji$ with $i < j$ contribute independently to the fit. While neutral-current processes are sensitive to $\left| \tilde C_{ij} \right|^2 + \left| \tilde C_{ji} \right|^2$, charged-current processes such as $W \to \ell_i \nu_j$ receive contributions from only one of the two components, leading to slight differences in the resulting bounds.

\begin{figure}
    \centering
    \includegraphics[width=\textwidth]{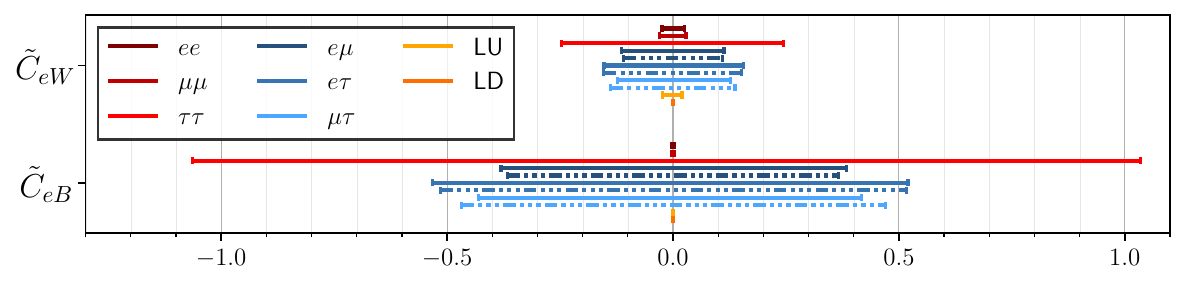}
    \caption{90\% credible intervals for the dipole coefficients $\tilde C_{eB}$ and $\tilde C_{eW}$ from global SMEFT fits. The solid lines show the bounds for the flavor components $ij$ with $i<j$, while the striped lines show the bounds for $ji$. }
    \label{fig:credible_dipole}
\end{figure}

\begin{figure}
    \centering
    \includegraphics[width=\textwidth]{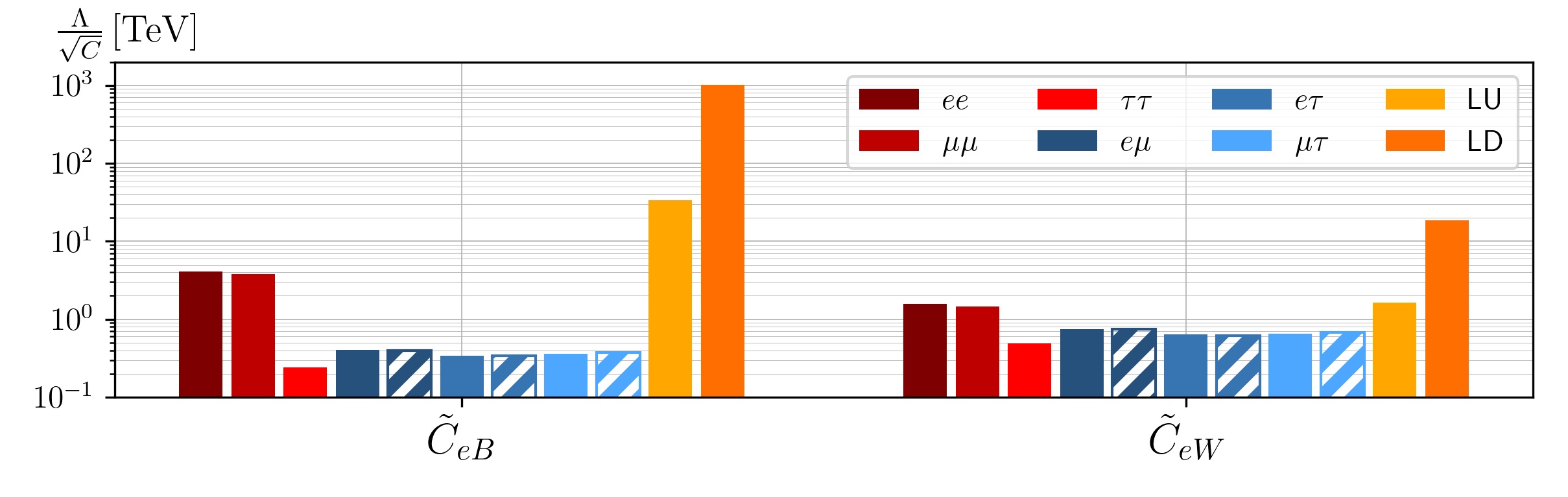}
    \caption{Lower bounds on $\Lambda/\sqrt{C}$ for the dipole coefficients $\tilde C_{eB}$ and $\tilde C_{eW}$ obtained from the 90\% credible intervals shown in Fig.~\ref{fig:credible_dipole}. The solid bars show the bounds for the flavor components $ij$ with $i<j$, while the striped bars show the bounds for $ji$.}
    \label{fig:scale_dipole}
\end{figure}

Fig.~\ref{fig:credible_dipole} displays the 90\% credible intervals for the Wilson coefficients obtained from the  global SMEFT fit, while Fig.~\ref{fig:scale_dipole} shows the corresponding lower bounds on $\Lambda / \sqrt{C}$. Solid lines and bars denote the bounds for the $ij$ coefficients with $i < j$, and striped lines and bars correspond to the $ji$ components. Numerical values of the credible intervals are given in Tab.~\ref{tab:credible_intervals_dp}.

The coefficient $\tilde C_{eW}$ is more tightly constrained than $\tilde C_{eB}$ for the $\tau\tau$ and LFV coefficients, as it contributes to a broader range of observables, including charged-current processes. It also dominates the linear combination $\tilde C_{eZ}$, which is strongly constrained by $Z$-pole observables. 
On the other hand,
in the $ee$ and $\mu\mu$ channels, we find  tighter limits for  $\tilde C_{eB}$ than for $\tilde C_{eW}$. This stems from  the AMMs of the electron and muon, which  are very  sensitive to the LEFT dipole coefficient $\mathcal{C}_{e\gamma}$, to which the primary matching contribution comes from $\tilde C_{eB}$.

\enlargethispage{\baselineskip}
Synergies across the different observables significantly enhance sensitivity in the LD scenario. The 90\% credible interval for $\tilde C_{eB}$  is $[-5.95, 0.11] \cdot 10^{-8}$, corresponding to a lower bound on the NP scale of $\Lambda / \sqrt{C} \sim 1000$\,TeV, representing the strongest limit obtained in this analysis. This marks an improvement of more than two orders of magnitude compared to the individual flavor-specific fits.

The origin of this enhancement is that in single-flavor fits at least one linear combination of Wilson coefficients remains comparatively weakly constrained. In particular, the scalar and tensor coefficients $\tilde C_{lequ}^{(1)}$ and $\tilde C_{lequ}^{(3)}$ mix into the dipole operators under RGE running~\cite{Jenkins:2013zja, Alonso:2013hga, Jenkins:2013wua}, which induces a degeneracy in most flavor-specific analyses. Direct constraints on these coefficients are currently available only from $t\bar{t}\ell\ell$ production, and thus limited to the $ee$ and $\mu\mu$ flavor channels.
By combining the information from all lepton flavors, however, this degeneracy can be lifted, leading to substantially stronger bounds on the dipole coefficients. The global analysis fully exploits the stringent limits from LFV $Z$ decays together with the precise measurements of the electron and muon anomalous magnetic moments.

\begin{figure}[h]
    \centering
    \includegraphics[width=\textwidth]{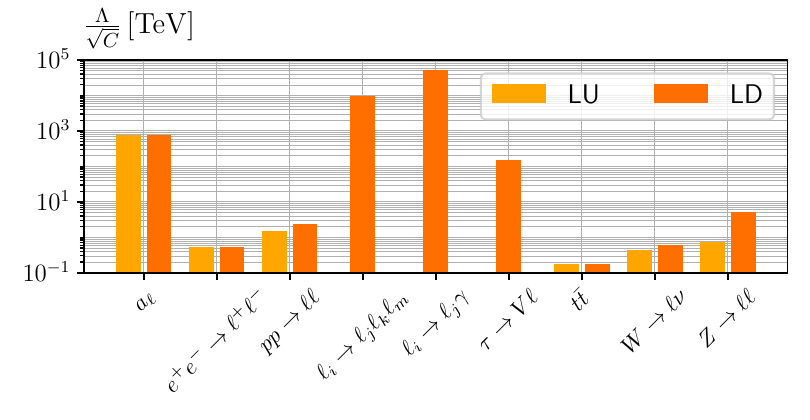}
    \caption{Lower bounds on $\Lambda/\sqrt{C}$ at the 90\% credible level for the Wilson coefficient $\tilde C_{eW}$ resulting from single-coefficient fits to individual sectors. We compare fits in the LU scenario (yellow) to the LD scenario (orange).}
    \label{fig:single_ew}
\end{figure}

Fig.~\ref{fig:single_ew} illustrates the impact of the individual sectors on the dipole bounds. We show the 90\% credible limits on $\Lambda/\sqrt{C}$ in a $\tilde C_{eW}$-only fit.
The strongest bounds come from radiative lepton decays, which are experimentally very well constrained. In the LU fit, the best limit results from the anomalous magnetic moments, with a reach of several hundred TeV.

\subsection{Scalar and Tensor Operators \label{sec:ST}}

The scalar and tensor operators $O_{lequ}^{(1)}$ and $O_{lequ}^{(3)}$ contribute only via their coupling to the top quark in our MFV framework. As a result, they are not directly constrained by DY production or flavor observables. The only direct experimental constraints come from $t\bar{t}  ee$ and $t\bar{t} \mu \mu$ production~\cite{ATLAS:2025yww}.
Beyond these direct constraints, $\tilde C_{lequ}^{(1)}$ and $\tilde C_{lequ}^{(3)}$ are constrained only via their RG-mixing  with the electroweak dipoles  $\tilde C_{eW}$ and $\tilde C_{eB}$. Since the tensor operator mixes more strongly with the dipoles than the scalar operator, the resulting constraints on $\tilde C_{lequ}^{(3)}$ are roughly an order of magnitude tighter than those on $\tilde C_{lequ}^{(1)}$.

\begin{figure}[h]
    \centering
    \includegraphics[width=\textwidth]{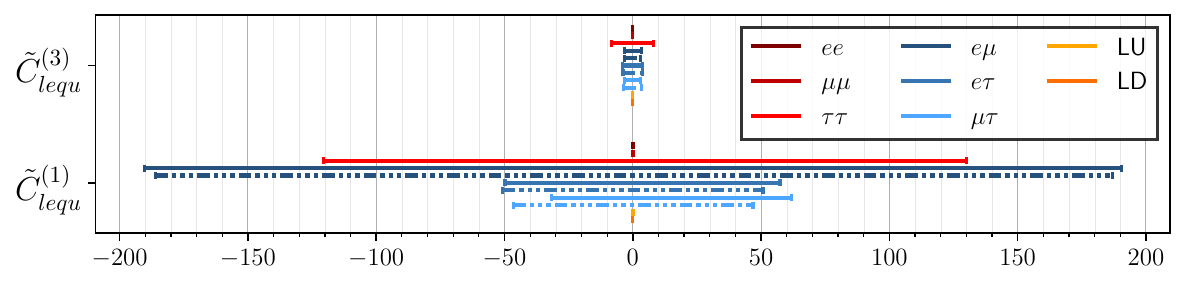}
    \caption{90\% credible intervals for  $\tilde C_{lequ}^{(1)}$ and $\tilde C_{lequ}^{(3)}$ from  global SMEFT fits. The solid lines show the bounds for the flavor components $ij$ with $i<j$, while the striped lines show the bounds for $ji$.}
    \label{fig:credible_lequ}
\end{figure}

\begin{figure}[h]
    \centering
    \includegraphics[width=\textwidth]{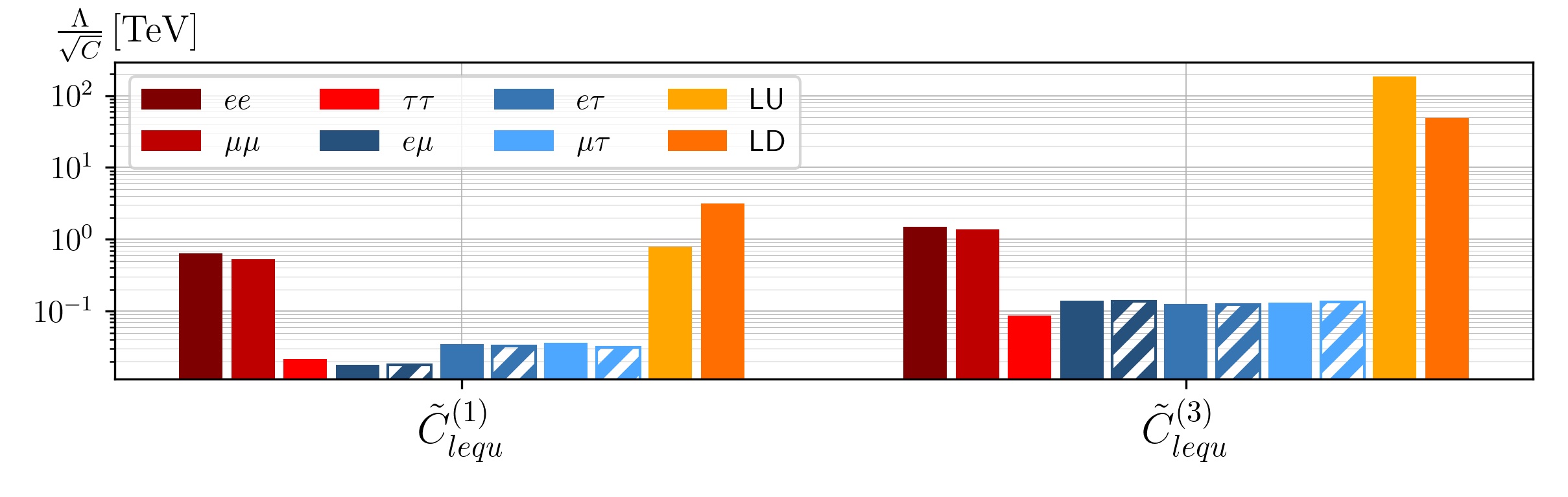}
    \caption{Lower bounds on $\Lambda/\sqrt{C}$ for  $\tilde C_{lequ}^{(1)}$ and $\tilde C_{lequ}^{(3)}$ from global SMEFT fits. The solid bars show the bounds for the flavor combinations $ij$ with $i<j$, while the striped bars show the bounds for $ji$. }
    \label{fig:scale_lequ}
\end{figure}

Fig.~\ref{fig:credible_lequ} shows the 90\% credible intervals for these Wilson coefficients, and Fig.~\ref{fig:scale_lequ} displays the corresponding lower bounds on $\Lambda/\sqrt{C}$. Numerical values are listed in Tab.~\ref{tab:credible_intervals_dp} in Appendix~\ref{app:credible_intervals}. The operators $O_{lequ}^{(1)}$ and $O_{lequ}^{(3)}$ are not Hermitian, so the flavor components $ij$ and $ji$ are treated as independent degrees of freedom. We indicate the bounds for $ij$ with $i<j$ using solid lines and bars, and those for $ji$ with striped ones.

As expected from the AMMs and $t\bar{t} \ell\ell$ production the
constraints on the $ee$ and $\mu\mu$ coefficients are significantly stronger than those on other flavor combinations.
Consequently, the bounds in the LU and LD scenarios also benefit from these constraints,
reaching scales for the tensor coupling $\tilde C_{{lequ}}^{(3)}$ of  $\Lambda / \sqrt{C} \sim 100$\,TeV and  $\Lambda / \sqrt{C} \sim 50$\,TeV, respectively.  In contrast, the LFV coefficient $\tilde C_{{lequ}}^{(1)}$ in the $e\mu$ fit receives the weakest constraints among all coefficients in our analysis, with a 90\% credible interval of $[-193, 193]$.

\subsection{Testing MFV  \label{sec:MFV}}

We discuss the fit results for the parameters in the MFV expansion  $r_L$, $r_R$ defined in Eq.~\eqref{eqn:MFV_degrees_of_freedom}. 
We include operator specific ratios $r_L$ for  coefficients with left-handed quarks, $C_{lq}^{+}$, $C_{lq}^{-}$ and $C_{qe}$, and  ratios $r_R$ for  $C_{lu}$ and $C_{eu}$ with right-handed up-type quarks.
The  ratios $r_L$ induce down-type FCNCs and non-universal couplings to the top and down-type quarks, while $r_R$ only affect the right-handed $t\bar{t}$-couplings.

We find that the posterior distributions for $r_R$ are flat and unbounded for both operators, indicating that the right-handed flavor structure is essentially unconstrained by the observables considered in this work. This is expected, as the only process probing $r_R$ is $t\bar{t}\ell\ell$ production, whose constraints are orders of magnitude weaker than those from flavor observables and Drell--Yan production.

\begin{figure}[h]
    \centering
    \includegraphics[width=0.32\textwidth]{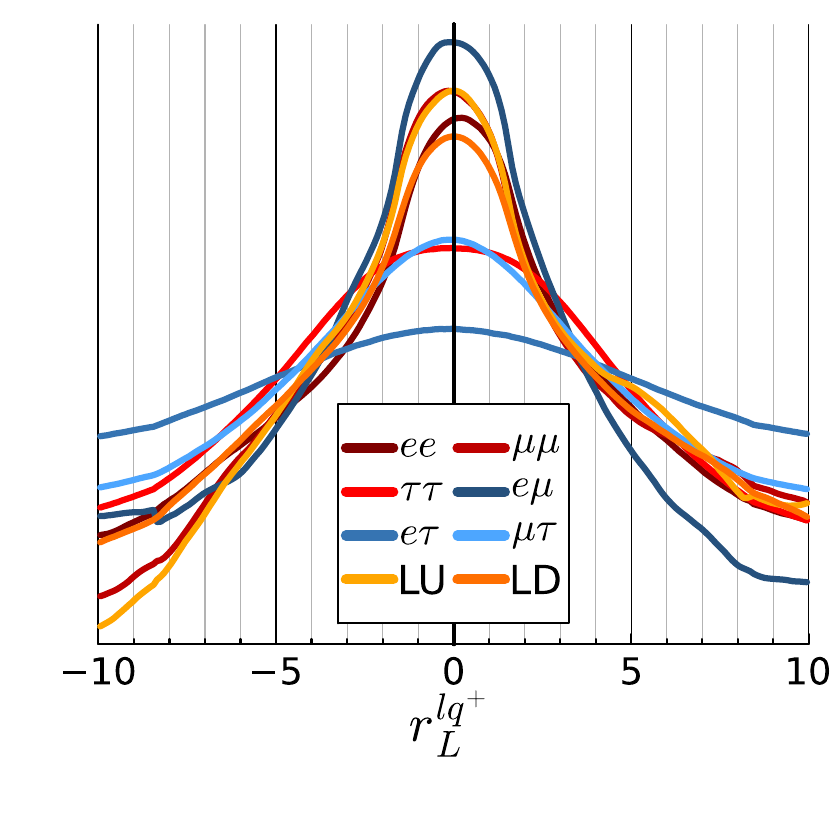}
    \includegraphics[width=0.32\textwidth]{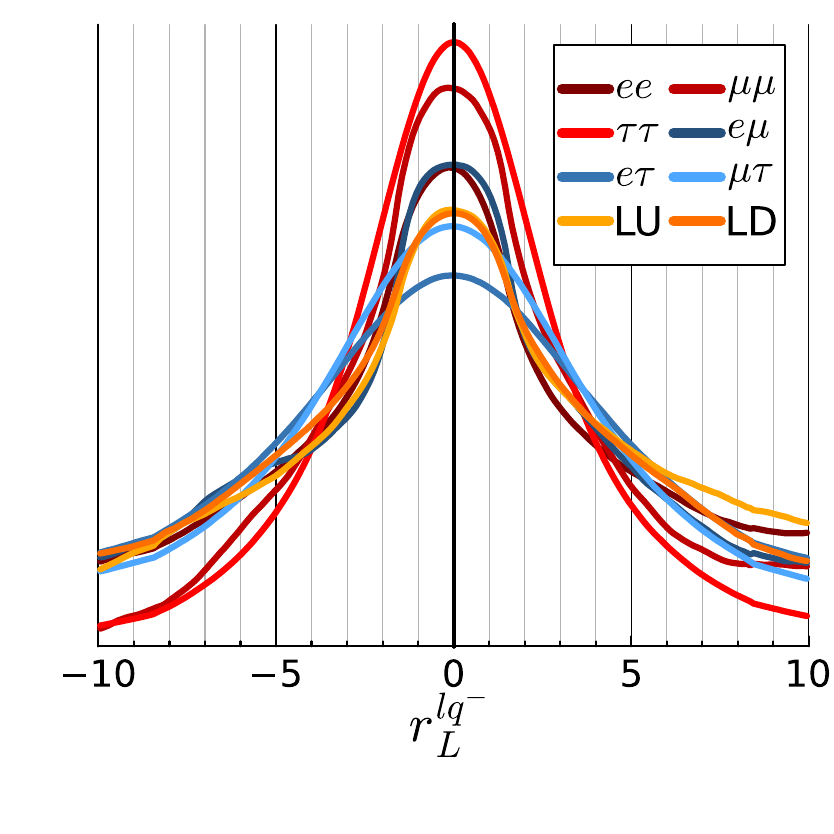}
    \includegraphics[width=0.32\textwidth]{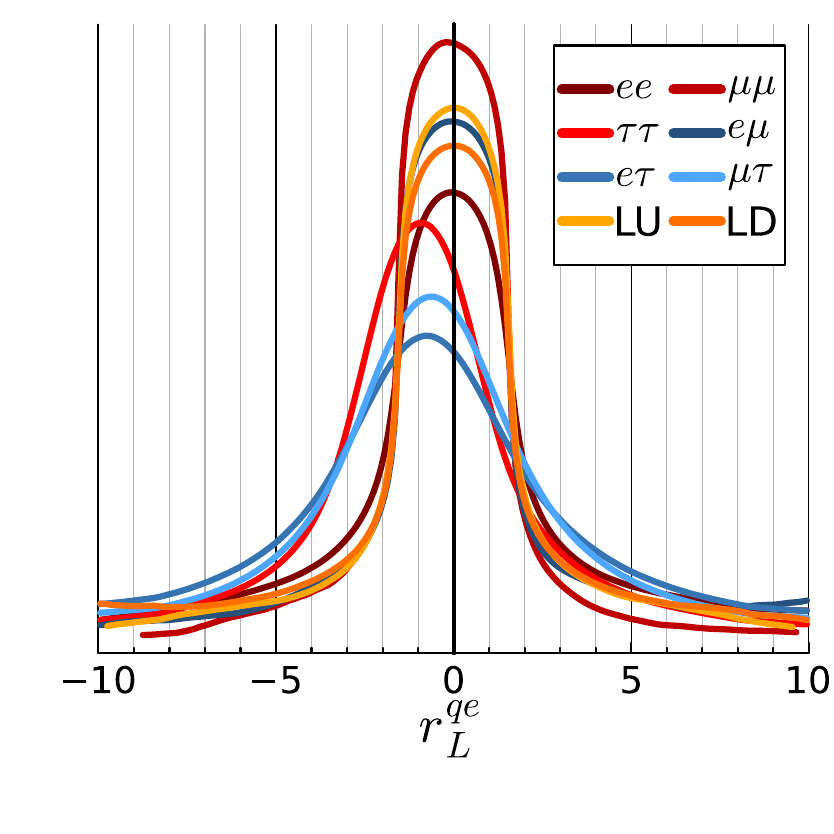}
    \caption{Posterior distributions for the MFV ratios $r_L^{lq^{+}}$ (left), $r_L^{lq^{-}}$ (middle) and $r_L^{qe}$ (right). The color coding indicates the different lepton-flavor scenarios.}
    \label{fig:MFV_ratios}
\end{figure}

In contrast, the posterior distributions for $r_L$  shown in Fig.~\ref{fig:MFV_ratios} for the operators $O_{lq}^{+}$ (left), $O_{lq}^{-}$ (center) and $O_{qe}$ (right) feature non-trivial, bells-type shapes.
They differ in sharpness  as a result of the sensitivity of the different lepton-flavor scenarios to the MFV structure. 
All distributions are  consistent with $r_L=0$, reflecting the fact that we currently do not see significant deviations from the SM  in the flavor data, which would require  next-to-leading 
terms in the MFV-expansion  \cite{Grunwald:2023nli}.
For the posterior of $r_L^{qe}$, note that the $\tau\tau$, $e\tau$ and $\mu\tau$  distributions  are slightly shifted towards negative values and peak around $r \sim -1$.

\subsection{CKM Parameters \label{sec:CKM}}

\begin{figure}[h]
    \centering
    \includegraphics[width=\textwidth]{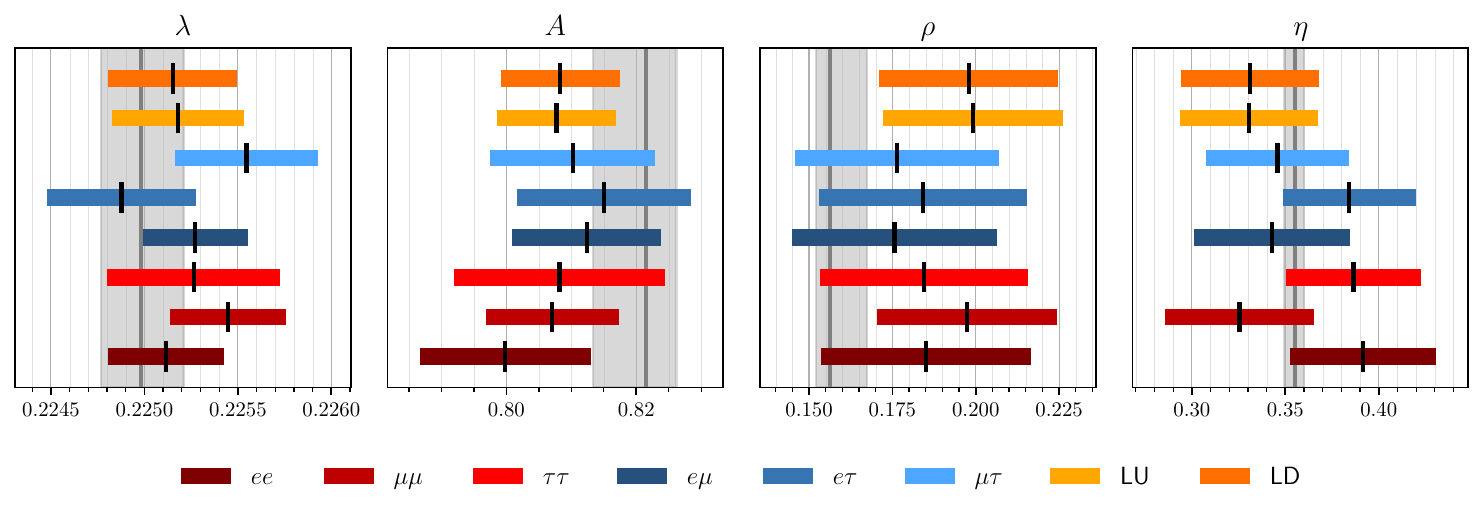}
    \caption{$1\sigma$ credible intervals (colored) and mean values (black) for the CKM parameters $\lambda$, $A$, $\rho$ and $\eta$ from the global SMEFT fits. Results are shown for lepton-flavor diagonal (red), lepton-flavor universal (orange), and lepton-flavor violating (blue) scenarios. The results of the CKMfitter group~\cite{CKMfitter:Summer23,Charles:2004jd} are shown in gray.}
    \label{fig:credible_ckm}
\end{figure}

The parameters $\lambda$, $A$, $\rho$, and $\eta$ of the Wolfenstein parameterization of the CKM matrix defined in Eq.~\eqref{eq:wolfenstein} are treated as nuisance parameters in our fit. In Fig.~\ref{fig:credible_ckm}, we show the mean values obtained from our global fits as black lines, along with the corresponding $1\sigma$ credible intervals shown as colored bands for the various lepton-flavor scenarios. For comparison, we show the results of the CKMfitter group~\cite{CKMfitter:Summer23,Charles:2004jd}, with mean values indicated by gray lines and the $1\sigma$ intervals by gray bands.

Our results are compatible  at 1 $\sigma$ with those from the CKMfitter collaboration, except for small deviations in $\rho$ in the LU, LD and $\mu\mu$ fits.
One also notices  that the ranges for the parameter $A$ from our fit tend towards smaller values compared to the CKMfitter results, while the values for $\rho$ are slightly larger. For $\lambda$ and $\eta$, on the other hand, we find the results vary more strongly across the different lepton-flavor scenarios.

\begin{figure}[h]
    \centering
    \includegraphics[width=\textwidth]{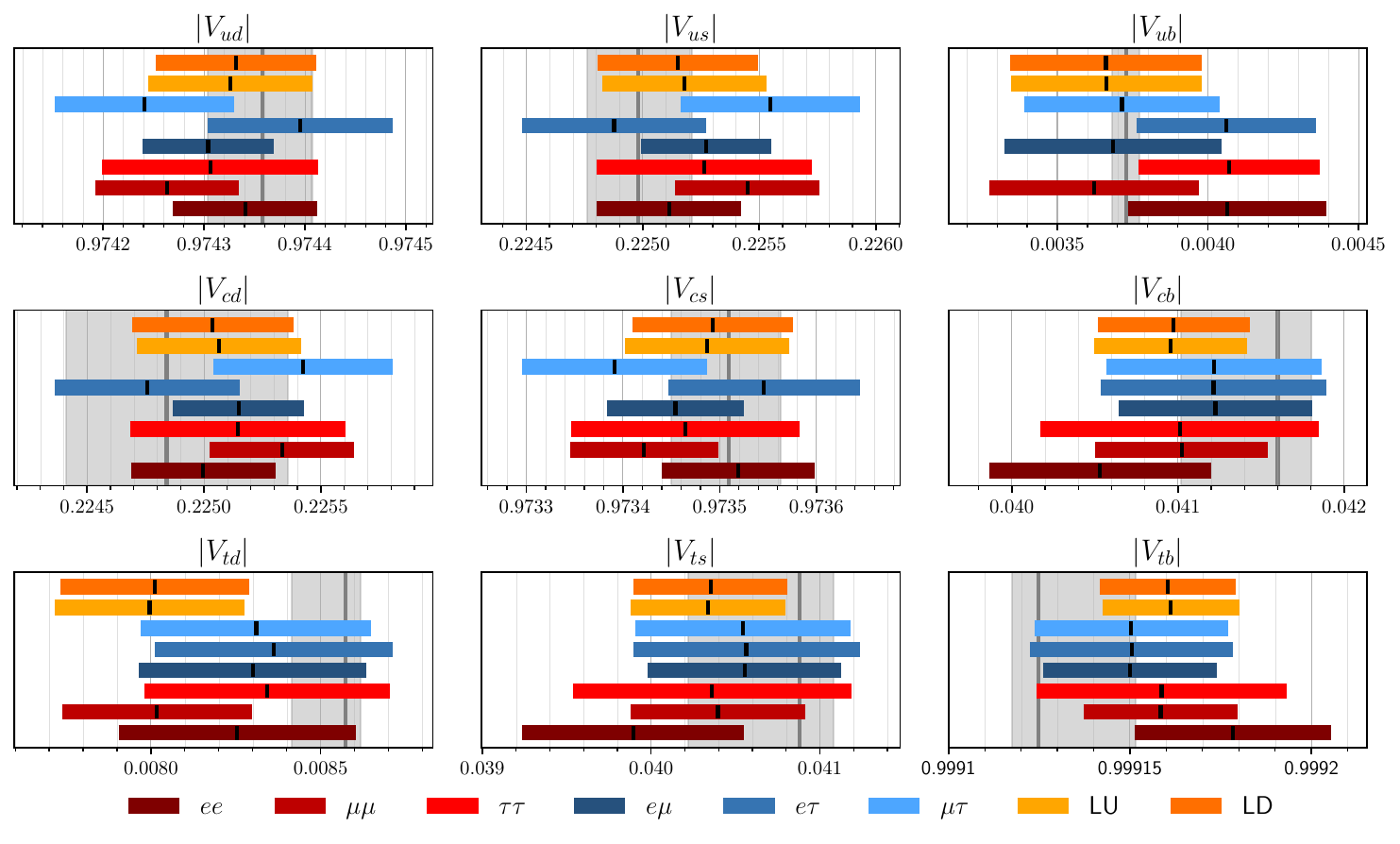}
    \caption{Absolute values of the CKM matrix elements derived from the posterior distributions of the Wolfenstein parameters $\lambda$, $A$, $\rho$, and $\eta$. Colored bands represent the $1\sigma$ credible intervals, with mean values shown as black lines. Results are compared to those from the CKMfitter group~\cite{CKMfitter:Summer23,Charles:2004jd} (gray lines and bands).}
    \label{fig:ckm_matrix}
\end{figure}

Using Eq.~\eqref{eq:wolfenstein}, we compute the CKM matrix elements corresponding to the posterior distributions. The results are presented in Fig.~\ref{fig:ckm_matrix}, where black lines denote the mean values and colored bands indicate the $1\sigma$ credible intervals. We again compare our results to those of the CKMfitter group, shown as gray lines and bands.
As for the Wolfenstein parameters, we find overall good agreement between our results for the CKM elements and the CKMfitter values, with somewhat larger uncertainties. This is expected, since our analysis relies on fewer observables and includes additional sources of uncertainty from the Wilson coefficients. 
For the CKM element $V_{cb}$, we obtain slightly smaller values compared to CKMfitter. This originates from different input, see Tab.~\ref{tab:ckm_inputs}.  Our analysis relies on exclusive determinations, whereas CKMfitter also includes inclusive measurements. The downward shift in $V_{cb}$ therefore follows the well-established trend that exclusive determinations yield lower central values than inclusive ones~\cite{FlavourLatticeAveragingGroupFLAG:2024oxs,ParticleDataGroup:2024cfk}, and it also leads to a smaller value of $A$ in our fits.
The Wolfenstein parameter $\rho$ obtained from our fits is larger than the CKMfitter results, mainly reflecting a shift in $V_{td}$. In particular, we observe a preference for smaller values of $\left|V_{td}\right|$, driven primarily by the neutral meson mass splitting $\Delta M_d$.

Across all parameters, we observe consistency among the different lepton-flavor scenarios, indicating that the impact of lepton-flavor universality on the CKM parameters is mild.

\section{Predicting Dineutrino Modes \label{sec:BKnunu}}

The only  dineutrino branching fraction of a $B$ meson decay evidenced  to date  is  ${{\cal{B}} (B^+ \to K^+ \nu \bar \nu)}$, by the Belle II collaboration~\cite{Belle-II:2023esi}. For all other dineutrino modes,  only upper limits exist at the level of  few $\cdot 10^{-5}$ \cite{Belle:2013tnz,Belle:2017oht}. Belle II is expected to significantly improve the sensitivity to these channels, with the goal of measuring the branching fractions of $B \to K^{(\star)} \nu \bar \nu$ with a precision of roughly 10\%~\cite{Belle-II:2018jsg}. 

\begin{figure}[h]
    \centering
    \includegraphics[width=0.8\textwidth]{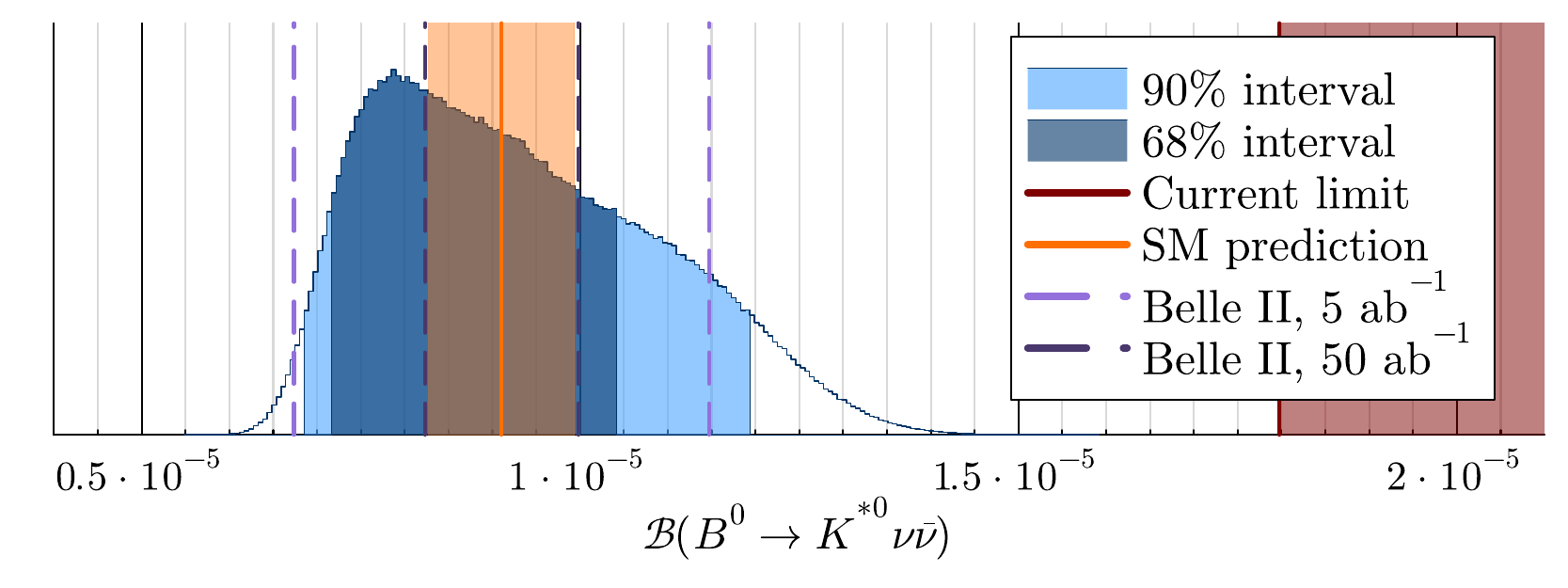}
    \includegraphics[width=0.8\textwidth]{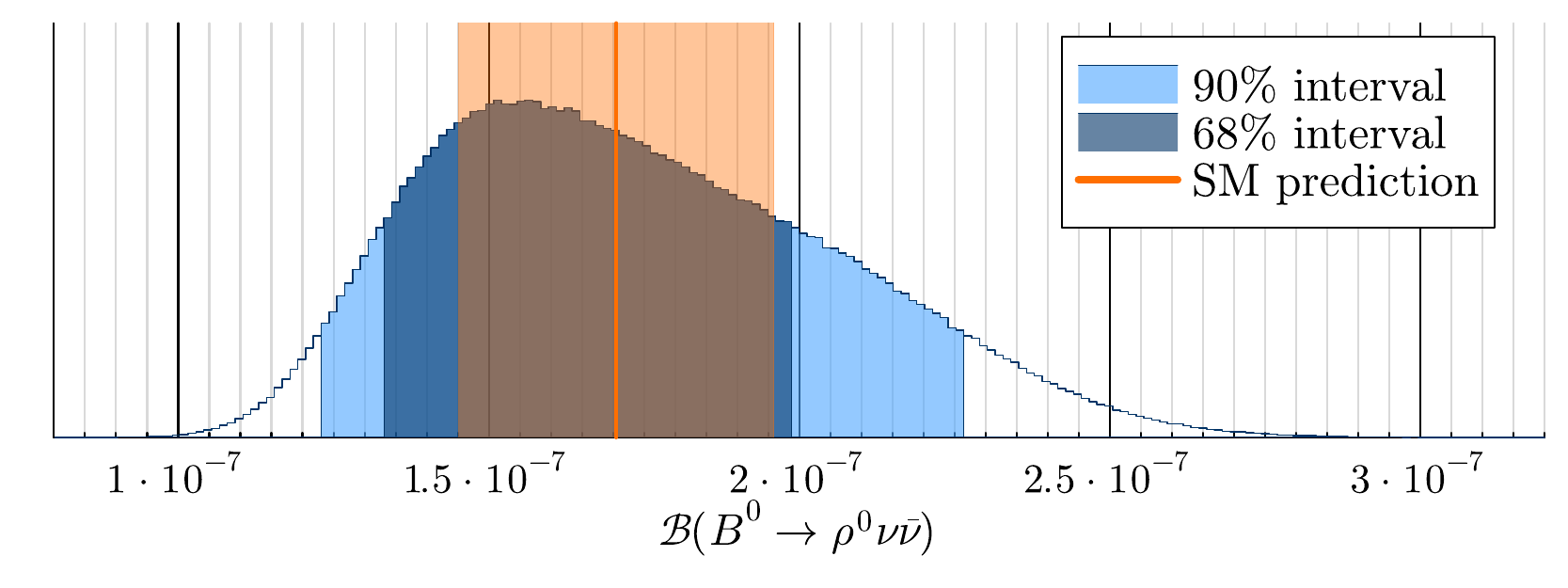}
    \caption{Predictions for the dineutrino decays $B^0 \to K^{*0} \nu \bar \nu$ (top) and $B^0 \to \rho^0 \nu \bar \nu$ (bottom) based on the global fit in the $\tau\tau$ scenario. Colored bands (blue) indicate the 68\% and 90\% credible intervals; the orange band denotes the 1$\sigma$ SM prediction. The red area represents the current experimental 90\% CL bound
    $\mathcal{B}(B^0 \to K^{*0} \nu \bar \nu)<1.8 \cdot 10^{-5}$ 
    \cite{Belle:2017oht}, while dashed purple lines show Belle II projected sensitivities  \cite{Belle-II:2018jsg}, if applicable.}
    \label{fig:predictions_dineutrino}
\end{figure}

Using the posterior distributions of the Wilson coefficients from the fits, we predict the branching fractions for these dineutrino modes within our MFV approach. Fig.~\ref{fig:predictions_dineutrino} illustrates the resulting probability distributions for the decays $B^0 \to K^{*0} \nu \bar \nu$ (top) and $B^0 \to \rho^0 \nu \bar \nu$ (bottom) based on the lepton-flavor specific fit with only the
 $\tau\tau$ coefficient switched on. We compare the predictions from the fit (blue) to the SM expectations (orange), the current experimental bounds (red), and the projected Belle II sensitivity~\cite{Belle-II:2018jsg}, if available. The  experimental upper limit  $\mathcal{B}(B^0 \to \rho^0 \nu \bar \nu)<4.0 \cdot 10^{-5}$ 
    \cite{Belle:2017oht} is not shown as it is outside the plot.
 Additional plots for the decays of charged $B$ mesons as well as $B^0 \to K^0 \nu \bar \nu$ and $B^0 \to \pi^0 \nu \bar \nu$ are given  in Appendix~\ref{app:predictions_dineutrino}.

The MFV-SMEFT  predictions agree  with the SM. They peak at slightly smaller values compared to the SM predictions, for which we used as CKM input the SM-only CKM fit results given in App.~\ref{app:CKM},
while for the SMEFT predictions we fit the CKM parameters simultaneously. 
The uncertainties of the predictions based on our posterior distributions are larger than those of the SM predictions, due to the additional uncertainties introduced by the SMEFT Wilson coefficients.

Predictions for the short-distance contributions to the  branching fractions of $B \to K^{(*)} \nu \bar \nu$ and 
${B \to (\pi,\rho) \nu \bar \nu}$ are given in Tabs.~\ref{tab:predictions_B_Knunu} and~\ref{tab:predictions_B_Pinu}, respectively, across different lepton-flavor scenarios abd 1$\sigma$ uncertainties
in parentheses.
In charged meson decays $B^+ \to M^+ \nu \bar{\nu}$, with $M$  a pseudoscalar or vector meson, an additional long-distance contribution arises from cascade decays  ${B^+ \to \tau^+ (\to M^+ \bar{\nu}) \nu}$.
For $b \to s$ FCNCs, these resonant decays are at the order of $10\%$, whereas  for $b \to d$ the $\tau$-background is Cabibbo-enhanced and exceeds the 
SM short-distance contribution in  $B^+ \to \pi^+ \nu \bar \nu$ decays by nearly two orders of magnitude \cite{Kamenik:2009kc,Parrott:2022zte, Bause:2021cna}. 
Although clearly important, in our prediction we only give the short-distance contributions as the precise value of  the background depends on cuts and  experimental aspects.
We note that in principle also the $\tau$-background could be modified by NP, however, experimental determinations of ${\cal B}(B^+ \to \tau^+ \nu)$  \cite{ParticleDataGroup:2024cfk}  and ${\cal B}(\tau^+ \to M^+ \bar{\nu})$ (see Tab.\ref{tab:tau_SL_neutrinos}) are in good agreement with the SM predictions, thereby limiting  possible deviations.

\begin{table}[h]
  \centering
  \setlength{\tabcolsep}{12pt}
  \renewcommand{\arraystretch}{1.2}
  \begin{tabular}{l l l l }
    \toprule
    Fit & $\mathcal{B}^0 \to K^{0} \nu \bar{\nu}$ & $\mathcal{B}(B^0 \to K^{*0} \nu \bar{\nu})$ & $\mathcal{B}(B^+ \to K^{*+} \nu \bar{\nu})$ \\
    \midrule
    $ee$ & $4.09(0.44)\cdot 10^{-6}$ & $9.07(0.98)\cdot 10^{-6}$ & $9.8(1.1)\cdot 10^{-6}$ \\
    $\mu\mu$ & $4.22(0.44)\cdot 10^{-6}$ & $9.36(0.97)\cdot 10^{-6}$ & $10.1(1.0)\cdot 10^{-6}$ \\
    $\tau\tau$ & $4.19(0.71)\cdot 10^{-6}$ & $9.3(1.6)\cdot 10^{-6}$ & $10.0(1.7)\cdot 10^{-6}$ \\
    $e \mu$ & $4.19(0.12)\cdot 10^{-6}$ & $9.31(0.26)\cdot 10^{-6}$ & $10.0(0.3)\cdot 10^{-6}$ \\
    $e \tau$ & $4.21(0.14)\cdot 10^{-6}$ & $9.33(0.31)\cdot 10^{-6}$ & $10.1(0.3)\cdot 10^{-}$ \\
    $\mu \tau$ & $4.20(0.14)\cdot 10^{-6}$ & $9.33(0.30)\cdot 10^{-6}$ & $10.1(0.3)\cdot 10^{-6}$ \\
    LD & $4.15(0.15)\cdot 10^{-6}$ & $9.20(0.33)\cdot 10^{-6}$ & $9.93(0.35)\cdot 10^{-6}$ \\
  \bottomrule
  \end{tabular}
  \caption{Predictions for the branching fractions of the short-distance contributions to $B \to K^{(*)}\nu \bar \nu $ decays in the various lepton-flavor scenarios. The numbers in parentheses indicate the $1\sigma$ uncertainties. }
  \label{tab:predictions_B_Knunu}
\end{table}

\begin{table}[h]
  \centering
  \setlength{\tabcolsep}{12pt}
  \renewcommand{\arraystretch}{1.2}
  \begin{tabular}{l l l l l }
    \toprule
    Fit & $\mathcal{B}(B^0 \to \pi^{0} \nu \bar{\nu})$ & $\mathcal{B}(B^+ \to \pi^{+} \nu \bar{\nu})$ & $\mathcal{B}(B^0 \to \rho^{0} \nu \bar{\nu})$ & $\mathcal{B}(B^+ \to \rho^{+} \nu \bar{\nu})$ \\
    \midrule
    $ee$ & $5.88(0.78)\cdot 10^{-8}$ & $1.27(0.17)\cdot 10^{-7}$ & $1.70(0.22)\cdot 10^{-7}$ & $3.66(0.49)\cdot 10^{-7}$ \\
    $\mu\mu$ & $5.59(0.70)\cdot 10^{-8}$ & $1.20(0.15)\cdot 10^{-7}$ & $1.61(0.20)\cdot 10^{-7}$ & $3.48(0.44)\cdot 10^{-7}$ \\
    $\tau\tau$ & $6.0(1.1)\cdot 10^{-8}$ & $1.30(0.24)\cdot 10^{-7}$ & $1.74(0.32)\cdot 10^{-7}$ & $3.75(0.70)\cdot 10^{-7}$ \\
    $e \mu$ & $5.90(0.48)\cdot 10^{-8}$ & $1.27(0.10)\cdot 10^{-7}$ & $1.70(0.14)\cdot 10^{-7}$ & $3.68(0.30)\cdot 10^{-7}$ \\
    $e \tau$ & $6.00(0.50)\cdot 10^{-8}$ & $1.29(0.11)\cdot 10^{-7}$ & $1.73(0.15)\cdot 10^{-7}$ & $3.74(0.31)\cdot 10^{-7}$ \\
    $\mu \tau$ & $5.93(0.48)\cdot 10^{-8}$ & $1.28(0.10)\cdot 10^{-7}$ & $1.71(0.14)\cdot 10^{-7}$ & $3.70(0.30)\cdot 10^{-7}$ \\
    LD & $5.49(0.41)\cdot 10^{-8}$ & $1.18(0.09)\cdot 10^{-7}$ & $1.58(0.12)\cdot 10^{-7}$ & $3.42(0.25)\cdot 10^{-7}$ \\
  \bottomrule
  \end{tabular}
  \caption{Predictions for the branching ratios of the short distance contributions to $B \to (\pi,\rho) \nu \bar \nu $ decays, see Tab.~\ref{tab:predictions_B_Knunu}. }
   \label{tab:predictions_B_Pinu}
\end{table}

The mean values of the branching fractions are similar across the different flavor scenarios. However, scenarios involving LFC coefficients tend to exhibit larger uncertainties, since these coefficients interfere with the SM amplitudes, thereby amplifying the parameter-induced spread in the predictions. The largest uncertainties are observed in the $\tau\tau$ scenario, since the corresponding Wilson coefficients are least constrained.

A distinctive feature of the MFV setup is that FCNC dineutrino decays are maximally correlated, since right-handed currents do not contribute to either the leptonic or quark currents. Consequently, only a single Wilson coefficient, ${\cal{C}}_{\nu d}^{V,LL}$, governs these processes. The ratios of branching fractions are therefore independent of SMEFT effects and depend solely on form factors, kinematics, and CKM matrix elements within the MFV paradigm:
\begin{equation}
    \begin{aligned}
        &\frac{\mathcal{B}(B^0 \to K^0 \nu \bar\nu)}
         {\mathcal{B}(B^0 \to K^{*0} \nu \bar\nu)}
     = 0.451(38)\,, \quad
     \frac{\mathcal{B}(B^0 \to \pi^0 \nu \bar\nu)}
         {\mathcal{B}(B^0 \to \rho^0 \nu \bar\nu)}
     = 0.346(85)\,, \quad \\
    &\frac{\mathcal{B}(B^0 \to \pi^0 \nu \bar\nu)}
         {\mathcal{B}(B^0 \to K^0 \nu \bar\nu)} 
        = 0.335(70) \cdot |V_{td}|^2/|V_{ts}|^2
        = 0.0144(32)\,, 
    \\
    &\frac{\mathcal{B}(B^0 \to \rho^0 \nu \bar\nu)}
         {\mathcal{B}(B^0 \to K^{*0} \nu \bar\nu)}
        = 0.436(68) \cdot |V_{td}|^2/|V_{ts}|^2
        = 0.0187(33)\,.
    \end{aligned}
    \label{eqn:ratios_dineutrinos}
\end{equation}
Here we used $|V_{td}|^2/|V_{ts}|^2 = 0.0428(34)$ obtained from the  global $\tau\tau$ fit as input for the CKM parameters.
Ratios involving decays of $B^+$ mesons need to be amended by ${B^+ \to \tau^+ (\to M^+ \bar{\nu}) \nu}$ induced contributions.
The ratios (\ref{eqn:ratios_dineutrinos}) probe the MFV structure in $b \to d$ and $b \to s$ transitions. Deviations would indicate the breakdown of the SM and MFV, that could arise
from right-handed currents \cite{Buras:2014fpa} or light NP.

\section{Conclusions \label{sec:conclusions}}

We perform a  comprehensive global analysis of processes involving leptons and neutrinos from low and high-$p_T$ in the SMEFT, covering
semileptonic and purely leptonic four-fermion operators,  electroweak dipoles and  leptonic  Higgs-current ("penguin") operators.
Our main target is the lepton flavor structure and exploring  its impact on the fit.
Specifically, we  analyze the data in three ways, (i) lepton-flavor specifically, that is, with only one pair of lepton  flavors switched on at a time, for all LFC and LFV  combinations,
(ii) assuming lepton universality,  and  (iii) assuming all couplings in the coefficient matrix to be equal.

We find that the strongest limit  on NP in the global fit is on the electroweak dipole operator $\mathcal{O}_{eB}$ in the  flavor democratic LD-fit at $\sim 1000$ TeV, followed by the tensor
$\mathcal{O}_{l e q u}^{(3)}$ in the flavor universal LU-fit at $\sim 100$ TeV,  the four-fermion operator  $\mathcal{O}_{\ell q}^{(3)}$ in the LD-fit  and the Higgs-current one $\mathcal{O}_{\varphi e}$ at $\sim 30$ TeV.
These limits are obtained in fits combining different lepton flavors, except for the latter, which results from the LFV, flavor-specific $e \mu$-fit. 
The results  of the fit are consistent with the SM and MFV. 
90 \% credible intervals are given in App.~\ref{app:credible_intervals}.

We also identify in single-coefficient fits, shown in Figs~\ref{fig:single_chl3_collider}, \ref{fig:single_chl3_flavor},  \ref{fig:single_clq3} and \ref{fig:single_ew},  the impact of sets of observables. We summarize the results for the LFC, universal case and the LFV, democratic case in Tab.~\ref{tab:single-best}, where we  give the observables with the strongest limits.  Improving precision in these observables, theoretical and experimental,  is hence  
contributing to  improvements in the global fit. This invites explorations at a future Tera-$Z$-factory, such as FCC-ee \cite{FCC:2025lpp}, CEPC \cite{Ai:2024nmn}, a  super-$\tau$-charm factory \cite{Achasov:2023gey}, the HL-LHC and precision study in rare $B$-decays, as ongoing at the LHC-experiments and Belle II.  However, also other observables, while not performing best in the single-coefficient fits, are key for the global fit   by resolving ambiguities and flat directions, or
sharpening the flavor picture. Two important examples,
$t \bar t$ production with leptons, and $B$ decays to dineutrinos, are highlighted below. A detailed assessment of the individual benefits of the $\mathcal{O}(170)$ decay modes included in the global analysis, and their impact considering possible improvements in theory and experimental uncertainties  on future SMEFT fits, is beyond the scope of this work.

\begin{table}[h]
  \setlength{\tabcolsep}{15pt}
  \renewcommand{\arraystretch}{1.2}
  \centering
  \begin{tabular}{cccc}
      \toprule
      operator & lepton universality (LFC)  & lepton democracy  (LFV)  \\
      \midrule
      $\mathcal{O}_{\varphi \ell}^{(3)}$  & $Z \to \ell \ell$, $ \ell \to \ell \nu \bar \nu$ &  $Z \to \ell \ell$, $ \ell \to \ell \ell \ell$, $\tau \to M \ell$ \\
     $\mathcal{O}_{\ell q}^{(3)}$  & $B \to V  \ell \ell$, $pp \to \ell \ell $ & $B \to V  \ell \ell$,  $pp \to \ell \ell$, $\tau \to V \ell $ \\
        $\mathcal{O}_{e W}$  & $a_\ell $    & $ \ell \to \ell \gamma $ , $ \ell \to \ell \ell \ell $ \\
      \bottomrule
  \end{tabular}
  \caption{Observables with strongest impact in single-coefficient fits in LU and LD flavor scenarios.}
  \label{tab:single-best}
\end{table}

As the majority of observables are tied with the complexities of quark flavor we fit SMEFT coefficients simultaneously with the CKM elements and the coefficients in the MFV-ansatz,
which we employ for the quarks. We also test the MFV-ansatz including higher order terms in the global fit.
While data are consistent with MFV, the higher order MFV structure of the Wilson coefficients, parameterized  by  $r_{L,R}$ Eq.~\eqref{eqn:MFV_degrees_of_freedom} is  not well probed,
see  Fig.~\ref{fig:MFV_ratios}.
This could be improved by a better sensitivity to rare $B$-decays, also in the $ee$ channels and with other leptons complementing the flagship $B \to K^* \mu \mu$ mode. Moreover,  the $tt \ell\ell$ vertex is key to improve the sensitivity to the right-handed ratios $r_R$, which are currently essentially unconstrained. This could be achieved, for example, by further study of  the $t\bar{t} \ell\ell$ production, ideally in lepton flavor specific ways~\cite{ATLAS:2025yww}  or future measurements of $t\bar t$ production at a lepton collider. 
More generally, recall that NP signatures in top or charm FCNCs indicate BSM physics beyond MFV. 

Within MFV the scalar and tensor operators $O_{lequ}^{(1)}$ and $O_{lequ}^{(3)}$ are directly probed only  in  $tt \ell\ell$ production. While the top sector
alone is not causing the strong limits displayed in Fig.~\ref{fig:scale_lequ}, it provides an independent,  complementary constraint in parameter space to the indirect ones from the AMMs,
and resolves ambiguities.
This underlines the importance of further experimental study, also in other channels than $ee$ and $\mu \mu$.

We also obtain predictions for rare $B$-decays into dineutrinos from the fit-posterior, see Sec.~\ref{sec:BKnunu}.
Branching ratios are given in Tabs.~\ref{tab:predictions_B_Knunu} and~\ref{tab:predictions_B_Pinu}. We highlight the predictive power of MFV for
$b \to (d,s) \nu \bar\nu$ transistions for the  ratios of branching fractions  (\ref{eqn:ratios_dineutrinos}). If these ratios deviate significantly from the SM predictions, it would indicate the presence of NP  that is not captured by the MFV or SMEFT framework, such as sizeable right-handed FCNCs or new light degrees of freedom.
Such possibilities together with lepton-nonuniversality are also suggested by the enhanced  $B^+ \to K^+ \nu \bar \nu$  branching ratio~\cite{Bause:2023mfe}. Therefore
further study of dineutrino modes is strongly encouraged.

Our explorative works contributes to the advancement of flavorful SMEFT fits. We look forward to pursue flavorful global fits  with future data.

{\bf Note added:} During the completion of this work, the preprints 2507.06191~\cite{deBlas:2025xhe}  and 2507.13141~\cite{Chattopadhyay:2025air} 
appeared on the arXiv, which also consider multi-observable SMEFT fits but differ in the following from our works: The former~\cite{deBlas:2025xhe} focusses on $U(3)^5$ and $U(2)^5$ flavor symmetry, hence does not fit lepton-flavor specifically;  the latter~\cite{Chattopadhyay:2025air} analyses  LFV four-fermion operators outside of  MFV, also presenting only single-coefficient fit results.

\begin{acknowledgments}
    LN is supported by the doctoral scholarship program of the  {\it Studienstiftung des Deutschen Volkes}.
\end{acknowledgments}

\appendix

\section{Numerical Input Parameters}
\label{app:input_parameters}

$G_F$ and $v$  are  expressed in terms of  the input parameters $\{\aem, m_W, m_Z \}$  as
\begin{equation}
    G_F = \frac{\pi \aem m_Z^2}{\sqrt{2} m_W^2 (m_Z^2 - m_W^2)}\,(1 + \delta_R) \,, 
    \label{eq:GF_tree}
\end{equation}
where $\delta_R = 0.03686(21)$~\cite{ParticleDataGroup:2024cfk} and $ v = (\sqrt{2} G_F)^{-1/2}$.
The weak mixing angle $\theta_W$ is given as
\begin{equation}
    \sin^2 \theta_W = 1 - \frac{m_W^2}{m_Z^2}(1 +\delta \sin^2 \theta_W) \,,
    \label{eq:sin2thetaW}
\end{equation}
with $\delta \sin^2 \theta_W = 0.0349(3)$~\cite{ParticleDataGroup:2024cfk}.

Tab.~\ref{tab:input_parameters} lists the input parameters used in this analysis, from~\cite{ParticleDataGroup:2024cfk}. The values of the decay constants 
and form factors are given in Tab.~\ref{tab:meson_decay_constants} and Tab.~\ref{tab:form_factors}, respectively.
\begin{table}[H]
  \centering
    \setlength{\tabcolsep}{5pt}
    \renewcommand{\arraystretch}{1.2}
  \begin{tabular}{c l @{\,\hspace{40pt}\,} c l}
      \toprule
       Parameter & Value & Parameter & Value \\
      \midrule
      $\alpha_{\text{em}}$ & 0.0072973525693(11) & $\alpha_s (m_Z)$ & 0.1180 (9)  \\
      $m_W$ & 80.369(13)\,GeV  & $m_Z$ & 91.1880(20)\,GeV \\
      $m_e$ & 510.99895000(15)\,keV  & 
      $m_{\mu}$ & 105.6583755(23)\,MeV \\
      $m_{\tau}$ & 1.77693(9)\,GeV  & $m_b$ & 4.183(7)\,GeV \\
      $m_t$ & 72.57(29)\,GeV  & & \\
      \bottomrule
  \end{tabular}
  \caption{Input parameters used throughout this analysis, taken from~\cite{ParticleDataGroup:2024cfk} .}
  \label{tab:input_parameters}
\end{table}

\begin{table}[H]
  \centering
  \setlength{\tabcolsep}{8pt}
  \renewcommand{\arraystretch}{1.2}
  \begin{tabular}{l l @{\,\hspace{40pt}\,} l l }
      \hline
      $f_M$ & Value & $f_M$ & Value  \\
      \hline
      $f_{\pi}$ & 130.2(0.8)\,MeV~\cite{FlavourLatticeAveragingGroupFLAG:2024oxs} &
      $f_{\eta}^q$ & 108(3)\,MeV~\cite{Beneke:2002jn} \\ 
      $f_{\eta}^s$ & -111(6)\,MeV~\cite{Beneke:2002jn} & $f_{\eta^{\prime}}^q$ & 89(3)\,MeV~\cite{Beneke:2002jn} \\ 
      $f_{\eta^{\prime}}^s$ & 136(6)\,MeV~\cite{Beneke:2002jn} & 
      $f_{\rho/\omega}$ & 199(4) MeV~\cite{Braun:2016wnx} \\
      $f_{\phi}$ & 241(9)\,MeV~\cite{Chen:2020qma} & $f_K$ & 155.7(0.3)\,MeV~\cite{FlavourLatticeAveragingGroupFLAG:2024oxs} \\ 
      $f_{K^*}$ & 241\,MeV~\cite{Bhagwat:2006pu} &
      $f_{J/\Psi}$ & 418(9)\,MeV~\cite{Becirevic:2013bsa} \\ 
      $f_D$ & 210.4(1.5)\,MeV~\cite{FlavourLatticeAveragingGroupFLAG:2024oxs} & $f_{D_s}$ & 247.7(1.2)\,MeV~\cite{FlavourLatticeAveragingGroupFLAG:2024oxs} \\
      $f_{\Upsilon}$ & 677.2(9.7)\,MeV~\cite{Hatton:2021dvg} & $f_B$ & 190.0(1.3)\,MeV~\cite{FlavourLatticeAveragingGroupFLAG:2024oxs} \\
      $f_{B_s}$ & 230.3(1.3)\,MeV~\cite{FlavourLatticeAveragingGroupFLAG:2024oxs} & & \\
      \hline
  \end{tabular}
  \caption{Decay constants of the mesons used in this analysis.}
  \label{tab:meson_decay_constants}
\end{table}

\begin{table}
    \setlength{\tabcolsep}{5pt}
    \renewcommand{\arraystretch}{1.2}
    \centering
    \begin{tabular}{l l @{\,\hspace{40pt}\,} l l @{\,\hspace{40pt}\,} l l} 
        \toprule
        Decay & Ref. & Decay & Ref. & Decay & Ref. \\
        \midrule
        $D \to \pi$ & \cite{FermilabLattice:2022gku} &
        $D \to \eta$ & \cite{Wu:2006rd} & 
        $D \to \eta^{\prime}$ & \cite{Bali:2014pva} \\
        $D \to K$ & \cite{FlavourLatticeAveragingGroupFLAG:2024oxs} &
        $B \to \pi$ & \cite{FlavourLatticeAveragingGroupFLAG:2024oxs} &
        $B \to K$ & \cite{Grunwald:2023nli} \\
        $B \to D$ & \cite{FlavourLatticeAveragingGroupFLAG:2024oxs} & 
        $B_s \to K$ & \cite{FlavourLatticeAveragingGroupFLAG:2024oxs} &
        $B_s \to D_s$ & \cite{McLean:2019qcx} \\
        $D \to \rho$ & \cite{Wu:2006rd} & 
        $D \to \omega$ & \cite{Wu:2006rd} & 
        $D \to K^*$ & \cite{Wu:2006rd} \\
        $D_s \to \phi$ & \cite{Wu:2006rd} &
        $D_s \to K^*$ & \cite{Wu:2006rd} & 
        $B \to \rho$ & \cite{Bharucha:2015bzk} \\ 
        $B \to K^*$ & \cite{Bharucha:2015bzk} &
        $B \to D^*$ & \cite{Gubernari:2018wyi} & 
        $B_s \to \phi$ & \cite{Bharucha:2015bzk} \\
        $B_s \to K^*$ & \cite{Bharucha:2015bzk} & 
        $B_s \to D_s^*$ & \cite{FlavourLatticeAveragingGroupFLAG:2024oxs} & & \\
        \bottomrule
    \end{tabular}
    \caption{Form factors  for  semileptonic decays of mesons throughout this analysis. The form factors are obtained from lattice QCD calculations or light-cone sum rules.}
    \label{tab:form_factors}
\end{table}

\section{Treatment of the CKM matrix}
\label{app:CKM}

We parameterize the CKM matrix in terms of the Wolfenstein parameters $\lambda$, $A$, $\rho$ and $\eta$
\begin{equation}
  V_{\text{CKM}} =
  \left( \begin{array}{c@{\hskip -1pt}c@{\hskip -5pt}c}  
    1 - \frac{\lambda^2}{2} - \frac{\lambda^4}{8} & \lambda & A \lambda^3 \left(1+ \frac{\lambda^2}{2}\right) \left(\rho - i \eta\right) \\
    -\lambda + A^2 \lambda^5 \left(\frac{1}{2} - \rho -i\eta \right) & 1 - \frac{\lambda^2}{2} - \frac{\lambda^4}{8} \left(1+ 4A^2\right) & A \lambda^2 \\
    A \lambda^3 \left(1 - \rho - i \eta\right) & -A \lambda^2 + A \lambda^4 \left( \frac{1}{2} - \rho - i\eta \right) & 1 - \frac{\lambda^4}{2} A^2
  \end{array} \right)  \, .
  \label{eq:wolfenstein}
\end{equation}
In the global fit, we treat the CKM parameters as nuisance parameters.

For the MFV expansion  in terms of the quark yukawas which involve  CKM elements in the mass basis, we employ fixed (central) values for the CKM matrix elements, which we obtain from a fit of  ${\Gamma (K\to \mu \nu_{\mu})/ \Gamma(\pi \to \mu \nu_{\mu})}$, ${\Gamma(B\to \tau \nu_{\tau}), \ \Delta M_{d}, \ \Delta M_{s}}$~\cite{Descotes-Genon:2018foz},  neglecting BSM effects. 
Tab.~\ref{tab:ckm_inputs} lists the observables, experimental inputs and SM predictions. We obtain
\begin{equation}
    \{ \lambda,  A,  \rho,  \eta\} = \{0.22532(47), 0.804(17), 0.184(32), 0.388(52)\} \,,
\end{equation}
corresponding to the absolute values of the CKM matrix elements
\begin{equation}
    {\renewcommand{\arraystretch}{1.5}
    V_{\text{CKM}} = \begin{pmatrix}
    0.97429(11) & 0.22532(47) & 0.00405(49) \\
    0.22520(47) & 0.97346(11) & 0.04081(85) \\
    0.00832(21) & 0.04016(87) & 0.999167(35) 
  \end{pmatrix} \,.}
  \label{eqn:CKM_matrix_numerical}
\end{equation}
We use these results  from  "SM-only CKM  fit" to compute the SM predictions of various observables, such as those in Tab.\ref{tab:semileptonic_P_modes_dineutrinos}. 

\begin{table}[H]
    \centering
    \setlength{\tabcolsep}{10pt}
    \renewcommand{\arraystretch}{1.2}
    \begin{tabular}{l l l}
        \toprule
        Observable & Measurement & SM Prediction \\
        \midrule
        ${\cal B} (B^+ \to \tau^+ \bar \nu_{\tau})$ & $1.09(24)\cdot 10^{-4}$~\cite{ParticleDataGroup:2024cfk} & $6.361(89) \cdot \left| V_{ub} \right|^2$\\
        $\Gamma (K^+ \to \mu^+ \nu_{\mu}) / \Gamma(\pi^+ \to \mu^+ \nu_{\mu})$ & $1.3367(32)$~\cite{ParticleDataGroup:2024cfk} & $ 24.993(90) \cdot \left| V_{us}\right|^2/\left| V_{ud}\right|^2$ \\
        $\Delta M_{d}$ & $0.5069 (19) \cdot 10^{12} \,\text{s}^{-1}$~\cite{ParticleDataGroup:2024cfk} & $ 4.81(25)\cdot 10^{15} \cdot \left|V_{td}V_{tb}^*\right|^2\,\text{s}^{-1}$ \\
        $ \Delta M_{s}$ & $17.765 (5) \cdot 10^{12}\, \text{s}^{-1}$~\cite{ParticleDataGroup:2024cfk} & $7.23(32)\cdot 10^{15} \cdot \left|V_{ts}V_{tb}^*\right|^2\,\text{s}^{-1}$ \\
        \bottomrule
    \end{tabular}
    \caption{Observables used to determine the CKM parameters in the SM-only CKM fit. The SM predictions are computed following Ref.~\cite{Descotes-Genon:2018foz}.}
    \label{tab:ckm_inputs}
\end{table}

\section{Matching Conditions}
\label{app:matching_conditions}

We match the SMEFT coefficients of  the operators in Tab.~\ref{tab:operators} to the LEFT coefficients at the scale ${\mu = m_W}$  at tree-level using~\cite{Jenkins:2017jig}. The matching conditions 
for the dipole operators are given in Eq.~(\ref{eqn:Gamma_dipole_coefficient}). 

For  the four-fermion operators, the matching for the purely left-handed operators reads
\begin{align}
  &\begin{aligned}
    {\cal{C}}_{\underset{ijkl}{ee}}^{V,LL} =&\: \frac{1}{v^2} \Bigg( \tilde C_{\underset{ijkl}{ll}} - \frac{1}{4} \bigg( \Big(\delta_{ij} (2\szW - 1) -\tilde C_{\underset{ij}{\varphi l}}^{+}\Big)\Big(\delta_{kl} (2\szW - 1) -\tilde C_{\underset{kl}{\varphi l}}^{+}\Big) \\&+ \Big(\delta_{il} (2\szW - 1) -\tilde C_{\underset{il}{\varphi l}}^{+}\Big)\Big(\delta_{kj} (2\szW - 1) -\tilde C_{\underset{kj}{\varphi l}}^{+}\Big) \bigg) \Bigg) \,,
  \end{aligned} \\
  &\begin{aligned}
    {\cal{C}}_{\underset{ijkl}{\nu e}}^{V,LL} =&\: \frac{1}{v^2} \Bigg( \Big( \tilde C_{\underset{ijkl}{ll}} + \tilde C_{\underset{klij}{ll}} \Big) -  \bigg( \Big(\delta_{ij} -\tilde C_{\underset{ij}{\varphi l}}^{-}\Big)\Big(\delta_{kl} (2\szW - 1) -\tilde C_{\underset{kl}{\varphi l}}^{+}\Big) \\&- 2 \Big(\delta_{il} + \tilde C_{\underset{il}{\varphi l}}^{(3)}\Big)\Big(\delta_{kj} + \tilde C_{\underset{kj}{\varphi l}}^{(3)}\Big) \bigg) \Bigg) \,,
  \end{aligned} \\
  &\begin{aligned}
    {\cal{C}}_{\underset{ijkl}{\nu u}}^{V,LL} =&\: \frac{1}{v^2} \Bigg(\tilde C_{\underset{ijkl}{lq}}^{+}  -   \Big(\delta_{ij} -\tilde C_{\underset{ij}{\varphi l}}^{-}\Big)\Big(\delta_{kl} (1 - \frac{4}{3}\szW) \Big) \Bigg)\,,
  \end{aligned} \\
  &\begin{aligned}
    {\cal{C}}_{\underset{ijkl}{\nu d}}^{V,LL} =&\: \frac{1}{v^2} \Bigg( \tilde C_{\underset{ijkl}{lq}}^{-}  -  \Big(\delta_{ij} -\tilde C_{\underset{ij}{\varphi l}}^{-}\Big)\Big(\delta_{kl} (-1 + \frac{2}{3}\szW) \Big) \Bigg) \,,
  \end{aligned} \\
  &\begin{aligned}
    {\cal{C}}_{\underset{ijkl}{eu}}^{V,LL} =&\: \frac{1}{v^2} \Bigg( \tilde C_{\underset{ijkl}{lq}}^{-}  - \Big(\delta_{ij} (2\szW - 1) -\tilde C_{\underset{ij}{\varphi l}}^{+}\Big)\Big(\delta_{kl} (1 - \frac{4}{3}\szW) \Big)  \Bigg) \,,
  \end{aligned} \\
  &\begin{aligned}
    {\cal{C}}_{\underset{ijkl}{ed}}^{V,LL} =&\: \frac{1}{v^2} \Bigg( \tilde C_{\underset{ijkl}{lq}}^{+}  -  \Big(\delta_{ij} (2\szW - 1) -\tilde C_{\underset{ij}{\varphi l}}^{+}\Big)\Big(\delta_{kl} (-1 + \frac{2}{3}\szW) \Big) \Bigg) \,,
  \end{aligned} \\
  &\begin{aligned}
    {\cal{C}}_{\underset{ijkl}{\nu edu}}^{V,LL} + \text{h.c.}=&\: \frac{2}{v^2} \Bigg( \tilde C_{\underset{ijkl}{lq}}^{(3)} - \Big(\delta_{ij} + \tilde C_{\underset{ij}{\varphi l}}^{(3)}\Big) V_{lk } \Bigg) \,.
  \end{aligned} 
\end{align}
For the purely right-handed operators, the matching conditions are given by
\begin{align}
  &\begin{aligned}
    {\cal{C}}_{\underset{ijkl}{ee}}^{V,RR} =&\: \frac{1}{v^2} \Bigg( \tilde C_{\underset{ijkl}{ee}} - \frac{1}{4} \bigg( \Big(\delta_{ij} 2\szW -\tilde C_{\underset{ij}{\varphi e}}\Big)\Big(\delta_{kl} 2 \szW -\tilde C_{\underset{kl}{\varphi e}}\Big) \\&+ \Big(\delta_{il} 2\szW -\tilde C_{\underset{il}{\varphi e}}\Big)\Big(\delta_{kj} 2\szW  -\tilde C_{\underset{kj}{\varphi e}}\Big) \bigg) \Bigg) \,,
  \end{aligned} \\
  &\begin{aligned}
    {\cal{C}}_{\underset{ijkl}{eu}}^{V,RR} =&\: \frac{1}{v^2} \Bigg( \tilde C_{\underset{ijkl}{eu}} -   \Big(\delta_{ij} 2\szW -\tilde C_{\underset{ij}{\varphi e}}\Big)\Big( - \delta_{kl} \frac{4}{3} \szW \Big) \Bigg) \,,
  \end{aligned} \\
  &\begin{aligned}
    {\cal{C}}_{\underset{ijkl}{ed}}^{V,RR} =&\: \frac{1}{v^2} \Bigg( \tilde C_{\underset{ijkl}{ed}} - \Big(\delta_{ij} 2\szW -\tilde C_{\underset{ij}{\varphi e}}\Big)\Big(\delta_{kl} \frac{2}{3} \szW \Big)  \Bigg) \,,
  \end{aligned} \\
\end{align}
and for the mixed chirality operators, 
\begin{align}
  &\begin{aligned}
    {\cal{C}}_{\underset{ijkl}{ee}}^{V,LR} =&\: \frac{1}{v^2} \Bigg( \tilde C_{\underset{ijkl}{le}} -  \Big(\delta_{ij} (2\szW - 1) -\tilde C_{\underset{ij}{\varphi l}}^{+}\Big)\Big(\delta_{kl} 2 \szW -\tilde C_{\underset{kl}{\varphi e}}\Big) \Bigg) \,,
  \end{aligned} \\
  &\begin{aligned}
    {\cal{C}}_{\underset{ijkl}{\nu e}}^{V,LR} =&\: \frac{1}{v^2} \Bigg( \tilde C_{\underset{ijkl}{le}} -  \Big(\delta_{ij} -\tilde C_{\underset{ij}{\varphi l}}^{-}\Big)\Big(\delta_{kl} 2 \szW -\tilde C_{\underset{kl}{\varphi e}}\Big)  \Bigg) \,,
  \end{aligned} \\
  &\begin{aligned}
    {\cal{C}}_{\underset{ijkl}{\nu u}}^{V,LR} =&\: \frac{1}{v^2} \Bigg( \tilde C_{\underset{ijkl}{lu}} -   \Big(\delta_{ij} -\tilde C_{\underset{ij}{\varphi l}}^{-}\Big)\Big(-\delta_{kl}\frac{4}{3}\szW\Big)  \Bigg) \,,
  \end{aligned} \\
  &\begin{aligned}
    {\cal{C}}_{\underset{ijkl}{\nu d}}^{V,LR} =&\: \frac{1}{v^2} \Bigg( \tilde C_{\underset{ijkl}{ld}} -   \Big(\delta_{ij} -\tilde C_{\underset{ij}{\varphi l}}^{-}\Big)\Big(\delta_{kl}\frac{2}{3}\szW\Big)  \Bigg) \,,
  \end{aligned} \\
  &\begin{aligned}
    {\cal{C}}_{\underset{ijkl}{eu}}^{V,LR} =&\: \frac{1}{v^2} \Bigg( \tilde C_{\underset{ijkl}{lu}} -  \Big(\delta_{ij} (2\szW - 1) -\tilde C_{\underset{ij}{\varphi l}}^{+}\Big)\Big(-\delta_{kl}\frac{4}{3}\szW\Big)  \Bigg) \,,
  \end{aligned} \\
  &\begin{aligned}
    {\cal{C}}_{\underset{ijkl}{ed}}^{V,LR} =&\: \frac{1}{v^2} \Bigg( \tilde C_{\underset{ijkl}{ld}} -  \Big(\delta_{ij} (2\szW - 1) -\tilde C_{\underset{ij}{\varphi l}}^{+}\Big)\Big(\delta_{kl}\frac{2}{3}\szW\Big)  \Bigg) \,,
  \end{aligned} \\
  &\begin{aligned}
    {\cal{C}}_{\underset{ijkl}{ue}}^{V,LR} =&\: \frac{1}{v^2} \Bigg( \tilde C_{\underset{ijkl}{qe}} -   \Big(\delta_{ij}(1- \frac{4}{3}\szW)\Big)\Big(\delta_{kl} 2 \szW -\tilde C_{\underset{kl}{\varphi e}}\Big)  \Bigg) \,,
  \end{aligned} \\
  &\begin{aligned}
    {\cal{C}}_{\underset{ijkl}{de}}^{V,LR} =&\: \frac{1}{v^2} \Bigg( \tilde C_{\underset{ijkl}{qe}} -  \Big(\delta_{ij}(-1+ \frac{2}{3}\szW)\Big)\Big(\delta_{kl} 2 \szW -\tilde C_{\underset{kl}{\varphi e}}\Big)  \Bigg) \,.
  \end{aligned} 
\end{align}

\section{Credible intervals of the global fit}
\label{app:credible_intervals}

We show the 90\% credible intervals for the SMEFT coefficients for the global fit in Tabs.~\ref{tab:credible_intervals_hl}-\ref{tab:credible_intervals_lequ} for the different operator classes.

\begin{table}[h]
  \centering
  \setlength{\tabcolsep}{12pt}
  \renewcommand{\arraystretch}{1.2}
  \begin{tabular}{l l l l }
    \toprule
    Lepton & $\tilde C_{\varphi l}^{(1)}$ & $\tilde C_{\varphi l}^{(3)}$ & $\tilde C_{\varphi e}$ \\
    \midrule
    $ee$ & $[-0.59,5.52]\cdot 10^{-3}$ & $[-2.18,0.70]\cdot 10^{-3}$ & $[-2.59,5.83]\cdot 10^{-3}$ \\
    $\mu\mu$ & $[-3.83,9.65]\cdot 10^{-3}$ & $[-0.80,2.18]\cdot 10^{-3}$ & $[-4.88,9.06]\cdot 10^{-3}$ \\
    $\tau\tau$ & $[-0.73,1.83]\cdot 10^{-2}$ & $[-1.49,3.22]\cdot 10^{-3}$ & $[-2.83,0.85]\cdot 10^{-2}$ \\
    $e\mu$ & $[-2.11,3.95]\cdot 10^{-3}$ & $[-3.69,2.03]\cdot 10^{-3}$ & $[-4.83,4.32]\cdot 10^{-5}$ \\
    $e\tau$ & $[-3.07,2.39]\cdot 10^{-2}$ & $[-2.21,2.90]\cdot 10^{-2}$ & $[-8.54,8.51]\cdot 10^{-4}$ \\
    $\mu\tau$ & $[-2.50,3.67]\cdot 10^{-2}$ & $[-3.40,2.38]\cdot 10^{-2}$ & $[-7.49,7.51]\cdot 10^{-4}$ \\
    LU & $[0.21,3.08]\cdot 10^{-3}$ & $[-1.82,1.07]\cdot 10^{-3}$ & $[-1.20,3.80]\cdot 10^{-3}$ \\
    LD & $[-5.43,5.96]\cdot 10^{-4}$ & $[-4.99,5.41]\cdot 10^{-4}$ & $[-1.96,1.32]\cdot 10^{-4}$ \\
  \bottomrule
  \end{tabular}
  \caption{90\% credible intervals for the leptonic Higgs-current coefficients.}
  \label{tab:credible_intervals_hl}
\end{table}

\begin{table}[h]
  \centering
  \setlength{\tabcolsep}{12pt}
  \renewcommand{\arraystretch}{1.2}
  \begin{tabular}{l l l l }
    \toprule
    Lepton & $\tilde C_{lq}^{(1)}$ & $\tilde C_{lq}^{(3)}$ & $\tilde C_{qe}$ \\
    \midrule
    $ee$ & $[-0.95,1.10]\cdot 10^{-3}$ & $[-0.32,1.26]\cdot 10^{-3}$ & $[-6.17,9.79]\cdot 10^{-4}$ \\
    $\mu\mu$ & $[-0.46,1.64]\cdot 10^{-3}$ & $[-1.02,0.88]\cdot 10^{-3}$ & $[-2.04,0.54]\cdot 10^{-3}$ \\
    $\tau\tau$ & $[-4.33,3.35]\cdot 10^{-3}$ & $[-1.24,2.43]\cdot 10^{-3}$ & $[-9.94,1.46]\cdot 10^{-3}$ \\
    $e\mu$ & $[-2.80,3.15]\cdot 10^{-4}$ & $[-0.01,3.19]\cdot 10^{-4}$ & $[-1.72,1.13]\cdot 10^{-4}$ \\
    $e\tau$ & $[-3.43,3.84]\cdot 10^{-4}$ & $[-0.27,4.49]\cdot 10^{-4}$ & $[-3.70,3.69]\cdot 10^{-4}$ \\
    $\mu\tau$ & $[-3.12,3.56]\cdot 10^{-4}$ & $[0.78,5.74]\cdot 10^{-4}$ & $[-3.23,3.23]\cdot 10^{-4}$ \\
        LU & $[-4.72,9.14]\cdot 10^{-4}$ & $[-2.85,1.32]\cdot 10^{-4}$ & $[-3.07,6.81]\cdot 10^{-4}$ \\
    LD & $[-1.77,1.78]\cdot 10^{-4}$ & $[-0.86,5.91]\cdot 10^{-5}$ & $[-0.87,1.10]\cdot 10^{-4}$ \\
  \bottomrule
  \end{tabular}
  \caption{90\% credible intervals for the semileptonic four-fermion coefficients with left-handed quarks.}
  \label{tab:credible_intervals_lq_l}
\end{table}

\begin{table}[h]
  \centering
  \setlength{\tabcolsep}{8pt}
  \renewcommand{\arraystretch}{1.2}
  \begin{tabular}{l l l l l }
    \toprule
    Lepton & $\tilde C_{lu}$ & $\tilde C_{ld}$ & $\tilde C_{eu}$ & $\tilde C_{ed}$ \\
    \midrule
    $ee$ & $[-2.38,1.72]\cdot 10^{-3}$ & $[-2.22,2.54]\cdot 10^{-3}$ & $[-1.67,3.90]\cdot 10^{-3}$ & $[-3.18,2.00]\cdot 10^{-3}$ \\
    $\mu\mu$ & $[-2.77,5.82]\cdot 10^{-3}$ & $[-3.27,1.77]\cdot 10^{-3}$ & $[-8.17,3.24]\cdot 10^{-3}$ & $[-2.12,4.21]\cdot 10^{-3}$ \\
    $\tau\tau$ & $[-1.75,1.24]\cdot 10^{-2}$ & $[-6.79,9.17]\cdot 10^{-3}$ & $[-3.85,0.42]\cdot 10^{-2}$ & $[-0.27,1.98]\cdot 10^{-2}$ \\
    $e\mu$ & $[-4.98,4.88]\cdot 10^{-4}$ & $[-5.50,5.57]\cdot 10^{-4}$ & $[-2.63,2.68]\cdot 10^{-4}$ & $[-4.78,4.75]\cdot 10^{-4}$ \\
    $e\tau$ & $[-5.56,6.60]\cdot 10^{-4}$ & $[-4.90,4.07]\cdot 10^{-4}$ & $[-5.52,5.51]\cdot 10^{-4}$ & $[-4.48,4.47]\cdot 10^{-4}$ \\
    $\mu\tau$ & $[-4.94,6.30]\cdot 10^{-4}$ & $[-4.74,3.59]\cdot 10^{-4}$ & $[-5.12,5.09]\cdot 10^{-4}$ & $[-4.12,4.15]\cdot 10^{-4}$ \\
        LU & $[-0.69,1.86]\cdot 10^{-3}$ & $[-1.99,1.24]\cdot 10^{-3}$ & $[0.12,2.44]\cdot 10^{-3}$ & $[-2.39,0.80]\cdot 10^{-3}$ \\
    LD & $[-2.87,3.69]\cdot 10^{-4}$ & $[-2.51,2.50]\cdot 10^{-4}$ & $[-2.66,2.74]\cdot 10^{-4}$ & $[-2.39,2.19]\cdot 10^{-4}$ \\
  \bottomrule
  \end{tabular}
  \caption{90\% credible intervals for the semileptonic four-fermion coefficients with right-handed quarks.}
  \label{tab:credible_intervals_lq_r}
\end{table}

\begin{table}[h]
  \centering
  \setlength{\tabcolsep}{12pt}
  \renewcommand{\arraystretch}{1.2}
  \begin{tabular}{l l l l }
    \toprule
    Lepton & $\tilde C_{ll}$ & $\tilde C_{ee}$ & $\tilde C_{le}$ \\
    \midrule
    $ee$ & $[-0.01,0.19]$ & $[-0.21,0.14]$ & $[-0.24,0.08]$ \\
    $\mu\mu$ & $[-0.13,0.11]$ & $[-0.25,0.88]$ & $[-1.39,0.91]$ \\
    $\tau\tau$ & $[-0.04,0.41]$ & $[-0.89,3.51]$ & $[-3.62,4.60]$ \\
    $e\mu$ & $[-0.28,0.29]$ & & $[-0.49,0.56]$ \\
    $e\tau$ & $[-0.39,0.39]$ &  & $[-0.70,0.75]$ \\
    $\mu\tau$ & $[-0.38,0.35]$ &  & $[-0.62,0.72]$ \\
    LU & $[-4.84,1.49]\cdot 10^{-3}$ & $[-4.37,0.98]\cdot 10^{-2}$ & $[-2.55,0.42]\cdot 10^{-2}$ \\
    LD & $[-4.21,6.14]\cdot 10^{-5}$ & $[-4.34,2.87]\cdot 10^{-5}$ & $[-9.08,6.10]\cdot 10^{-5}$ \\
  \bottomrule
  \end{tabular}
  \caption{90\% credible intervals for the four-lepton coefficients.}
  \label{tab:credible_intervals_ll}
\end{table}

\begin{table}[h]
  \centering
  \setlength{\tabcolsep}{12pt}
  \renewcommand{\arraystretch}{1.2}
  \begin{tabular}{l l l }
    \toprule
    Lepton & $\tilde C_{eB}$ & $\tilde C_{eW}$ \\
    \midrule
    $ee$ & $[-3.49,3.55]\cdot 10^{-3}$ & $[-2.44,2.46]\cdot 10^{-2}$ \\
    $\mu\mu$ & $[-4.12,4.21]\cdot 10^{-3}$ & $[-2.94,2.87]\cdot 10^{-2}$ \\
    $\tau\tau$ & $[-1.06,1.04]$ & $[-0.25,0.24]$ \\
    $e\mu$ & $[-0.38,0.38]$ & $[-0.11,0.11]$ \\
    $\mu e$ & $[-0.37,0.37]$ & $[-0.11,0.11]$ \\
    $e\tau$ & $[-0.53,0.52]$ & $[-0.15,0.16]$ \\
    $\tau e$ & $[-0.52,0.52]$ & $[-0.15,0.15]$ \\
    $\mu\tau$ & $[-0.43,0.42]$ & $[-0.12,0.13]$ \\
    $\tau\mu$ & $[-0.47,0.47]$ & $[-0.14,0.14]$ \\
      LU & $[-5.47,5.46]\cdot 10^{-5}$ & $[-2.29,1.99]\cdot 10^{-2}$ \\
    LD & $[-5.95,0.11]\cdot 10^{-8}$ & $[-1.73,1.71]\cdot 10^{-4}$ \\
  \bottomrule
  \end{tabular}
  \caption{90\% credible intervals for the leptonic dipole coefficients.}
  \label{tab:credible_intervals_dp}
\end{table}

\begin{table}[h]
  \centering
  \setlength{\tabcolsep}{12pt}
  \renewcommand{\arraystretch}{1.2}
  \begin{tabular}{l l l }
    \toprule
    Lepton & $\tilde C_{lequ}^{(1)}$ & $\tilde C_{lequ}^{(3)}$ \\
    \midrule
    $ee$ & $[-0.14,0.15]$ & $[-2.68,2.63]\cdot 10^{-2}$ \\
    $\mu\mu$ & $[-0.22,0.20]$ & $[-3.11,3.18]\cdot 10^{-2}$ \\
    $\tau\tau$ & $[-121,130]$ & $[-8.16,7.98]$ \\
    $e\mu$ & $[-190,190]$ & $[-3.20,3.24]$ \\
    $\mu e$ & $[-186,187]$ & $[-3.10,3.08]$ \\
    $e\tau$ & $[-49.8,57.4]$ & $[-4.04,3.91]$ \\
    $\tau e$ & $[-50.7,50.7]$ & $[-3.82,3.84]$ \\
    $\mu\tau$ & $[-31.8,61.7]$ & $[-3.32,3.10]$ \\
    $\tau\mu$ & $[-46.6,46.9]$ & $[-3.48,3.50]$ \\
     LU & $[-0.10,0.10]$ & $[-1.76,0.50]\cdot 10^{-6}$ \\
    LD & $[-5.89,6.17]\cdot 10^{-3}$ & $[-2.49,2.34]\cdot 10^{-5}$ \\
  \bottomrule
  \end{tabular}
  \caption{90\% credible intervals for the scalar and tensor coefficients.}
  \label{tab:credible_intervals_lequ}
\end{table}

\section{Further predictions for $B \to (K,K^*,\pi,\rho) \nu \bar \nu$ }
\label{app:predictions_dineutrino}

\begin{figure}[h]
    \centering
    \includegraphics[width=0.8\textwidth]{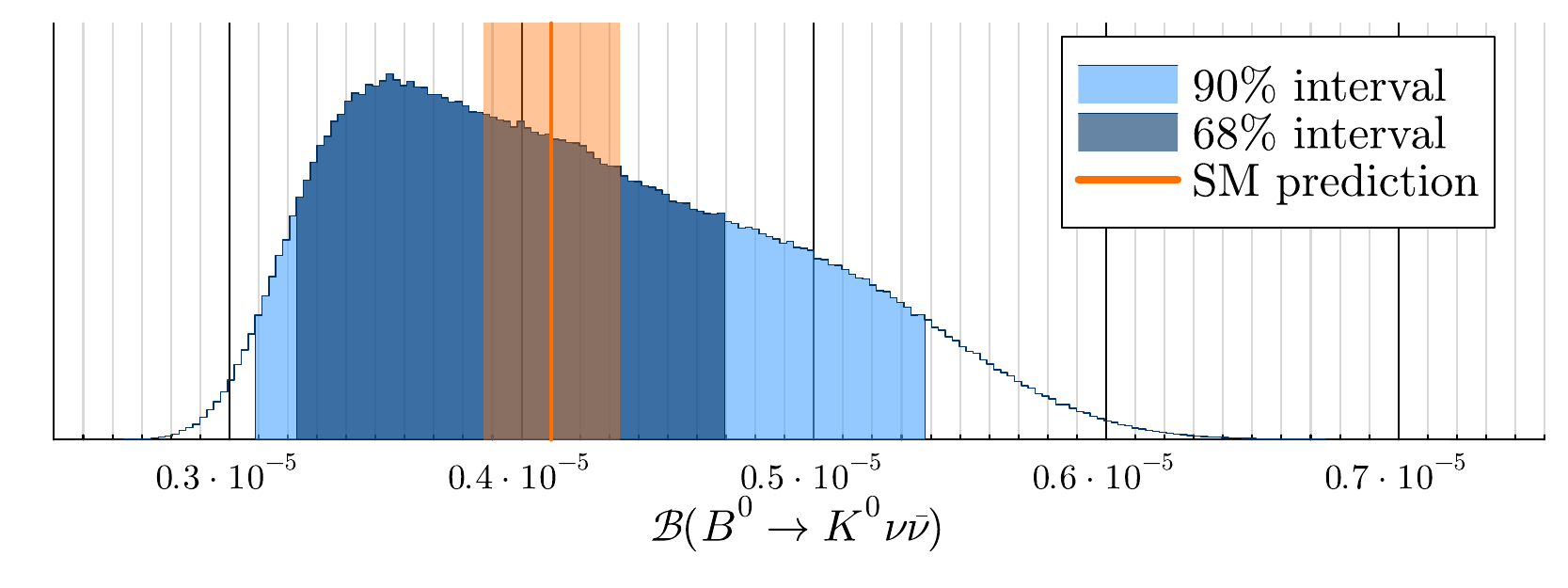}
    \includegraphics[width=0.8\textwidth]{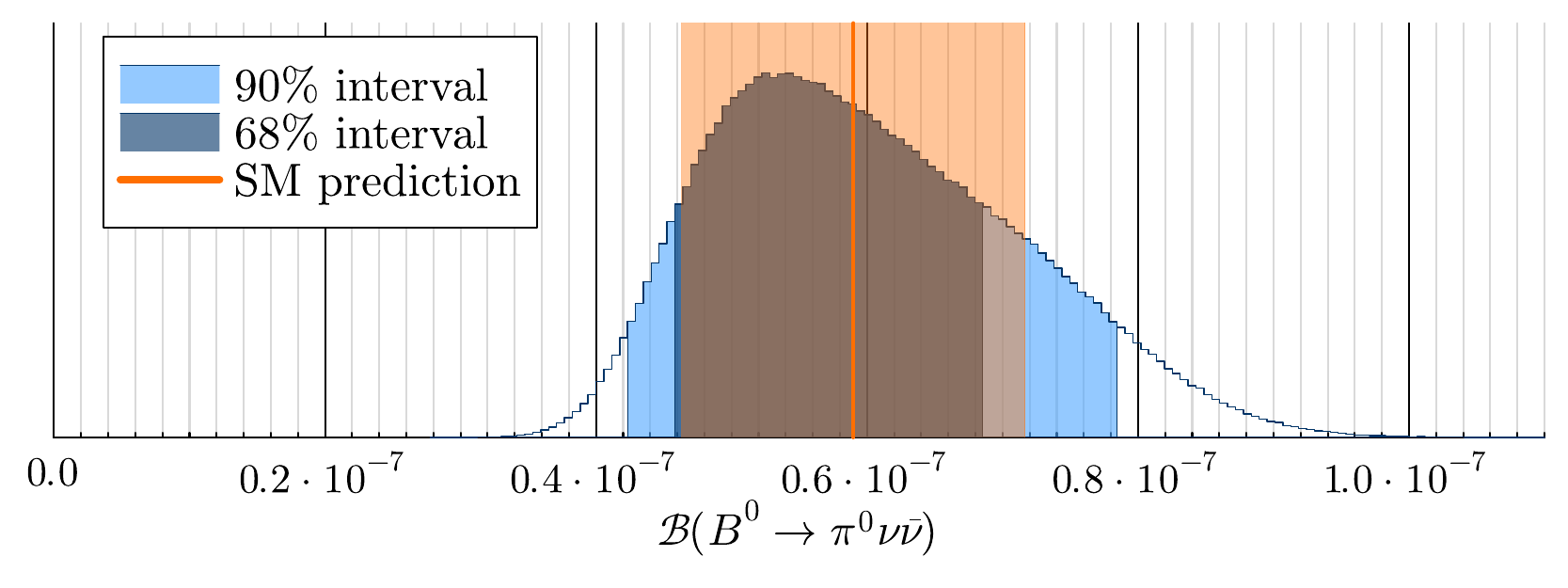}
    \caption{Predictions for $B^0 \to K^{0} \nu \bar \nu$ (top) and $B^0 \to \pi^0 \nu \bar \nu$ (bottom) based on the global fit in the $\tau\tau$ scenario. Colored bands indicate the 68\% and 90\% credible intervals; the orange band denotes the 1$\sigma$ SM prediction.}
    \label{fig:dineutrino_neutral}
\end{figure}

\begin{figure}[h]
    \centering
        \includegraphics[width=0.8\textwidth]{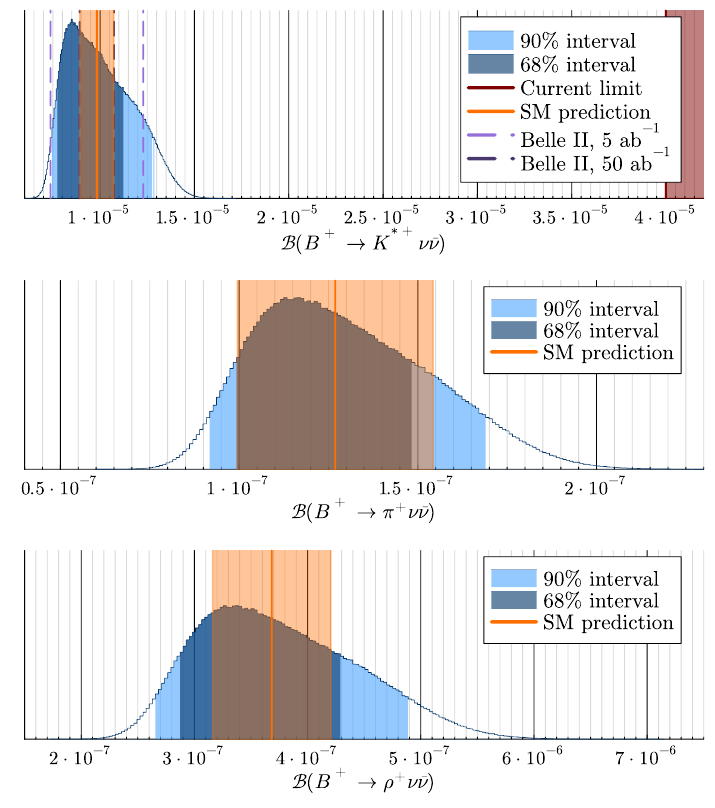}
    \caption{Predictions for the short-distance contributions to the branching ratios of $B^+ \to K^{*+} \nu \bar \nu$ (top), $B^+ \to \pi^+ \nu \bar \nu$ (middle) and $B^+ \to \rho^+ \nu \bar \nu$ (bottom) based on the global fit in the $\tau\tau$ scenario. Blue-colored bands indicate the 68\% (light) and 90\% (dark) credible intervals; the orange band denotes the 1$\sigma$ SM prediction. The red area represents the current 90\% CL bound 
    $\mathcal{B}(B^+ \to K^{*+} \nu \bar \nu) < 4.0  \cdot 10^{-5}$ \cite{Belle:2013tnz}, while dashed purple lines show Belle II projected sensitivities \cite{Belle-II:2018jsg}, if applicable.}
    \label{fig:dineutrino_charged}
\end{figure}

In Fig.~\ref{fig:dineutrino_neutral} we show  ${\cal B}(B^0 \to K^0 \nu \bar \nu)$ (top) and ${\cal B}(B^0 \to \pi^0 \nu \bar \nu)$ (bottom) based on the posterior probability distribution of the global fit in the $\tau\tau$ scenario. Fig.~\ref{fig:dineutrino_charged} shows the predictions  for the short distance contributions to the dineutrino decay rates of charged $B$ mesons.
For the latter, resonant-contributions from the $\tau$~ \cite{Kamenik:2009kc} are not included here.

\FloatBarrier

\clearpage

\bibliography{references}
\bibliographystyle{JHEP}
\end{document}

%% file: feynman/doubleInsertion.tex
\begin{tikzpicture}[scale=1, transform shape]

    \begin{scope}[shift={(0,0)}]
        \begin{feynman}
            \vertex (li) at (-1.5,1) {\(\ell^-\)};
            \vertex (lbari) at (-1.5,-1) {\(\ell^+\)};
            \vertex (ww1) at (0,0);
            \vertex (ww2) at (1.5,0);
            \vertex (lf) at (3,1) {\(\ell^-\)};
            \vertex (lbarf) at (3,-1) {\(\ell^+\)};
            \diagram* {
            (li) -- [thick, fermion] (ww1) -- [thick, fermion] (lbari),
            (ww1) -- [thick, boson, edge label=\(Z\), momentum' = \(q\),] (ww2),
            (lbarf) -- [thick, fermion] (ww2) -- [thick, fermion] (lf),
            };
        \end{feynman}

        \draw[fill=dOrange, dOrange] (0,0) circle (0.12) {};

        \draw[fill=dOrange, dOrange] (1.5,0) circle (0.12) {};

        \coordinate (arrowStart) at (3.3, 0) {};
        \coordinate (arrowEnd) at (5.2, 0) {};
        \node at (4.2,0.3) {{\(q^2 \ll m_Z^2\)}}; 

        \node at (-0.9,0.05) {{\({\tilde C}_{\varphi e}\)}}; 
        \node at (2.35,0) {{\({\tilde C}_{\varphi e}\)}}; 


        \draw [thick, decoration={markings,
        mark=at position 1 with {\arrow[scale=1,>=Triangle]{Triangle}}},
        postaction={decorate}] (arrowStart) -- (arrowEnd);

    \end{scope}

    \begin{scope}[shift={(7,0)}]
        \begin{feynman}
            \vertex (li) at (-1.5,1) {\(\ell^-\)};
            \vertex (lbari) at (-1.5,-1) {\(\ell^+\)};
            \vertex (ww) at (0,0);
            \vertex (lf) at (1.5,1) {\(\ell^-\)};
            \vertex (lbarf) at (1.5,-1) {\(\ell^+\)};
            \diagram* {
            (li) -- [thick, fermion] (ww) -- [thick, fermion] (lbari),
            (lbarf) -- [thick, fermion] (ww) -- [thick, fermion] (lf),
            };
        \end{feynman}

        \draw[fill=dOrange, dOrange] (0,0) circle (0.12) {};

        \node at (0.1,0.6) {{\({\cal C}^{V,RR}_{ee}\)}}; 
        \node at (0.9,0) {{\(\sim \Lambda^{-4}\)}}; 


    \end{scope}

\end{tikzpicture}

%% file: references.bib
@article{Buchmuller:1985jz,
  author       = {Buchmuller, W. and Wyler, D.},
  title        = {{Effective Lagrangian Analysis of New Interactions and Flavor Conservation}},
  reportnumber = {CERN-TH-4254/85},
  doi          = {10.1016/0550-3213(86)90262-2},
  journal      = {Nucl. Phys. B},
  volume       = {268},
  pages        = {621--653},
  year         = {1986}
}

@article{Li:2020lba,
    author = "Li, Tong and Ma, Xiao-Dong and Schmidt, Michael A.",
    title = "{General neutrino interactions with sterile neutrinos in light of coherent neutrino-nucleus scattering and meson invisible decays}",
    eprint = "2005.01543",
    archivePrefix = "arXiv",
    primaryClass = "hep-ph",
    doi = "10.1007/JHEP07(2020)152",
    journal = "JHEP",
    volume = "07",
    pages = "152",
    year = "2020"
}

@article{DAmbrosio:2002vsn,
    author = "D'Ambrosio, G. and Giudice, G. F. and Isidori, G. and Strumia, A.",
    title = "{Minimal flavor violation: An Effective field theory approach}",
    eprint = "hep-ph/0207036",
    archivePrefix = "arXiv",
    reportNumber = "CERN-TH-2002-147, IFUP-TH-2002-17",
    doi = "10.1016/S0550-3213(02)00836-2",
    journal = "Nucl. Phys. B",
    volume = "645",
    pages = "155--187",
    year = "2002"
}

@article{Achasov:2023gey,
    author = "Achasov, M. and others",
    title = "{STCF conceptual design report (Volume 1): Physics {\&} detector}",
    eprint = "2303.15790",
    archivePrefix = "arXiv",
    primaryClass = "hep-ex",
    doi = "10.1007/s11467-023-1333-z",
    journal = "Front. Phys. (Beijing)",
    volume = "19",
    number = "1",
    pages = "14701",
    year = "2024"
}

@article{Brivio:2017vri,
    author = "Brivio, Ilaria and Trott, Michael",
    title = "{The Standard Model as an Effective Field Theory}",
    eprint = "1706.08945",
    archivePrefix = "arXiv",
    primaryClass = "hep-ph",
    doi = "10.1016/j.physrep.2018.11.002",
    journal = "Phys. Rept.",
    volume = "793",
    pages = "1--98",
    year = "2019"
}

@article{FCC:2025lpp,
    author = "Benedikt, M. and others",
    collaboration = "FCC",
    title = "{Future Circular Collider Feasibility Study Report: Volume 1, Physics, Experiments, Detectors}",
    eprint = "2505.00272",
    archivePrefix = "arXiv",
    primaryClass = "hep-ex",
    reportNumber = "CERN-FCC-PHYS-2025-0002",
    doi = "10.17181/CERN.9DKX.TDH9",
    month = "4",
    year = "2025"
}

@article{Ai:2024nmn,
    author = "Ai, Xiaocong and others",
    title = "{Flavor physics at the CEPC: a general perspective*}",
    eprint = "2412.19743",
    archivePrefix = "arXiv",
    primaryClass = "hep-ex",
    doi = "10.1088/1674-1137/adf1f0",
    journal = "Chin. Phys.",
    volume = "49",
    number = "10",
    pages = "103003",
    year = "2025"
}

@article{Aebischer:2021uvt,
    author = "Aebischer, Jason and Dekens, Wouter and Jenkins, Elizabeth E. and Manohar, Aneesh V. and Sengupta, Dipan and Stoffer, Peter",
    title = "{Effective field theory interpretation of lepton magnetic and electric dipole moments}",
    eprint = "2102.08954",
    archivePrefix = "arXiv",
    primaryClass = "hep-ph",
    reportNumber = "UWThPh 2021-2",
    doi = "10.1007/JHEP07(2021)107",
    journal = "JHEP",
    volume = "07",
    pages = "107",
    year = "2021"
}

@article{Kamenik:2009kc,
    author = "Kamenik, Jernej F. and Smith, Christopher",
    title = "{Tree-level contributions to the rare decays B+ ---{\ensuremath{>}} pi+ nu anti-nu, B+ ---{\ensuremath{>}} K+ nu anti-nu, and B+ ---{\ensuremath{>}} K*+ nu anti-nu in the Standard Model}",
    eprint = "0908.1174",
    archivePrefix = "arXiv",
    primaryClass = "hep-ph",
    reportNumber = "TTP09-27, SFB-CPP-09-71",
    doi = "10.1016/j.physletb.2009.09.041",
    journal = "Phys. Lett. B",
    volume = "680",
    pages = "471--475",
    year = "2009"
}

@article{Belle:2017oht,
    author = "Grygier, J. and others",
    collaboration = "Belle",
    title = "{Search for $\boldsymbol{B\to h\nu\bar{\nu}}$ decays with semileptonic tagging at Belle}",
    eprint = "1702.03224",
    archivePrefix = "arXiv",
    primaryClass = "hep-ex",
    doi = "10.1103/PhysRevD.96.091101",
    journal = "Phys. Rev. D",
    volume = "96",
    number = "9",
    pages = "091101",
    year = "2017",
    note = "[Addendum: Phys.Rev.D 97, 099902 (2018)]"
}

@article{Belle:2013tnz,
    author = "Lutz, O. and others",
    collaboration = "Belle",
    title = "{Search for $B \to h^{(*)} \nu \bar{\nu}$ with the full Belle $\Upsilon(4S)$ data sample}",
    eprint = "1303.3719",
    archivePrefix = "arXiv",
    primaryClass = "hep-ex",
    reportNumber = "BELLE-PREPRINT-2013-1, KEK-PREPRINT-2012-37",
    doi = "10.1103/PhysRevD.87.111103",
    journal = "Phys. Rev. D",
    volume = "87",
    number = "11",
    pages = "111103",
    year = "2013"
}

@article{Crivellin:2013hpa,
  author        = {Crivellin, Andreas and Najjari, Saereh and Rosiek, Janusz},
  title         = {{Lepton Flavor Violation in the Standard Model with general Dimension-Six Operators}},
  eprint        = {1312.0634},
  archiveprefix = {arXiv},
  primaryclass  = {hep-ph},
  doi           = {10.1007/JHEP04(2014)167},
  journal       = {JHEP},
  volume        = {04},
  pages         = {167},
  year          = {2014}
}

@article{Plakias:2023esq,
  author        = {Plakias, I. and Sumensari, O.},
  title         = {{Lepton Flavor Violation in Semileptonic Observables}},
  eprint        = {2312.14070},
  archiveprefix = {arXiv},
  primaryclass  = {hep-ph},
  month         = {12},
  year          = {2023}
}

@article{ATLAS:2020tre,
  author        = {Aad, Georges and others},
  collaboration = {ATLAS},
  title         = {{Search for lepton-flavour violation in high-mass dilepton final states using 139 fb$^{-1}$ of pp collisions at $ \sqrt{s} $ = 13 TeV with the ATLAS detector}},
  eprint        = {2307.08567},
  archiveprefix = {arXiv},
  primaryclass  = {hep-ex},
  reportnumber  = {CERN-EP-2023-089},
  doi           = {10.1007/JHEP10(2023)082},
  journal       = {JHEP},
  volume        = {23},
  pages         = {082},
  year          = {2020}
}

@article{ATLAS:2019lsy,
  author        = {Aad, Georges and others},
  collaboration = {ATLAS},
  title         = {{Search for a heavy charged boson in events with a charged lepton and missing transverse momentum from $pp$ collisions at $\sqrt{s} = 13$ TeV with the ATLAS detector}},
  eprint        = {1906.05609},
  archiveprefix = {arXiv},
  primaryclass  = {hep-ex},
  reportnumber  = {CERN-EP-2019-100},
  doi           = {10.1103/PhysRevD.100.052013},
  journal       = {Phys. Rev. D},
  volume        = {100},
  number        = {5},
  pages         = {052013},
  year          = {2019}
}

@article{CMS:2022ncp,
  author        = {Tumasyan, A. and others},
  collaboration = {CMS},
  title         = {{Search for new physics in the \ensuremath{\tau} lepton plus missing transverse momentum final state in proton-proton collisions at $ \sqrt{s} $ = 13 TeV}},
  eprint        = {2212.12604},
  archiveprefix = {arXiv},
  primaryclass  = {hep-ex},
  reportnumber  = {CMS-EXO-21-009, CERN-EP-2022-268},
  doi           = {10.1007/JHEP09(2023)051},
  journal       = {JHEP},
  volume        = {09},
  pages         = {051},
  year          = {2023}
}

@article{ATLAS:2022uhq,
  author        = {Aad, Georges and others},
  collaboration = {ATLAS},
  title         = {{Search for the charged-lepton-flavor-violating decay $Z\rightarrow e\mu$ in $pp$ collisions at $\sqrt{s}=13$ TeV with the ATLAS detector}},
  eprint        = {2204.10783},
  archiveprefix = {arXiv},
  primaryclass  = {hep-ex},
  reportnumber  = {CERN-EP-2022-007},
  doi           = {10.1103/PhysRevD.108.032015},
  journal       = {Phys. Rev. D},
  volume        = {108},
  pages         = {032015},
  year          = {2023}
}

@article{ATLAS:2021bdj,
  author        = {Aad, Georges and others},
  collaboration = {ATLAS},
  title         = {{Search for lepton-flavor-violation in $Z$-boson decays with $\tau$-leptons with the ATLAS detector}},
  eprint        = {2105.12491},
  archiveprefix = {arXiv},
  primaryclass  = {hep-ex},
  reportnumber  = {CERN-EP-2021-067},
  doi           = {10.1103/PhysRevLett.127.271801},
  journal       = {Phys. Rev. Lett.},
  volume        = {127},
  pages         = {271801},
  year          = {2022}
}

@article{SINDRUM:1987nra,
  author        = {Bellgardt, U. and others},
  collaboration = {SINDRUM},
  title         = {{Search for the Decay $\mu^+ \to e^+ e^+ e^-$}},
  reportnumber  = {SIN-PR-87-09},
  doi           = {10.1016/0550-3213(88)90462-2},
  journal       = {Nucl. Phys. B},
  volume        = {299},
  pages         = {1--6},
  year          = {1988}
}

@article{Hayasaka:2010np,
  author        = {Hayasaka, K. and others},
  title         = {{Search for Lepton Flavor Violating Tau Decays into Three Leptons with 719 Million Produced Tau+Tau- Pairs}},
  eprint        = {1001.3221},
  archiveprefix = {arXiv},
  primaryclass  = {hep-ex},
  doi           = {10.1016/j.physletb.2010.03.037},
  journal       = {Phys. Lett. B},
  volume        = {687},
  pages         = {139--143},
  year          = {2010}
}

@article{Belle:2007cio,
  author        = {Miyazaki, Y. and others},
  collaboration = {Belle},
  title         = {{Search for lepton flavor violating tau- decays into l- eta, l- eta-prime and l- pi0}},
  eprint        = {hep-ex/0703009},
  archiveprefix = {arXiv},
  reportnumber  = {BELLE-PRERPINT-2007-13, KEK-PRERPINT-2006-78},
  doi           = {10.1016/j.physletb.2007.03.027},
  journal       = {Phys. Lett. B},
  volume        = {648},
  pages         = {341--350},
  year          = {2007}
}

@article{BaBar:2006jhm,
  author        = {Aubert, Bernard and others},
  collaboration = {BaBar},
  title         = {{Search for Lepton Flavor Violating Decays $\tau^\pm \to \ell^\pm \pi^0$, $\ell^\pm \eta$, $\ell^\pm \eta^\prime$}},
  eprint        = {hep-ex/0610067},
  archiveprefix = {arXiv},
  reportnumber  = {SLAC-PUB-12170, BABAR-PUB-06-061},
  doi           = {10.1103/PhysRevLett.98.061803},
  journal       = {Phys. Rev. Lett.},
  volume        = {98},
  pages         = {061803},
  year          = {2007}
}

@article{BaBar:2009hkt,
  author        = {Aubert, Bernard and others},
  collaboration = {BaBar},
  title         = {{Searches for Lepton Flavor Violation in the Decays tau+- ---\ensuremath{>} e+- gamma and tau+- ---\ensuremath{>} mu+- gamma}},
  eprint        = {0908.2381},
  archiveprefix = {arXiv},
  primaryclass  = {hep-ex},
  reportnumber  = {SLAC-PUB-13753, BABAR-PUB-09-026},
  doi           = {10.1103/PhysRevLett.104.021802},
  journal       = {Phys. Rev. Lett.},
  volume        = {104},
  pages         = {021802},
  year          = {2010}
}

@article{Belle:2021ysv,
  author        = {Abdesselam, A. and others},
  collaboration = {Belle},
  title         = {{Search for lepton-flavor-violating tau-lepton decays to $\ell\gamma$ at Belle}},
  eprint        = {2103.12994},
  archiveprefix = {arXiv},
  primaryclass  = {hep-ex},
  doi           = {10.1007/JHEP10(2021)019},
  journal       = {JHEP},
  volume        = {10},
  pages         = {19},
  year          = {2021}
}

@article{MEG:2016leq,
  author        = {Baldini, A. M. and others},
  collaboration = {MEG},
  title         = {{Search for the lepton flavour violating decay $\mu ^+ \rightarrow \mathrm {e}^+ \gamma $ with the full dataset of the MEG experiment}},
  eprint        = {1605.05081},
  archiveprefix = {arXiv},
  primaryclass  = {hep-ex},
  doi           = {10.1140/epjc/s10052-016-4271-x},
  journal       = {Eur. Phys. J. C},
  volume        = {76},
  number        = {8},
  pages         = {434},
  year          = {2016}
}

@article{Angelescu:2020uug,
  author        = {Angelescu, Andrei and Faroughy, Darius A. and Sumensari, Olcyr},
  title         = {{Lepton Flavor Violation and Dilepton Tails at the LHC}},
  eprint        = {2002.05684},
  archiveprefix = {arXiv},
  primaryclass  = {hep-ph},
  reportnumber  = {ZU-TH 02/20},
  doi           = {10.1140/epjc/s10052-020-8210-5},
  journal       = {Eur. Phys. J. C},
  volume        = {80},
  number        = {7},
  pages         = {641},
  year          = {2020}
}

@article{Dedes:2023zws,
  author        = {Dedes, A. and Rosiek, J. and Ryczkowski, M. and Suxho, K. and Trifyllis, L.},
  title         = {{SmeftFR v3 \textendash{} Feynman rules generator for the Standard Model Effective Field Theory}},
  eprint        = {2302.01353},
  archiveprefix = {arXiv},
  primaryclass  = {hep-ph},
  doi           = {10.1016/j.cpc.2023.108943},
  journal       = {Comput. Phys. Commun.},
  volume        = {294},
  pages         = {108943},
  year          = {2024}
}

@article{Descotes-Genon:2018foz,
  author        = {Descotes-Genon, S\'ebastien and Falkowski, Adam and Fedele, Marco and Gonz\'alez-Alonso, Mart\'\i{}n and Virto, Javier},
  title         = {{The CKM parameters in the SMEFT}},
  eprint        = {1812.08163},
  archiveprefix = {arXiv},
  primaryclass  = {hep-ph},
  reportnumber  = {LPT Orsay 18-92, CERN-TH-2018-276, TUM-HEP-1178/18, MIT-CTP/5081,
                   NIOBE-2018-01},
  doi           = {10.1007/JHEP05(2019)172},
  journal       = {JHEP},
  volume        = {05},
  pages         = {172},
  year          = {2019}
}

@article{ParticleDataGroup:2024cfk,
  author        = {Navas, S. and others},
  collaboration = {Particle Data Group},
  title         = {{Review of particle physics}},
  doi           = {10.1103/PhysRevD.110.030001},
  journal       = {Phys. Rev. D},
  volume        = {110},
  number        = {3},
  pages         = {030001},
  year          = {2024}
}

@article{ALEPH:2005ab,
  author        = {Schael, S. and others},
  collaboration = {ALEPH, DELPHI, L3, OPAL, SLD, LEP Electroweak Working Group, SLD Electroweak Group, SLD Heavy Flavour Group},
  title         = {{Precision electroweak measurements on the $Z$ resonance}},
  eprint        = {hep-ex/0509008},
  archiveprefix = {arXiv},
  reportnumber  = {SLAC-R-774},
  doi           = {10.1016/j.physrep.2005.12.006},
  journal       = {Phys. Rept.},
  volume        = {427},
  pages         = {257--454},
  year          = {2006}
}

@article{CMS:2022mhs,
  author        = {Tumasyan, Armen and others},
  collaboration = {CMS},
  title         = {{Precision measurement of the W boson decay branching fractions in proton-proton collisions at $\sqrt{s}$ = 13 TeV}},
  eprint        = {2201.07861},
  archiveprefix = {arXiv},
  primaryclass  = {hep-ex},
  reportnumber  = {CMS-SMP-18-011, CERN-EP-2021-240},
  doi           = {10.1103/PhysRevD.105.072008},
  journal       = {Phys. Rev. D},
  volume        = {105},
  number        = {7},
  pages         = {072008},
  year          = {2022}
}

@article{Aoyama:2019ryr,
  author  = {Aoyama, Tatsumi and Kinoshita, Toichiro and Nio, Makiko},
  title   = {{Theory of the Anomalous Magnetic Moment of the Electron}},
  doi     = {10.3390/atoms7010028},
  journal = {Atoms},
  volume  = {7},
  number  = {1},
  pages   = {28},
  year    = {2019}
}

@article{Eidelman:2007sb,
  author        = {Eidelman, S. and Passera, M.},
  title         = {{Theory of the tau lepton anomalous magnetic moment}},
  eprint        = {hep-ph/0701260},
  archiveprefix = {arXiv},
  doi           = {10.1142/S0217732307022694},
  journal       = {Mod. Phys. Lett. A},
  volume        = {22},
  pages         = {159--179},
  year          = {2007}
}

@article{ATLAS:2022ryk,
  author        = {Aad, Georges and others},
  collaboration = {ATLAS},
  title         = {{Observation of the \ensuremath{\gamma}\ensuremath{\gamma}\textrightarrow{}\ensuremath{\tau}\ensuremath{\tau} Process in Pb+Pb Collisions and Constraints on the \ensuremath{\tau}-Lepton Anomalous Magnetic Moment with the ATLAS Detector}},
  eprint        = {2204.13478},
  archiveprefix = {arXiv},
  primaryclass  = {hep-ex},
  reportnumber  = {CERN-EP-2022-079},
  doi           = {10.1103/PhysRevLett.131.151802},
  journal       = {Phys. Rev. Lett.},
  volume        = {131},
  number        = {15},
  pages         = {151802},
  year          = {2023}
}

@article{Grzadkowski:2010es,
  author        = {Grzadkowski, B. and Iskrzynski, M. and Misiak, M. and Rosiek, J.},
  title         = {{Dimension-Six Terms in the Standard Model Lagrangian}},
  eprint        = {1008.4884},
  archiveprefix = {arXiv},
  primaryclass  = {hep-ph},
  reportnumber  = {IFT-9-2010, TTP10-35},
  doi           = {10.1007/JHEP10(2010)085},
  journal       = {JHEP},
  volume        = {10},
  pages         = {085},
  year          = {2010}
}

@article{Grunwald:2023nli,
  author        = {Grunwald, Cornelius and Hiller, Gudrun and Kr\"oninger, Kevin and Nollen, Lara},
  title         = {{More synergies from beauty, top, Z and Drell-Yan measurements in SMEFT}},
  eprint        = {2304.12837},
  archiveprefix = {arXiv},
  primaryclass  = {hep-ph},
  reportnumber  = {DO-TH 23/03},
  doi           = {10.1007/JHEP11(2023)110},
  journal       = {JHEP},
  volume        = {11},
  pages         = {110},
  year          = {2023}
}

@article{Greljo:2022cah,
  author        = {Greljo, Admir and Palavri\'c, Ajdin and Thomsen, Anders Eller},
  title         = {{Adding Flavor to the SMEFT}},
  eprint        = {2203.09561},
  archiveprefix = {arXiv},
  primaryclass  = {hep-ph},
  doi           = {10.1007/JHEP10(2022)005},
  journal       = {JHEP},
  volume        = {10},
  pages         = {010},
  year          = {2022}
}

@article{Jenkins:2013zja,
  author        = {Jenkins, Elizabeth E. and Manohar, Aneesh V. and Trott, Michael},
  title         = {{Renormalization Group Evolution of the Standard Model Dimension Six Operators I: Formalism and lambda Dependence}},
  eprint        = {1308.2627},
  archiveprefix = {arXiv},
  primaryclass  = {hep-ph},
  doi           = {10.1007/JHEP10(2013)087},
  journal       = {JHEP},
  volume        = {10},
  pages         = {087},
  year          = {2013}
}

@article{Jenkins:2013wua,
  author        = {Jenkins, Elizabeth E. and Manohar, Aneesh V. and Trott, Michael},
  title         = {{Renormalization Group Evolution of the Standard Model Dimension Six Operators II: Yukawa Dependence}},
  eprint        = {1310.4838},
  archiveprefix = {arXiv},
  primaryclass  = {hep-ph},
  reportnumber  = {CERN-PH-TH/2015-247},
  doi           = {10.1007/JHEP01(2014)035},
  journal       = {JHEP},
  volume        = {01},
  pages         = {035},
  year          = {2014}
}

@article{Alonso:2013hga,
  author        = {Alonso, Rodrigo and Jenkins, Elizabeth E. and Manohar, Aneesh V. and Trott, Michael},
  title         = {{Renormalization Group Evolution of the Standard Model Dimension Six Operators III: Gauge Coupling Dependence and Phenomenology}},
  eprint        = {1312.2014},
  archiveprefix = {arXiv},
  primaryclass  = {hep-ph},
  reportnumber  = {CERN-PH-TH-2013-305, CERN-PH-TH/2013-305},
  doi           = {10.1007/JHEP04(2014)159},
  journal       = {JHEP},
  volume        = {04},
  pages         = {159},
  year          = {2014}
}

@article{Aebischer:2018bkb,
  author        = {Aebischer, Jason and Kumar, Jacky and Straub, David M.},
  title         = {{Wilson: a Python package for the running and matching of Wilson coefficients above and below the electroweak scale}},
  eprint        = {1804.05033},
  archiveprefix = {arXiv},
  primaryclass  = {hep-ph},
  doi           = {10.1140/epjc/s10052-018-6492-7},
  journal       = {Eur. Phys. J. C},
  volume        = {78},
  number        = {12},
  pages         = {1026},
  year          = {2018}
}

@article{Farina:2016rws,
  author        = {Farina, Marco and Panico, Giuliano and Pappadopulo, Duccio and Ruderman, Joshua T. and Torre, Riccardo and Wulzer, Andrea},
  title         = {{Energy helps accuracy: electroweak precision tests at hadron colliders}},
  eprint        = {1609.08157},
  archiveprefix = {arXiv},
  primaryclass  = {hep-ph},
  reportnumber  = {CERN-TH-2016-205},
  doi           = {10.1016/j.physletb.2017.06.043},
  journal       = {Phys. Lett. B},
  volume        = {772},
  pages         = {210--215},
  year          = {2017}
}

@article{CMS:2021ctt,
  author        = {Sirunyan, Albert M and others},
  collaboration = {CMS},
  title         = {{Search for resonant and nonresonant new phenomena in high-mass dilepton final states at $ \sqrt{s} $ = 13 TeV}},
  eprint        = {2103.02708},
  archiveprefix = {arXiv},
  primaryclass  = {hep-ex},
  reportnumber  = {CMS-EXO-19-019, CERN-EP-2021-026},
  doi           = {10.1007/JHEP07(2021)208},
  journal       = {JHEP},
  volume        = {07},
  pages         = {208},
  year          = {2021}
}

@article{Jenkins:2017jig,
  author        = {Jenkins, Elizabeth E. and Manohar, Aneesh V. and Stoffer, Peter},
  title         = {{Low-Energy Effective Field Theory below the Electroweak Scale: Operators and Matching}},
  eprint        = {1709.04486},
  archiveprefix = {arXiv},
  primaryclass  = {hep-ph},
  doi           = {10.1007/JHEP03(2018)016},
  journal       = {JHEP},
  volume        = {03},
  pages         = {016},
  year          = {2018},
  note          = {[Erratum: JHEP 12, 043 (2023)]}
}

@article{Jenkins:2017dyc,
  author        = {Jenkins, Elizabeth E. and Manohar, Aneesh V. and Stoffer, Peter},
  title         = {{Low-Energy Effective Field Theory below the Electroweak Scale: Anomalous Dimensions}},
  eprint        = {1711.05270},
  archiveprefix = {arXiv},
  primaryclass  = {hep-ph},
  doi           = {10.1007/JHEP01(2018)084},
  journal       = {JHEP},
  volume        = {01},
  pages         = {084},
  year          = {2018},
  note          = {[Erratum: JHEP 12, 042 (2023)]}
}

@article{FlavourLatticeAveragingGroupFLAG:2024oxs,
  author        = {Aoki, Y. and others},
  collaboration = {Flavour Lattice Averaging Group (FLAG)},
  title         = {{FLAG Review 2024}},
  eprint        = {2411.04268},
  archiveprefix = {arXiv},
  primaryclass  = {hep-lat},
  reportnumber  = {CERN-TH-2024-192, FERMILAB-PUB-24-0785-T},
  month         = {11},
  year          = {2024}
}

@article{Arroyo-Urena:2021nil,
  author        = {Arroyo-Ure\~na, M. A. and Hern\'andez-Tom\'e, G. and L\'opez-Castro, G. and Roig, P. and Rosell, I.},
  title         = {{Radiative corrections to \ensuremath{\tau}\textrightarrow{}\ensuremath{\pi}(K)\ensuremath{\nu}\ensuremath{\tau}[\ensuremath{\gamma}]: A reliable new physics test}},
  eprint        = {2107.04603},
  archiveprefix = {arXiv},
  primaryclass  = {hep-ph},
  doi           = {10.1103/PhysRevD.104.L091502},
  journal       = {Phys. Rev. D},
  volume        = {104},
  number        = {9},
  pages         = {L091502},
  year          = {2021}
}

@article{HeavyFlavorAveragingGroupHFLAV:2024ctg,
  author        = {Banerjee, Swagato and others},
  collaboration = {Heavy Flavor Averaging Group (HFLAV)},
  title         = {{Averages of $b$-hadron, $c$-hadron, and $\tau$-lepton properties as of 2023}},
  eprint        = {2411.18639},
  archiveprefix = {arXiv},
  primaryclass  = {hep-ex},
  month         = {11},
  year          = {2024}
}

@article{Erler:2002mv,
  author        = {Erler, Jens},
  title         = {{Electroweak radiative corrections to semileptonic tau decays}},
  eprint        = {hep-ph/0211345},
  archiveprefix = {arXiv},
  journal       = {Rev. Mex. Fis.},
  volume        = {50},
  pages         = {200--202},
  year          = {2004}
}

@article{Beneke:2002jn,
  author        = {Beneke, Martin and Neubert, Matthias},
  title         = {{Flavor singlet B decay amplitudes in QCD factorization}},
  eprint        = {hep-ph/0210085},
  archiveprefix = {arXiv},
  reportnumber  = {CLNS-02-1802, PITHA-02-14},
  doi           = {10.1016/S0550-3213(02)01091-X},
  journal       = {Nucl. Phys. B},
  volume        = {651},
  pages         = {225--248},
  year          = {2003}
}

@article{Belle:2023ziz,
  author        = {Tsuzuki, N. and others},
  collaboration = {Belle},
  title         = {{Search for lepton-flavor-violating \ensuremath{\tau} decays into a lepton and a vector meson using the full Belle data sample}},
  eprint        = {2301.03768},
  archiveprefix = {arXiv},
  primaryclass  = {hep-ex},
  reportnumber  = {Belle Preprint 2022-33; KEK Preprint 2022-46},
  doi           = {10.1007/JHEP06(2023)118},
  journal       = {JHEP},
  volume        = {06},
  pages         = {118},
  year          = {2023}
}

@article{Becirevic:2013bsa,
  author        = {Be\v{c}irevi\'c, Damir and Duplan\v{c}i\'c, Goran and Klajn, Bruno and Meli\'c, Bla\v{z}enka and Sanfilippo, Francesco},
  title         = {{Lattice QCD and QCD sum rule determination of the decay constants of $\eta_c$, J/$\psi$ and $h_c$ states}},
  eprint        = {1312.2858},
  archiveprefix = {arXiv},
  primaryclass  = {hep-ph},
  reportnumber  = {LPT-13-90},
  doi           = {10.1016/j.nuclphysb.2014.03.024},
  journal       = {Nucl. Phys. B},
  volume        = {883},
  pages         = {306--327},
  year          = {2014}
}

@article{Braun:2016wnx,
  author        = {Braun, Vladimir M. and others},
  title         = {{The \ensuremath{\rho}-meson light-cone distribution amplitudes from lattice QCD}},
  eprint        = {1612.02955},
  archiveprefix = {arXiv},
  primaryclass  = {hep-lat},
  reportnumber  = {LTH-1108},
  doi           = {10.1007/JHEP04(2017)082},
  journal       = {JHEP},
  volume        = {04},
  pages         = {082},
  year          = {2017}
}

@article{Chen:2020qma,
  author        = {Chen, Ying and Chiu, Wei-Feng and Gong, Ming and Liu, Zhaofeng and Ma, Yunheng},
  collaboration = {\ensuremath{\chi}QCD},
  title         = {{Charmed and $\phi$ meson decay constants from 2+1-flavor lattice QCD}},
  eprint        = {2008.05208},
  archiveprefix = {arXiv},
  primaryclass  = {hep-lat},
  doi           = {10.1088/1674-1137/abcd8f},
  journal       = {Chin. Phys. C},
  volume        = {45},
  number        = {2},
  pages         = {023109},
  year          = {2021}
}

@article{Bhagwat:2006pu,
  author        = {Bhagwat, M. S. and Maris, P.},
  title         = {{Vector meson form factors and their quark-mass dependence}},
  eprint        = {nucl-th/0612069},
  archiveprefix = {arXiv},
  doi           = {10.1103/PhysRevC.77.025203},
  journal       = {Phys. Rev. C},
  volume        = {77},
  pages         = {025203},
  year          = {2008}
}

@article{Buras:2014fpa,
    author = "Buras, Andrzej J. and Girrbach-Noe, Jennifer and Niehoff, Christoph and Straub, David M.",
    title = "{$ B\to {K}^{\left(\ast \right)}\nu \overline{\nu} $ decays in the Standard Model and beyond}",
    eprint = "1409.4557",
    archivePrefix = "arXiv",
    primaryClass = "hep-ph",
    reportNumber = "FLAVOUR(267104)-ERC-80",
    doi = "10.1007/JHEP02(2015)184",
    journal = "JHEP",
    volume = "02",
    pages = "184",
    year = "2015"
}

@article{Bause:2023mfe,
    author = "Bause, Rigo and Gisbert, Hector and Hiller, Gudrun",
    title = "{Implications of an enhanced B{\textrightarrow}K{\ensuremath{\nu}}{\ensuremath{\nu}}{\textasciimacron} branching ratio}",
    eprint = "2309.00075",
    archivePrefix = "arXiv",
    primaryClass = "hep-ph",
    doi = "10.1103/PhysRevD.109.015006",
    journal = "Phys. Rev. D",
    volume = "109",
    number = "1",
    pages = "015006",
    year = "2024"
}

@article{Hatton:2021dvg,
  author        = {Hatton, D. and Davies, C. T. H. and Koponen, J. and Lepage, G. P. and Lytle, A. T.},
  title         = {{Bottomonium precision tests from full lattice QCD: Hyperfine splitting, Upsilon{} leptonic width, and b quark contribution to $e^+e^- \rightarrow$ hadrons}},
  eprint        = {2101.08103},
  archiveprefix = {arXiv},
  primaryclass  = {hep-lat},
  doi           = {10.1103/PhysRevD.103.054512},
  journal       = {Phys. Rev. D},
  volume        = {103},
  number        = {5},
  pages         = {054512},
  year          = {2021}
}

@article{Belle:2008tjs,
  author        = {Wei, J. -T. and others},
  collaboration = {Belle},
  title         = {{Search for B ---\ensuremath{>} pi l+ l- Decays at Belle}},
  eprint        = {0804.3656},
  archiveprefix = {arXiv},
  primaryclass  = {hep-ex},
  reportnumber  = {BELLE-2008-10, KEK-2008-4},
  doi           = {10.1103/PhysRevD.78.011101},
  journal       = {Phys. Rev. D},
  volume        = {78},
  pages         = {011101},
  year          = {2008}
}

@article{LHCb:2015hsa,
  author        = {Aaij, Roel and others},
  collaboration = {LHCb},
  title         = {{First measurement of the differential branching fraction and $C\!P$ asymmetry of the $B^\pm\to\pi^\pm\mu^+\mu^-$ decay}},
  eprint        = {1509.00414},
  archiveprefix = {arXiv},
  primaryclass  = {hep-ex},
  reportnumber  = {LHCB-PAPER-2015-035, CERN-PH-EP-2015-219},
  doi           = {10.1007/JHEP10(2015)034},
  journal       = {JHEP},
  volume        = {10},
  pages         = {034},
  year          = {2015}
}

@article{BaBar:2016wgb,
  author        = {Lees, J. P. and others},
  collaboration = {BaBar},
  title         = {{Search for $B^{+}\rightarrow K^{+} \tau^{+}\tau^{-}$ at the BaBar experiment}},
  eprint        = {1605.09637},
  archiveprefix = {arXiv},
  primaryclass  = {hep-ex},
  doi           = {10.1103/PhysRevLett.118.031802},
  journal       = {Phys. Rev. Lett.},
  volume        = {118},
  number        = {3},
  pages         = {031802},
  year          = {2017}
}

@article{DELPHI:2005wxt,
  author        = {Abdallah, J. and others},
  collaboration = {DELPHI},
  title         = {{Measurement and interpretation of fermion-pair production at LEP energies above the Z resonance}},
  eprint        = {hep-ex/0512012},
  archiveprefix = {arXiv},
  reportnumber  = {CERN-PH-EP-2005-045},
  doi           = {10.1140/epjc/s2005-02461-0},
  journal       = {Eur. Phys. J. C},
  volume        = {45},
  pages         = {589--632},
  year          = {2006}
}

@misc{Mathematica,
  author = {Wolfram Research{,} Inc.},
  title  = {Mathematica, {V}ersion 14.2},
  url    = {https://www.wolfram.com/mathematica},
  note   = {Champaign, IL, 2024}
}

@article{Shtabovenko:2016sxi,
  author        = {Shtabovenko, Vladyslav and Mertig, Rolf and Orellana, Frederik},
  title         = {{New Developments in FeynCalc 9.0}},
  eprint        = {1601.01167},
  archiveprefix = {arXiv},
  primaryclass  = {hep-ph},
  reportnumber  = {TUM-EFT-71-15},
  doi           = {10.1016/j.cpc.2016.06.008},
  journal       = {Comput. Phys. Commun.},
  volume        = {207},
  pages         = {432--444},
  year          = {2016}
}

@article{Shtabovenko:2020gxv,
  author        = {Shtabovenko, Vladyslav and Mertig, Rolf and Orellana, Frederik},
  title         = {{FeynCalc 9.3: New features and improvements}},
  eprint        = {2001.04407},
  archiveprefix = {arXiv},
  primaryclass  = {hep-ph},
  reportnumber  = {P3H-20-002, TTP19-020, TUM-EFT 130/19},
  doi           = {10.1016/j.cpc.2020.107478},
  journal       = {Comput. Phys. Commun.},
  volume        = {256},
  pages         = {107478},
  year          = {2020}
}

@article{Shtabovenko:2023idz,
  author        = {Shtabovenko, Vladyslav and Mertig, Rolf and Orellana, Frederik},
  title         = {{FeynCalc 10: Do multiloop integrals dream of computer codes?}},
  eprint        = {2312.14089},
  archiveprefix = {arXiv},
  primaryclass  = {hep-ph},
  reportnumber  = {P3H-23-089, TTP23-056, SI-HEP-2023-27},
  doi           = {10.1016/j.cpc.2024.109357},
  journal       = {Comput. Phys. Commun.},
  volume        = {306},
  pages         = {109357},
  year          = {2025}
}

@article{Alwall:2014hca,
  author        = {Alwall, J. and Frederix, R. and Frixione, S. and Hirschi, V. and Maltoni, F. and Mattelaer, O. and Shao, H. -S. and Stelzer, T. and Torrielli, P. and Zaro, M.},
  title         = {{The automated computation of tree-level and next-to-leading order differential cross sections, and their matching to parton shower simulations}},
  eprint        = {1405.0301},
  archiveprefix = {arXiv},
  primaryclass  = {hep-ph},
  reportnumber  = {CERN-PH-TH-2014-064, CP3-14-18, LPN14-066, MCNET-14-09, ZU-TH-14-14},
  doi           = {10.1007/JHEP07(2014)079},
  journal       = {JHEP},
  volume        = {07},
  pages         = {079},
  year          = {2014}
}

@article{NNPDF:2021njg,
  author        = {Ball, Richard D. and others},
  collaboration = {NNPDF},
  title         = {{The path to proton structure at 1\% accuracy}},
  eprint        = {2109.02653},
  archiveprefix = {arXiv},
  primaryclass  = {hep-ph},
  reportnumber  = {Edinburgh 2021/12, Nikhef-2021-013, TIF-UNIMI-2021-11},
  doi           = {10.1140/epjc/s10052-022-10328-7},
  journal       = {Eur. Phys. J. C},
  volume        = {82},
  number        = {5},
  pages         = {428},
  year          = {2022}
}

@article{Cruz-Martinez:2024cbz,
  author        = {Cruz-Martinez, Juan and Forte, Stefano and Laurenti, Niccolo and Rabemananjara, Tanjona R. and Rojo, Juan},
  title         = {{LO, NLO, and NNLO parton distributions for LHC event generators}},
  eprint        = {2406.12961},
  archiveprefix = {arXiv},
  primaryclass  = {hep-ph},
  reportnumber  = {CERN-TH-2024-080, TIF-UNIMI-2024-4},
  doi           = {10.1007/JHEP09(2024)088},
  journal       = {JHEP},
  volume        = {09},
  pages         = {088},
  year          = {2024}
}

@article{Buckley:2014ana,
  author        = {Buckley, Andy and Ferrando, James and Lloyd, Stephen and Nordstr\"om, Karl and Page, Ben and R\"ufenacht, Martin and Sch\"onherr, Marek and Watt, Graeme},
  title         = {{LHAPDF6: parton density access in the LHC precision era}},
  eprint        = {1412.7420},
  archiveprefix = {arXiv},
  primaryclass  = {hep-ph},
  reportnumber  = {GLAS-PPE-2014-05, MCNET-14-29, IPPP-14-111, DCPT-14-222},
  doi           = {10.1140/epjc/s10052-015-3318-8},
  journal       = {Eur. Phys. J. C},
  volume        = {75},
  pages         = {132},
  year          = {2015}
}

@article{Bierlich:2022pfr,
  author        = {Bierlich, Christian and others},
  title         = {{A comprehensive guide to the physics and usage of PYTHIA 8.3}},
  eprint        = {2203.11601},
  archiveprefix = {arXiv},
  primaryclass  = {hep-ph},
  reportnumber  = {LU-TP 22-16, MCNET-22-04, FERMILAB-PUB-22-227-SCD},
  doi           = {10.21468/SciPostPhysCodeb.8},
  journal       = {SciPost Phys. Codeb.},
  volume        = {2022},
  pages         = {8},
  year          = {2022}
}

@article{deFavereau:2013fsa,
  author        = {de Favereau, J. and Delaere, C. and Demin, P. and Giammanco, A. and Lema\^\i{}tre, V. and Mertens, A. and Selvaggi, M.},
  collaboration = {DELPHES 3},
  title         = {{DELPHES 3, A modular framework for fast simulation of a generic collider experiment}},
  eprint        = {1307.6346},
  archiveprefix = {arXiv},
  primaryclass  = {hep-ex},
  doi           = {10.1007/JHEP02(2014)057},
  journal       = {JHEP},
  volume        = {02},
  pages         = {057},
  year          = {2014}
}

@article{Korner:1990ri,
  author       = {Korner, J. G. and Schilcher, K. and Wirbel, M. and Wu, Y. L.},
  title        = {{A Detailed Analysis of $D$, $D(s$) and $B$ Meson Transition Form-factors and the Determination of $V (c b$)}},
  reportnumber = {MZ-TH-90-09, DO-TH-90-7},
  doi          = {10.1007/BF01614702},
  journal      = {Z. Phys. C},
  volume       = {48},
  pages        = {663--672},
  year         = {1990}
}

@article{Wu:2006rd,
  author        = {Wu, Yue-Liang and Zhong, Ming and Zuo, Ya-Bing},
  title         = {{B(s), D(s) ---\ensuremath{>} pi, K, eta, rho, K*, omega, phi Transition Form Factors and Decay Rates with Extraction of the CKM parameters |V(ub)|, |V(cs)|, |V(cd)|}},
  eprint        = {hep-ph/0604007},
  archiveprefix = {arXiv},
  doi           = {10.1142/S0217751X06033209},
  journal       = {Int. J. Mod. Phys. A},
  volume        = {21},
  pages         = {6125--6172},
  year          = {2006}
}

@article{Felkl:2021uxi,
  author        = {Felkl, Tobias and Li, Sze Lok and Schmidt, Michael A.},
  title         = {{A tale of invisibility: constraints on new physics in b \textrightarrow{} s\ensuremath{\nu}\ensuremath{\nu}}},
  eprint        = {2111.04327},
  archiveprefix = {arXiv},
  primaryclass  = {hep-ph},
  reportnumber  = {CPPC-2021-13},
  doi           = {10.1007/JHEP12(2021)118},
  journal       = {JHEP},
  volume        = {12},
  pages         = {118},
  year          = {2021}
}

@article{Belle-II:2023esi,
  author        = {Adachi, I. and others},
  collaboration = {Belle-II},
  title         = {{Evidence for B+\textrightarrow{}K+\ensuremath{\nu}\ensuremath{\nu}\textasciimacron{} decays}},
  eprint        = {2311.14647},
  archiveprefix = {arXiv},
  primaryclass  = {hep-ex},
  reportnumber  = {Belle II Preprint 2023-017, KEK Preprint 2023-35},
  doi           = {10.1103/PhysRevD.109.112006},
  journal       = {Phys. Rev. D},
  volume        = {109},
  number        = {11},
  pages         = {112006},
  year          = {2024}
}

@article{Altmannshofer:2008dz,
  author        = {Altmannshofer, Wolfgang and Ball, Patricia and Bharucha, Aoife and Buras, Andrzej J. and Straub, David M. and Wick, Michael},
  title         = {{Symmetries and Asymmetries of $B \to K^{*} \mu^{+} \mu^{-}$ Decays in the Standard Model and Beyond}},
  eprint        = {0811.1214},
  archiveprefix = {arXiv},
  primaryclass  = {hep-ph},
  reportnumber  = {IPPP-08-58, DCPT-08-116, TUM-HEP-696-08},
  doi           = {10.1088/1126-6708/2009/01/019},
  journal       = {JHEP},
  volume        = {01},
  pages         = {019},
  year          = {2009}
}

@article{Matias:2012xw,
  author        = {Matias, Joaquim and Mescia, Federico and Ramon, Marc and Virto, Javier},
  title         = {{Complete Anatomy of $\bar{B}_d -> \bar{K}^{* 0} (-> K \pi)l^+l^-$ and its angular distribution}},
  eprint        = {1202.4266},
  archiveprefix = {arXiv},
  primaryclass  = {hep-ph},
  reportnumber  = {UAB-FT-706, ICCUB-12-076, ECM-UB-68},
  doi           = {10.1007/JHEP04(2012)104},
  journal       = {JHEP},
  volume        = {04},
  pages         = {104},
  year          = {2012}
}

@article{Descotes-Genon:2013vna,
  author        = {Descotes-Genon, Sebastien and Hurth, Tobias and Matias, Joaquim and Virto, Javier},
  title         = {{Optimizing the basis of $B\to K^*ll$ observables in the full kinematic range}},
  eprint        = {1303.5794},
  archiveprefix = {arXiv},
  primaryclass  = {hep-ph},
  reportnumber  = {UAB-FT-732, LPT-ORSAY-13-24, MITP-13-020},
  doi           = {10.1007/JHEP05(2013)137},
  journal       = {JHEP},
  volume        = {05},
  pages         = {137},
  year          = {2013}
}

@article{Straub:2018kue,
  author        = {Straub, David M.},
  title         = {{flavio: a Python package for flavour and precision phenomenology in the Standard Model and beyond}},
  eprint        = {1810.08132},
  archiveprefix = {arXiv},
  primaryclass  = {hep-ph},
  month         = {10},
  year          = {2018}
}

@article{BaBar:2013qaj,
  author        = {Lees, J. P. and others},
  collaboration = {BaBar},
  title         = {{Search for the rare decays $B → \pi\ell^+\ell^-$ and $B^0 → \eta\ell^+\ell^-$}},
  eprint        = {1303.6010},
  archiveprefix = {arXiv},
  primaryclass  = {hep-ex},
  reportnumber  = {BABAR-PUB-12-33, SLAC-PUB-15401, BABAR-PUB-12-033},
  doi           = {10.1103/PhysRevD.88.032012},
  journal       = {Phys. Rev. D},
  volume        = {88},
  number        = {3},
  pages         = {032012},
  year          = {2013}
}

@article{Soni:2020bvu,
  author        = {Soni, Nakul R. and Issadykov, Aidos and Gadaria, Akshay N. and Patel, Janaki J. and Pandya, Jignesh N.},
  title         = {{Rare $b \rightarrow d$ decays in covariant confined quark model}},
  eprint        = {2008.07202},
  archiveprefix = {arXiv},
  primaryclass  = {hep-ph},
  doi           = {10.1140/epja/s10050-022-00685-y},
  journal       = {Eur. Phys. J. A},
  volume        = {58},
  number        = {3},
  pages         = {39},
  year          = {2022}
}

@article{LHCb:2014cxe,
  author        = {Aaij, R. and others},
  collaboration = {LHCb},
  title         = {{Differential branching fractions and isospin asymmetries of $B \to K^{(*)} \mu^+ \mu^-$ decays}},
  eprint        = {1403.8044},
  archiveprefix = {arXiv},
  primaryclass  = {hep-ex},
  reportnumber  = {LHCB-PAPER-2014-006, CERN-PH-EP-2014-055},
  doi           = {10.1007/JHEP06(2014)133},
  journal       = {JHEP},
  volume        = {06},
  pages         = {133},
  year          = {2014}
}

@article{LHCb:2025pxz,
    author = "Aaij, R. and others",
    collaboration = "LHCb",
    title = "{Angular analysis of B$^{0}${\textrightarrow} K$^{*0}$e$^{+}$e$^{-}$ decays}",
    eprint = "2502.10291",
    archivePrefix = "arXiv",
    primaryClass = "hep-ex",
    reportNumber = "LHCb-PAPER-2024-022, CERN-EP-2025-001",
    doi = "10.1007/JHEP06(2025)140",
    journal = "JHEP",
    volume = "06",
    pages = "140",
    year = "2025"
}

@article{LHCb:2020lmf,
  author        = {Aaij, Roel and others},
  collaboration = {LHCb},
  title         = {{Measurement of $CP$-Averaged Observables in the $B^{0}\rightarrow K^{*0}\mu^{+}\mu^{-}$ Decay}},
  eprint        = {2003.04831},
  archiveprefix = {arXiv},
  primaryclass  = {hep-ex},
  reportnumber  = {LHCb-PAPER-2020-002, CERN-EP-2020-027},
  doi           = {10.1103/PhysRevLett.125.011802},
  journal       = {Phys. Rev. Lett.},
  volume        = {125},
  number        = {1},
  pages         = {011802},
  year          = {2020}
}

@article{Belle-II:2025lwo,
    author = "Adachi, I. and others",
    collaboration = "Belle-II",
    title = "{Search for B0{\textrightarrow}K*0{\ensuremath{\tau}}+{\ensuremath{\tau}}- Decays at the Belle II Experiment}",
    eprint = "2504.10042",
    archivePrefix = "arXiv",
    primaryClass = "hep-ex",
    reportNumber = "Belle II Preprint 2025-010; KEK Preprint 2025-8",
    doi = "10.1103/v1q3-9dy8",
    journal = "Phys. Rev. Lett.",
    volume = "135",
    number = "15",
    pages = "151801",
    year = "2025"
}

@article{DELPHI:1996ohp,
  author        = {Adam, W. and others},
  collaboration = {DELPHI},
  title         = {{Study of rare b decays with the DELPHI detector at LEP}},
  reportnumber  = {CERN-PPE-96-067, CERN-PPE-96-67},
  doi           = {10.1007/s002880050238},
  journal       = {Z. Phys. C},
  volume        = {72},
  pages         = {207--220},
  year          = {1996}
}

@article{Castro:2016jjv,
  author        = {Castro, Nuno and Erdmann, Johannes and Grunwald, Cornelius and Kr\"oninger, Kevin and Rosien, Nils-Arne},
  title         = {{EFTfitter---A tool for interpreting measurements in the context of effective field theories}},
  eprint        = {1605.05585},
  archiveprefix = {arXiv},
  primaryclass  = {hep-ex},
  doi           = {10.1140/epjc/s10052-016-4280-9},
  journal       = {Eur. Phys. J. C},
  volume        = {76},
  number        = {8},
  pages         = {432},
  year          = {2016}
}

@article{Caldwell:2008fw,
  author        = {Caldwell, Allen and Kollar, Daniel and Kroninger, Kevin},
  title         = {{BAT: The Bayesian Analysis Toolkit}},
  eprint        = {0808.2552},
  archiveprefix = {arXiv},
  primaryclass  = {physics.data-an},
  doi           = {10.1016/j.cpc.2009.06.026},
  journal       = {Comput. Phys. Commun.},
  volume        = {180},
  pages         = {2197--2209},
  year          = {2009}
}

@article{DAmbrosio:2022kvb,
  author        = {D'Ambrosio, G. and Iyer, A. M. and Mahmoudi, F. and Neshatpour, S.},
  title         = {{Anatomy of kaon decays and prospects for lepton flavour universality violation}},
  eprint        = {2206.14748},
  archiveprefix = {arXiv},
  primaryclass  = {hep-ph},
  reportnumber  = {CERN-TH-2022-101},
  doi           = {10.1007/JHEP09(2022)148},
  journal       = {JHEP},
  volume        = {09},
  pages         = {148},
  year          = {2022}
}

@article{Parrott:2022zte,
  author        = {Parrott, W. G. and Bouchard, C. and Davies, C. T. H.},
  collaboration = {HPQCD},
  title         = {{Standard Model predictions for B\textrightarrow{}K\ensuremath{\ell}+\ensuremath{\ell}-, B\textrightarrow{}K\ensuremath{\ell}1-\ensuremath{\ell}2+ and B\textrightarrow{}K\ensuremath{\nu}\ensuremath{\nu}\textasciimacron{} using form factors from Nf=2+1+1 lattice QCD}},
  eprint        = {2207.13371},
  archiveprefix = {arXiv},
  primaryclass  = {hep-ph},
  doi           = {10.1103/PhysRevD.107.014511},
  journal       = {Phys. Rev. D},
  volume        = {107},
  number        = {1},
  pages         = {014511},
  year          = {2023},
  note          = {[Erratum: Phys.Rev.D 107, 119903 (2023)]}
}

@article{Kindra:2018ayz,
  author        = {Kindra, Bharti and Mahajan, Namit},
  title         = {{Predictions of angular observables for $\bar{B}_s\to K^{\ast}\ell\ell$ and $\bar{B}\to \rho\ell\ell$ in the standard model}},
  eprint        = {1803.05876},
  archiveprefix = {arXiv},
  primaryclass  = {hep-ph},
  doi           = {10.1103/PhysRevD.98.094012},
  journal       = {Phys. Rev. D},
  volume        = {98},
  number        = {9},
  pages         = {094012},
  year          = {2018}
}

@article{LHCb:2018rym,
  author        = {Aaij, Roel and others},
  collaboration = {LHCb},
  title         = {{Evidence for the decay $ {B}_S^0\to {\overline{K}}^{\ast 0}{\mu}^{+}{\mu}^{-} $}},
  eprint        = {1804.07167},
  archiveprefix = {arXiv},
  primaryclass  = {hep-ex},
  reportnumber  = {LHCb-PAPER-2018-004, CERN-EP-2018-059, LHCB-PAPER-2018-004},
  doi           = {10.1007/JHEP07(2018)020},
  journal       = {JHEP},
  volume        = {07},
  pages         = {020},
  year          = {2018}
}

@article{LHCb:2021xxq,
  author        = {Aaij, Roel and others},
  collaboration = {LHCb},
  title         = {{Angular analysis of the rare decay $ {B}_s^0 $\textrightarrow{} \ensuremath{\phi}\ensuremath{\mu}$^{+}$\ensuremath{\mu}$^{-}$}},
  eprint        = {2107.13428},
  archiveprefix = {arXiv},
  primaryclass  = {hep-ex},
  reportnumber  = {LHCb-PAPER-2021-022, CERN-EP-2021-138},
  doi           = {10.1007/JHEP11(2021)043},
  journal       = {JHEP},
  volume        = {11},
  pages         = {043},
  year          = {2021}
}

@article{LHCb:2025rfy,
    author = "Aaij, Roel and others",
    collaboration = "LHCb",
    title = "{Angular analysis of the decay $ {B}_s^0 ${\textrightarrow} {\ensuremath{\phi}}e$^{+}$e$^{-}$}",
    eprint = "2504.06346",
    archivePrefix = "arXiv",
    primaryClass = "hep-ex",
    reportNumber = "LHCb-PAPER-2025-006, CERN-EP-2025-064",
    doi = "10.1007/JHEP07(2025)069",
    journal = "JHEP",
    volume = "07",
    pages = "069",
    year = "2025"
}

@article{Alguero:2023jeh,
  author        = {Alguer\'o, Marcel and Biswas, Aritra and Capdevila, Bernat and Descotes-Genon, S\'ebastien and Matias, Joaquim and Novoa-Brunet, Mart\'\i{}n},
  title         = {{To (b)e or not to (b)e: no electrons at LHCb}},
  eprint        = {2304.07330},
  archiveprefix = {arXiv},
  primaryclass  = {hep-ph},
  reportnumber  = {BARI-TH/23-747},
  doi           = {10.1140/epjc/s10052-023-11824-0},
  journal       = {Eur. Phys. J. C},
  volume        = {83},
  number        = {7},
  pages         = {648},
  year          = {2023}
}

@article{LHCb:2020gog,
  author        = {Aaij, Roel and others},
  collaboration = {LHCb},
  title         = {{Angular Analysis of the  $B^{+}\rightarrow K^{\ast+}\mu^{+}\mu^{-}$ Decay}},
  eprint        = {2012.13241},
  archiveprefix = {arXiv},
  primaryclass  = {hep-ex},
  reportnumber  = {LHCb-PAPER-2020-041, CERN-EP-2020-239},
  doi           = {10.1103/PhysRevLett.126.161802},
  journal       = {Phys. Rev. Lett.},
  volume        = {126},
  number        = {16},
  pages         = {161802},
  year          = {2021}
}

@article{Gubernari:2022hxn,
  author        = {Gubernari, Nico and Reboud, M\'eril and van Dyk, Danny and Virto, Javier},
  title         = {{Improved theory predictions and global analysis of exclusive $b \to s\mu^+\mu^-$ processes}},
  eprint        = {2206.03797},
  archiveprefix = {arXiv},
  primaryclass  = {hep-ph},
  reportnumber  = {SI-HEP-2022-12, P3H-22-059, TUM-HEP-1401/22, EOS-2022-02},
  doi           = {10.1007/JHEP09(2022)133},
  journal       = {JHEP},
  volume        = {09},
  pages         = {133},
  year          = {2022}
}

@article{Bause:2022rrs,
  author        = {Bause, Rigo and Gisbert, Hector and Golz, Marcel and Hiller, Gudrun},
  title         = {{Model-independent analysis of $b \rightarrow d$ processes}},
  eprint        = {2209.04457},
  archiveprefix = {arXiv},
  primaryclass  = {hep-ph},
  reportnumber  = {DO-TH 21/30},
  doi           = {10.1140/epjc/s10052-023-11586-9},
  journal       = {Eur. Phys. J. C},
  volume        = {83},
  number        = {5},
  pages         = {419},
  year          = {2023}
}

@article{FermilabLattice:2015cdh,
  author        = {Bailey, Jon A. and others},
  collaboration = {Fermilab Lattice, MILC},
  title         = {{$B\to\pi\ell\ell$ form factors for new-physics searches from lattice QCD}},
  eprint        = {1507.01618},
  archiveprefix = {arXiv},
  primaryclass  = {hep-ph},
  reportnumber  = {FERMILAB-PUB-15-288-T},
  doi           = {10.1103/PhysRevLett.115.152002},
  journal       = {Phys. Rev. Lett.},
  volume        = {115},
  number        = {15},
  pages         = {152002},
  year          = {2015}
}

@article{Schulz:2020ebm,
  author        = {Schulz, Oliver and Beaujean, Frederik and Caldwell, Allen and Grunwald, Cornelius and Hafych, Vasyl and Kr\"oninger, Kevin and La Cagnina, Salvatore and R\"ohrig, Lars and Shtembari, Lolian},
  title         = {{BAT.jl: A Julia-Based Tool for Bayesian Inference}},
  eprint        = {2008.03132},
  archiveprefix = {arXiv},
  primaryclass  = {stat.CO},
  doi           = {10.1007/s42979-021-00626-4},
  journal       = {SN Comput. Sci.},
  volume        = {2},
  number        = {3},
  pages         = {1--17},
  year          = {2021}
}

@article{Mertig:1990an,
  author       = {Mertig, R. and Bohm, M. and Denner, Ansgar},
  title        = {{FEYN CALC: Computer algebraic calculation of Feynman amplitudes}},
  reportnumber = {PRINT-90-0639 (WURZBURG)},
  doi          = {10.1016/0010-4655(91)90130-D},
  journal      = {Comput. Phys. Commun.},
  volume       = {64},
  pages        = {345--359},
  year         = {1991}
}

@article{Bharucha:2015bzk,
  author        = {Bharucha, Aoife and Straub, David M. and Zwicky, Roman},
  title         = {{$B\to V\ell^+\ell^-$ in the Standard Model from light-cone sum rules}},
  eprint        = {1503.05534},
  archiveprefix = {arXiv},
  primaryclass  = {hep-ph},
  reportnumber  = {TUM-HEP-957-14, CP3-Origins-2015-010, DIAS-2015-10},
  doi           = {10.1007/JHEP08(2016)098},
  journal       = {JHEP},
  volume        = {08},
  pages         = {098},
  year          = {2016}
}

@article{Gubernari:2018wyi,
  author        = {Gubernari, Nico and Kokulu, Ahmet and van Dyk, Danny},
  title         = {{$B\to P$ and $B\to V$ Form Factors from $B$-Meson Light-Cone Sum Rules beyond Leading Twist}},
  eprint        = {1811.00983},
  archiveprefix = {arXiv},
  primaryclass  = {hep-ph},
  reportnumber  = {EOS-2018-02, TUM-HEP-1172/18},
  doi           = {10.1007/JHEP01(2019)150},
  journal       = {JHEP},
  volume        = {01},
  pages         = {150},
  year          = {2019}
}

@article{McLean:2019qcx,
  author        = {McLean, E. and Davies, C. T. H. and Koponen, J. and Lytle, A. T.},
  title         = {{$B_s\to D_s \ell\nu$ Form Factors for the full $q^2$ range from Lattice QCD with non-perturbatively normalized currents}},
  eprint        = {1906.00701},
  archiveprefix = {arXiv},
  primaryclass  = {hep-lat},
  doi           = {10.1103/PhysRevD.101.074513},
  journal       = {Phys. Rev. D},
  volume        = {101},
  number        = {7},
  pages         = {074513},
  year          = {2020}
}

@article{Bali:2014pva,
  author        = {Bali, Gunnar S. and Collins, Sara and D\"urr, Stephan and Kanamori, Issaku},
  title         = {{$D_s \rightarrow \eta, \eta'$ semileptonic decay form factors with disconnected quark loop contributions}},
  eprint        = {1406.5449},
  archiveprefix = {arXiv},
  primaryclass  = {hep-lat},
  doi           = {10.1103/PhysRevD.91.014503},
  journal       = {Phys. Rev. D},
  volume        = {91},
  number        = {1},
  pages         = {014503},
  year          = {2015}
}

@article{FermilabLattice:2022gku,
  author        = {Bazavov, Alexei and others},
  collaboration = {Fermilab Lattice, MILC},
  title         = {{D-meson semileptonic decays to pseudoscalars from four-flavor lattice QCD}},
  eprint        = {2212.12648},
  archiveprefix = {arXiv},
  primaryclass  = {hep-lat},
  reportnumber  = {MIT-CTP/5513, FERMILAB-PUB-22-943-T},
  doi           = {10.1103/PhysRevD.107.094516},
  journal       = {Phys. Rev. D},
  volume        = {107},
  number        = {9},
  pages         = {094516},
  year          = {2023}
}

@article{Ali:2023kua,
  author        = {Ali, Md Isha and Chattopadhyay, Utpal and Rajeev, N. and Roy, Joydeep},
  title         = {{SMEFT analysis of charged lepton flavor violating B-meson decays}},
  eprint        = {2312.05071},
  archiveprefix = {arXiv},
  primaryclass  = {hep-ph},
  doi           = {10.1103/PhysRevD.109.075028},
  journal       = {Phys. Rev. D},
  volume        = {109},
  number        = {7},
  pages         = {075028},
  year          = {2024}
}

@article{ATLAS:2025yww,
  author        = {Aad, Georges and others},
  collaboration = {ATLAS},
  title         = {{Measurement of high-mass $t\bar{t}\ell^{+}\ell^{-}$ production and lepton flavour universality-inspired effective field theory interpretations at $\sqrt{s}=13$ TeV with the ATLAS detector}},
  eprint        = {2504.05919},
  archiveprefix = {arXiv},
  primaryclass  = {hep-ex},
  reportnumber  = {CERN-EP-2025-053},
  month         = {4},
  year          = {2025}
}

@article{Charles:2004jd,
  author        = {Charles, J. and Hocker, Andreas and Lacker, H. and Laplace, S. and Le Diberder, F. R. and Malcles, J. and Ocariz, J. and Pivk, M. and Roos, L.},
  collaboration = {CKMfitter Group},
  title         = {{CP violation and the CKM matrix: Assessing the impact of the asymmetric $B$ factories}},
  eprint        = {hep-ph/0406184},
  archiveprefix = {arXiv},
  reportnumber  = {CPT-2004-P-030, LAL-04-21, LAPP-EXP-2004-01, LPNHE-2004-01},
  doi           = {10.1140/epjc/s2005-02169-1},
  journal       = {Eur. Phys. J. C},
  volume        = {41},
  number        = {1},
  pages         = {1--131},
  year          = {2005}
}

@article{Buchner:2021cql,
  author        = {Buchner, Johannes},
  title         = {{UltraNest -- a robust, general purpose Bayesian inference engine}},
  eprint        = {2101.09604},
  archiveprefix = {arXiv},
  primaryclass  = {stat.CO},
  reportnumber  = {10.21105/joss.03001},
  month         = {1},
  year          = {2021}
}

@article{Vihola:2012,
  author     = {Vihola, Matti},
  title      = {Robust adaptive Metropolis algorithm with coerced acceptance rate},
  year       = {2012},
  issue_date = {September 2012},
  publisher  = {Kluwer Academic Publishers},
  address    = {USA},
  volume     = {22},
  number     = {5},
  issn       = {0960-3174},
  url        = {https://doi.org/10.1007/s11222-011-9269-5},
  doi        = {10.1007/s11222-011-9269-5},
  journal    = {Statistics and Computing},
  month      = sep,
  pages      = {997–1008},
  numpages   = {12}
}

@article{Belle-II:2018jsg,
  author        = {Altmannshofer, W. and others},
  editor        = {Kou, E. and Urquijo, P.},
  collaboration = {Belle-II},
  title         = {{The Belle II Physics Book}},
  eprint        = {1808.10567},
  archiveprefix = {arXiv},
  primaryclass  = {hep-ex},
  reportnumber  = {KEK Preprint 2018-27, BELLE2-PUB-PH-2018-001, FERMILAB-PUB-18-398-T, JLAB-THY-18-2780, INT-PUB-18-047, UWThPh 2018-26},
  doi           = {10.1093/ptep/ptz106},
  journal       = {PTEP},
  volume        = {2019},
  number        = {12},
  pages         = {123C01},
  year          = {2019},
  note          = {[Erratum: PTEP 2020, 029201 (2020)]}
}

@article{BESIII:2021slf,
  author        = {Ablikim, M. and others},
  collaboration = {BESIII},
  title         = {{Search for the decay $D^{0} \to \pi^{0} \nu \bar{\nu}$}},
  eprint        = {2112.14236},
  archiveprefix = {arXiv},
  primaryclass  = {hep-ex},
  doi           = {10.1103/PhysRevD.105.L071102},
  journal       = {Phys. Rev. D},
  volume        = {105},
  number        = {7},
  pages         = {L071102},
  year          = {2022}
}

@article{Bause:2021cna,
  author        = {Bause, Rigo and Gisbert, Hector and Golz, Marcel and Hiller, Gudrun},
  title         = {{Interplay of dineutrino modes with semileptonic rare B-decays}},
  eprint        = {2109.01675},
  archiveprefix = {arXiv},
  primaryclass  = {hep-ph},
  reportnumber  = {DO-TH 21/17},
  doi           = {10.1007/JHEP12(2021)061},
  journal       = {JHEP},
  volume        = {12},
  pages         = {061},
  year          = {2021}
}

@article{Aliberti:2025beg,
    author = "Aliberti, R. and others",
    title = "{The anomalous magnetic moment of the muon in the Standard Model: an update}",
    eprint = "2505.21476",
    archivePrefix = "arXiv",
    primaryClass = "hep-ph",
    reportNumber = "CERN-TH-2025-101, FERMILAB-PUB-25-0344-T, INT-PUB-25-015, IPARCOS-UCM-25-029, KEK Preprint 2025-22, LTH 1403, MITP-25-037, UWThPh 2025-15, UWThPh
  2025-15, ZU-TH 37/25, IPARCOS-UCM-25-029",
    doi = "10.1016/j.physrep.2025.08.002",
    journal = "Phys. Rept.",
    volume = "1143",
    pages = "1--158",
    year = "2025"
}

@misc{CKMfitter:Summer23,
  author       = {{CKMfitter Group}},
  title        = {{Numerical results: CKMfitter Summer 2023}},
  howpublished = {\url{http://ckmfitter.in2p3.fr/www/results/plots_summer23/num/ckmEval_results_summer23.html}},
  month        = {Jul},
  year         = {2023},
  note         = {Accessed 2025-06-30}
}

@article{Aoude:2020dwv,
    author = "Aoude, Rafael and Hurth, Tobias and Renner, Sophie and Shepherd, William",
    title = "{The impact of flavour data on global fits of the MFV SMEFT}",
    eprint = "2003.05432",
    archivePrefix = "arXiv",
    primaryClass = "hep-ph",
    doi = "10.1007/JHEP12(2020)113",
    journal = "JHEP",
    volume = "12",
    pages = "113",
    year = "2020"
}

@article{Bruggisser:2021duo,
    author = {Bruggisser, Sebastian and Sch{\"a}fer, Ruth and van Dyk, Danny and Westhoff, Susanne},
    title = "{The Flavor of UV Physics}",
    eprint = "2101.07273",
    archivePrefix = "arXiv",
    primaryClass = "hep-ph",
    doi = "10.1007/JHEP05(2021)257",
    journal = "JHEP",
    volume = "05",
    pages = "257",
    year = "2021"
}

@article{Greljo:2022jac,
    author = "Greljo, Admir and Salko, Jakub and Smolkovi{\v{c}}, Aleks and Stangl, Peter",
    title = "{Rare b decays meet high-mass Drell-Yan}",
    eprint = "2212.10497",
    archivePrefix = "arXiv",
    primaryClass = "hep-ph",
    reportNumber = "CERN-TH-2023-037",
    doi = "10.1007/JHEP05(2023)087",
    journal = "JHEP",
    volume = "05",
    pages = "087",
    year = "2023"
}

@article{Allwicher:2022gkm,
    author = "Allwicher, Lukas and Faroughy, Darius A. and Jaffredo, Florentin and Sumensari, Olcyr and Wilsch, Felix",
    title = "{Drell-Yan tails beyond the Standard Model}",
    eprint = "2207.10714",
    archivePrefix = "arXiv",
    primaryClass = "hep-ph",
    doi = "10.1007/JHEP03(2023)064",
    journal = "JHEP",
    volume = "03",
    pages = "064",
    year = "2023"
}

@article{Hiller:2025hpf,
    author = "Hiller, Gudrun and Nollen, Lara and Wendler, Daniel",
    title = "{Total Drell{\textendash}Yan in the flavorful SMEFT}",
    eprint = "2502.12250",
    archivePrefix = "arXiv",
    primaryClass = "hep-ph",
    doi = "10.1140/epjc/s10052-025-14349-w",
    journal = "Eur. Phys. J. C",
    volume = "85",
    number = "6",
    pages = "657",
    year = "2025"
}

@article{Muong-2:2025xyk,
    author = "Aguillard, D. P. and others",
    collaboration = "Muon g-2",
    title = "{Measurement of the Positive Muon Anomalous Magnetic Moment to 127~ppb}",
    eprint = "2506.03069",
    archivePrefix = "arXiv",
    primaryClass = "hep-ex",
    reportNumber = "FERMILAB-PUB-25-0364-PPD",
    doi = "10.1103/7clf-sm2v",
    journal = "Phys. Rev. Lett.",
    volume = "135",
    number = "10",
    pages = "101802",
    year = "2025"
}

@article{deBlas:2025xhe,
  author        = {de Blas, J. and Goncalves, A. and Miralles, V. and Reina, L. and Silvestrini, L. and Valli, M.},
  title         = {{Constraining new physics effective interactions via a global fit of electroweak, Drell-Yan, Higgs, top, and flavour observables}},
  eprint        = {2507.06191},
  archiveprefix = {arXiv},
  primaryclass  = {hep-ph},
  month         = {7},
  year          = {2025}
}

@article{Chattopadhyay:2025air,
  author        = {Chattopadhyay, Utpal and Das, Debottam and Puri, Rahul and Roy, Joydeep},
  title         = {{Constraining lepton flavor violating $2q 2\ell$ operators from low-energy cLFV processes}},
  eprint        = {2507.13141},
  archiveprefix = {arXiv},
  primaryclass  = {hep-ph},
  month         = {7},
  year          = {2025}
}
